\documentclass{aa}  
\usepackage{natbib}
\bibpunct{(}{)}{;}{a}{}{,} 
\usepackage{graphicx}
\usepackage{txfonts}
\begin{document}
   \title{X-ray spectral variability of seven LINER nuclei with \emph{XMM}-Newton and \emph{Chandra} data }

   \subtitle{}

   \author{Hern\'{a}ndez-Garc\'{i}a, L.\inst{1}; Gonz\'{a}lez-Mart\'{i}n, O.\inst{2}$^,$\inst{3}; M\'{a}rquez, I. \inst{1}; Masegosa, J.\inst{1}
          }

   \institute{Instituto de Astrof\'{i}sica de Andaluc\'{i}a, CSIC, Glorieta de la Astronom\'{i}a, s/n, 18008 Granada, Spain\\
              \email{lorena@iaa.es}
         \and
             Instituto de Astrof\'{i}sica de Canarias (IAC), C/V\'{i}a Lactea, s/n, 38205 La Laguna, Tenerife, Spain \\
        \vspace*{-0.35cm}
         \and
             Departamento de Astrof\'{i}sica, Universidad de La Laguna (ULL), 38205 La Laguna, Tenerife, Spain \\
             }

   \date{Received XXXX/ accepted YYYY}

\authorrunning{Hern\'{a}ndez-Garc\'{i}a et al.}
\titlerunning{X-ray variability in LINERs}

 
  \abstract
   {One of the most important features in active galactic nuclei (AGN) is the variability of their emission. Variability has been discovered at X-ray, UV, and radio frequencies on time scales from hours to years. Among the AGN family and according to theoretical studies, Low-Ionization Nuclear Emission Line Region (LINER) nuclei would be variable objects on long time scales. }
   {Our purpose is to investigate spectral X-ray variability in LINERs and to understand the nature of these kinds of objects, as well as their accretion mechanism.}
   {\emph{Chandra} and \emph{XMM}-Newton public archives were used to compile X-ray spectra of seven LINER nuclei at different epochs with time scales of years. To search for variability we fit all the spectra from the same object with a set of models, in order to identify the parameters responsible for the variability pattern. We also analyzed the light curves in order to search for short time scale (from hours to days) variability. Whenever possible, UV variability was also studied.}
   {We found spectral variability in four objects (namely NGC\,1052, NGC\,3226, NGC\,4278, and NGC\,4552), with variations mostly related to hard energies (2-10 keV). These variations are due to changes in the soft excess, and/or changes in the absorber, and/or intrinsic variations of the source. These variations were within years, the shorter time scale being found for NGC\,4278 (two months). Another two galaxies (NGC\,4261 and NGC\,5846) seem not to vary. Short time scale variations during individual observations were not found. Our analysis confirms	 the previously reported anticorrelation between the X-ray spectral index, $\Gamma$, and the Eddington ratio, $L_{bol}/L_{Edd}$, and also the correlation between the X-ray to UV flux ratio, $\alpha_{ox}$, and the Eddington ratio, $L_{bol}/L_{Edd}$. These results support an Advection Dominated Accretion Flow (ADAF) as the accretion mechanism in LINERs.}


   \maketitle
%

\section{Introduction}

The plethora of phenomena present in an active galactic nucleus (AGN) shows the existence of energetic processes in the nuclei of galaxies that cannot be attributed to stars \citep{bradley1997}. The unified model for AGN \citep{urrypadovani1995, antonucci1993} represents a scenario where the central, supermassive black hole (SMBH) is surrounded by a dusty torus; and depending on the line of sight of the observer, the AGN appears as type 1 (pole-on view) or type 2 (edge on view). However, even if, broadly speaking, the unified model is a good representation of AGN, there are a number of objects that still cannot be fitted under this scheme. This is the case of Low Ionisation Nuclear Emission Line Regions (LINERs), which are the main purpose of this research.

First defined by \citet{heckman1980}, the characterization of LINERs was made in the optical, where their spectra show strong low-ionization lines such as [OI] $\lambda$ 6300\AA~ and [SII] $\lambda \lambda$ 6717, 6731\AA~ \citep{veilleuxosterbrock1987}. Different options have been proposed to explain the ionization mechanism in LINERs, such as shock-heated gas \citep{dopitasutherland1995}, photoionization by hot stars \citep{terlevichmenlick1985}, by post main sequence stars (Cid-Fernandes et al 2011), or a low luminosity active galactic nuclei \citep{ho1993}. Today, the most accepted option is that they harbour AGN \citep[see][]{ho2008, marquez2012}.

X-ray data for LLAGN offer the most reliable probe of the high-energy spectrum, providing many AGN signatures \citep{donofriomarzianisulentic2012}. In LINERs it can be assessed that an AGN is present when a point-like source is detected at hard X-ray energies \citep{satyapal2004, satyapal2005, dudik2005, ho2008}. The most extensive work has been carried out by \cite{omaira2009a}, who analyzed 82 LINERs with \emph{Chandra} and/or \emph{XMM}-Newton data for single period observations. They found that 60\% of the sample showed a compact nuclear source in the 4.5-8 keV band; a multiwavelenght analysis resulted in about 80\% of the sample showing evidence of AGN-related properties, which is a lower limit since Compton-thick objects were not taken into account.

It is tempting to view LINERs as a scaled-down version of Seyfert galaxies, but in fact they are qualitatively different from their neighbouring class \citep{ho2008}. LINERs have lower luminosities (L$\rm{_{2-10 keV}\approx}$ $\rm{10^{39}-10^{42}}$ $\rm{erg \hspace*{0.1cm} s^{-1}}$), lower Eddington ratios (L$\rm{_{bol}}$/$\rm{L_{Edd}}$ $\rm{\approx}$ $\rm{10^ {-4}}$) and more massive black holes \citep[M$\rm{_{BH}}$ $\rm{\approx}$ $\rm{10^8 M\odot}$][]{eracleous2010b, masegosa2011}.

Variability is one of the main properties that characterizes AGN, most of these being at least mildly variable. When quasars were first discovered in the 1960s \citep{schmidt1963}, one of their key defining properties was their variability. These objects are variable over the entire wavelength range, many of them varying 0.3-0.5 magnitudes over time scales of a few months, and others varying significantly on time scales as short as a few days \citep{bradley1997}. Variability properties seem to correlate with AGN power; in quasars, variations likely result from both accretion disk instability and microlensing, while in Seyfert galaxies the brightness of the nucleus is thought to vary, the broad-line region (BLR) responding to these changes a few weeks later \citep{donofriomarzianisulentic2012}.

At X-ray frequencies many studies have been made to understand variability in Seyfert galaxies. \cite{risaliti2000} studied the highly variable Seyfert 1.8 galaxy NGC\,1365, which was observed for many years with different instruments; they also found variability in the Seyfert galaxy UGC\,4203 using \emph{Chandra} data \citep{risaliti2010}. They suggested a scenario in which the variability is produced by clouds intersecting the line of sight to the observer. \cite{evans2005} found that the 2-10 keV luminosity for NGC\,6251 varied a factor of $\approx$ 5 from 1991 to 2003. More recently, a study made by \cite{caballerogarcia2012} showed that the five Seyfert galaxies studied with \emph{Swift}/BAT showed flux variability on time-scales of 1-2 days. Narrow-Line Seyfert 1 galaxies (NLSy1) also show variability in X-rays \citep{panessa2011, risaliti2011}.

As stated by \cite{ptak1998}, ``LINERs tend to show little or no significant short-term variability (i.e., with timescales less than a day)'' \citep[see also][]{krolik1999}. The first clear evidence of variability in LINERs was reported by \cite{maoz2005} at UV frequencies, where all but three objects in their sample of 17 type 1 and 2 LINERs appeared variable. The works by \cite{pian2010} and \cite{younes2011} at X-ray energies for type 1 LINERs agrees with the consideration of variability being a common property of LINERs. \cite{omaira2011b} studied a \emph{Suzaku} observation of 80~ks of the \emph{Compton}-thick LINER NGC\,4102, who found two absorbers from the soft X-rays and the optical spectrum. They found variations of the soft-excess flux within $\approx$ 7 years comparing whith \emph{Chandra} data. This variation was described by a decrease in the normalisation of the power-law component and the thermal component by a factor of $\approx$7. Thus it is important to characterize the phenomenon fully, both the scale and magnitude of the variability. Taking the predictions by \cite{mchardy2006} of the time-scale variations scaling with black hole masses, $M_{BH}$, and bolometric luminosities, $L_{bol}$, \cite{omairavaughan2012} estimate time scales larger than tens of days in LINERs.

This paper is organized as follows: in Section \ref{sample} we present the sample and the data. The reduction of the data is explained in Section \ref{reduction}. Our methodolody is described in Section \ref{method}, where individual and simultaneous spectral fittings, comparisons when different appertures were used, flux variability in X-rays and UV frequencies, and short term variability subsections are explained. The results from this analysis are given in Section \ref{results}, and a discussion in Section \ref{discusion}. Finally our main results are summarized in Section \ref{conclusion}.

%
\section{\label{sample} The sample and the data}

From the sample of 82 type 1 and 2 LINERs of \cite{omaira2009a} we searched in the current literature for hints of variability by means of differences in luminosity  when different observations are considered. We used the HEASARC\footnote{http://heasarc.gsfc.nasa.gov/} archive to search for different observations with \emph{Chandra} and \emph{XMM}-Newton, with public available data until October 2012. This analysis provided us with 16 candidate variable sources. The four ULIRGs in this primary selection (NGC\,3690, NGC\,6240, IRAS\,17208-0014, and UGC\,08696) were discarded since our aim is to deal with typical pure LINERs \citep[see][]{ho2008}, not contaminated by strong star formation where other ionization sources could be at work. NGC\,4636 was also discarded because its X-ray emission is dominated by the cluster emission. We also rejected from the present analysis all the observations affected by pile-up larger than 10\% (\emph{Chandra} data observations for objects namely NGC\,4579, NGC\,3998, NGC\,4594, and NGC\,6251 and three \emph{Chandra} observations of NGC\,4278). Finally, to guarantee a proper spectral fitting we kept only observations with a minimum of 400 number counts. An exception was made with obsID 11269 for NGC\,4278, which met this criterion, but visual inspection revealed a very low count number in the hard band.

The final sample of LINERs contains seven objects. Table~\ref{general} shows the general properties of the target galaxies for this study and Table \ref{obs} the log of the observations. Number of counts and
hardness ratios, defined as HR = (H-S)/(H+S), where H is the number of
counts in the hard (2-10 keV) band and S is the number of counts in
the soft (0.5-2 keV) band, are also presented. For five sources, NGC\,1052, NGC\,3226, NGC\,4261, NGC\,4278, and NGC\,5846, observations at different epochs were taken with the same instrument, providing us with good examples for variability studies. In the other two sources, NGC\,3627, and NGC\,4552, we can estimate variability only by comparing \emph{XMM}-Newton with \emph{Chandra} data. These results should be viewed with caution due to the different apertures used by both instruments. A detailed study of the extended emission is made in these two objects (see Section \ref{method}).

%
%
\section{\label{reduction}Data reduction}

Data reduction was performed following the procedure described by \cite{omaira2009a}. We review the methodology here.

\subsection{Chandra data}

\emph{Chandra} observations were obtained with the ACIS instrument \citep{garmire2003}. The data reduction and analysis were carried out in a systematic, uniform way using CXC Chandra Interactive Analysis of Observations (CIAO\footnote{http://cxc.harvard.edu/ciao4.4/}), version 4.3. Level 2 event data were extracted by using the task { \sc acis-process-events}. We first cleaned the data from background flares (i.e. periods of high background) that could affect our analysis. These ``flares'' are due to low energy photons that interact with the detector. To clean them we use the task { \sc lc\_clean.sl}\footnote{http://cxc.harvard.edu/ciao/ahelp/lc\_clean. html}, that removes periods of anomalously low (or high) count rates from lightcurves, from source-free background regions of the CCD. This routine calculates a mean rate from which it deduces a minimum and maximum valid count rate, and creates a file with those periods which are considered by the algorithm to be good.

Nuclear spectra were extracted from a circular region centered at the positions given by NED\footnote{http://ned.ipac.caltech.edu/}. We chose circular radii, trying to get all the possible photons, but not including other sources or background effects. The radii are in the range between 3-8$\arcsec$ (or 6-16 pixels, see Tab.~\ref{obs}). The background selection was made taking regions free of sources, in the same chip as the target, and close to the source (5$\arcsec$ for NGC\,3627, NGC\,4278, NGC\,4261, and NGC\,4552 and 7$\arcsec$ for NGC~5846 and NGC\,3627), to minimize effects related to the spatial variations of the CCD response.

We used the {\sc dmextract} task to extract the spectra of the source and the background regions. The response matrix file (RMF) and ancillary reference file (ARF) were generated for each source region using the {\sc mkacisrmf} and {\sc mkwarf} tasks, respectively. Before the background subtraction, the spectra were binned to have a minimum of 20 counts per spectral bin, to be able to use the $\chi^2$-statistics. This was done by using the {\sc grppha} task included in {\sc ftools}.

\subsection{XMM-Newton data}

All of the \emph{XMM}-Newton observations were obtained from the EPIC pn camera\footnote{ EPIC pn is the most efficient camera because X-ray photons hit the detector from the rear side, avoiding cross-calibration problems between the pn and MOS cameras \citep{struder2001}}. The data were reduced in a systematic, uniform way using the Science Analysis Software (SAS \footnote{http://xmm.esa.int/sas/}), version 11.0.0. Before the extraction of the spectra, good-timing periods were selected (i.e. ``flares'' were excluded). The method used for this purpose maximizes the S/N ratio of the net source spectrum by applying a different constant count rate threshold on the single-events, E $>$ 10 keV field-of-view background light curve. The nuclear positions were taken from NED, while the extraction region was determined through circles of 25$\arcsec$ radius and the background was determined by using an algorithm that selects the best circular region around the source, free of other sources and as close as possible to the nucleus. This automatic selection was checked manually to ensure the best selection for the backgrounds.

The extraction of the source and the background regions were done by using  the {\sc evselect} task. Response matrix files (RMF) were generated using the {\sc rmfgen} task, and the ancillary response files (ARF) were generated using the {\sc arfgen} task. We then grouped the spectra to get at least 20 counts per spectral bin using the {\sc grppha} task, as is required to use the $\chi^2$-statistics.

\subsection{ \label{sigma} Light curves}

\begin{figure*}[H]
\caption{ \label{lightcurves1052} Lightcurves for NGC\,1052.}
\centering
\includegraphics[width=0.4\textwidth]{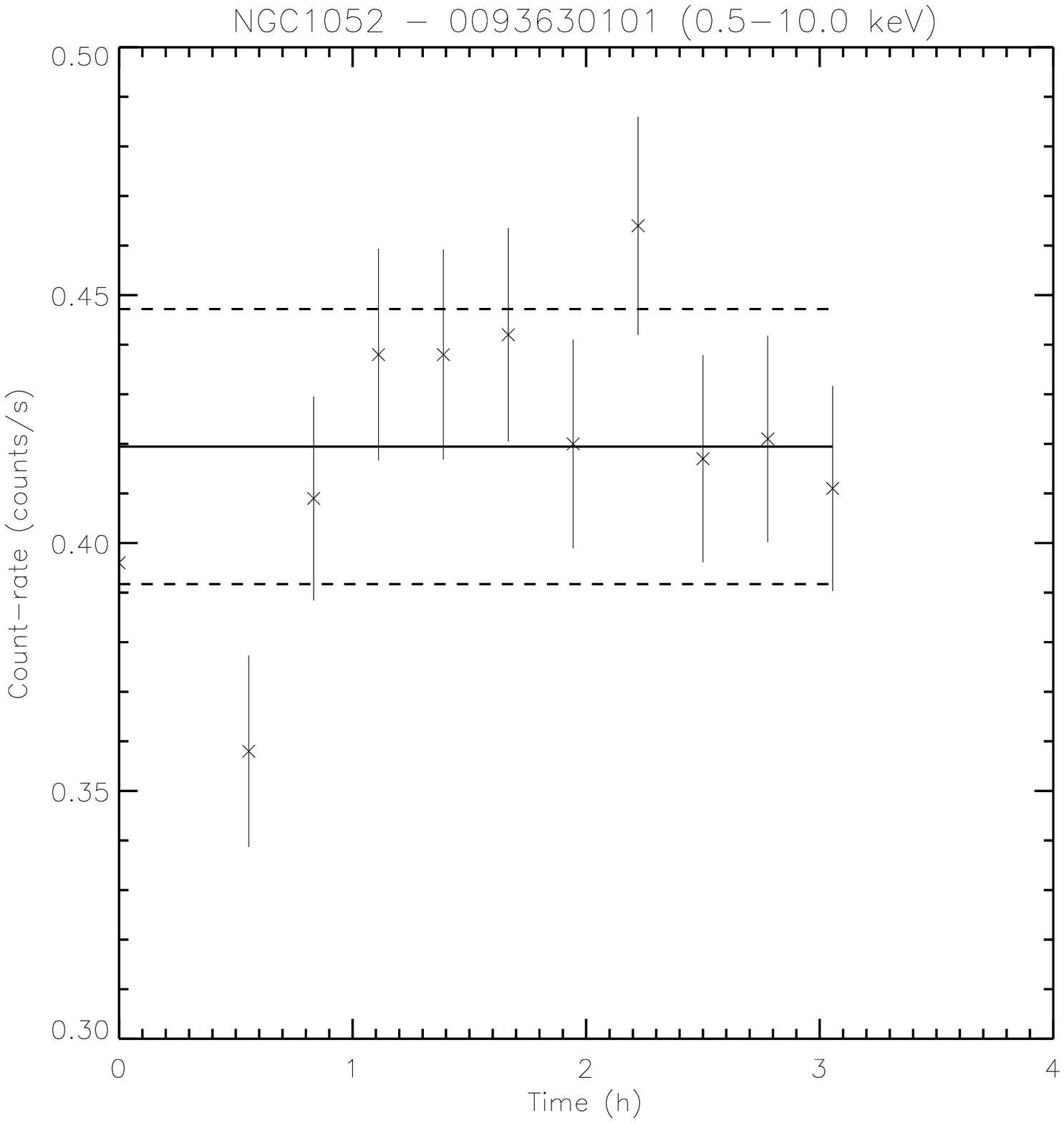} \includegraphics[width=0.4\textwidth]{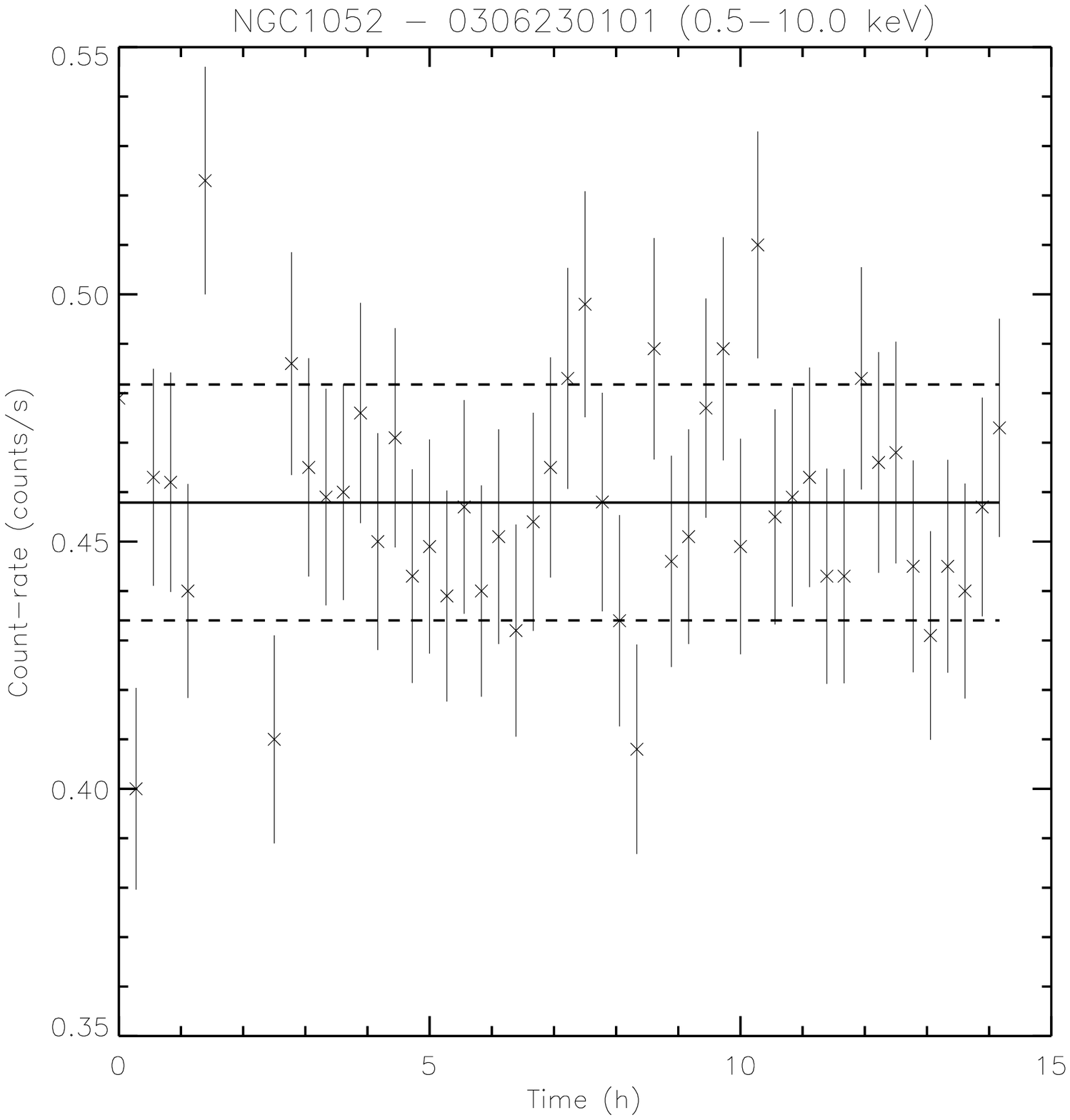}

\includegraphics[width=0.4\textwidth]{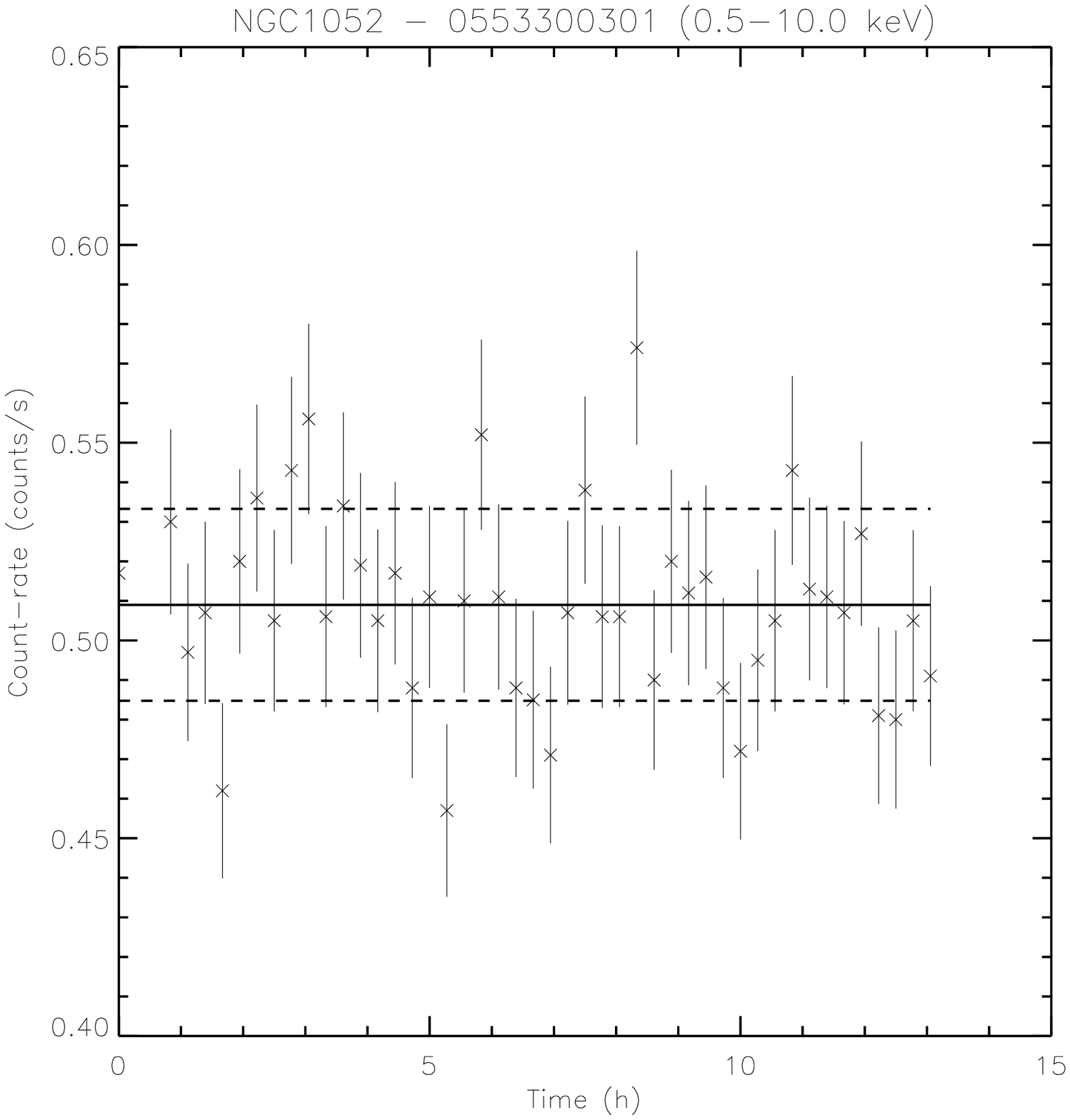} \includegraphics[width=0.4\textwidth]{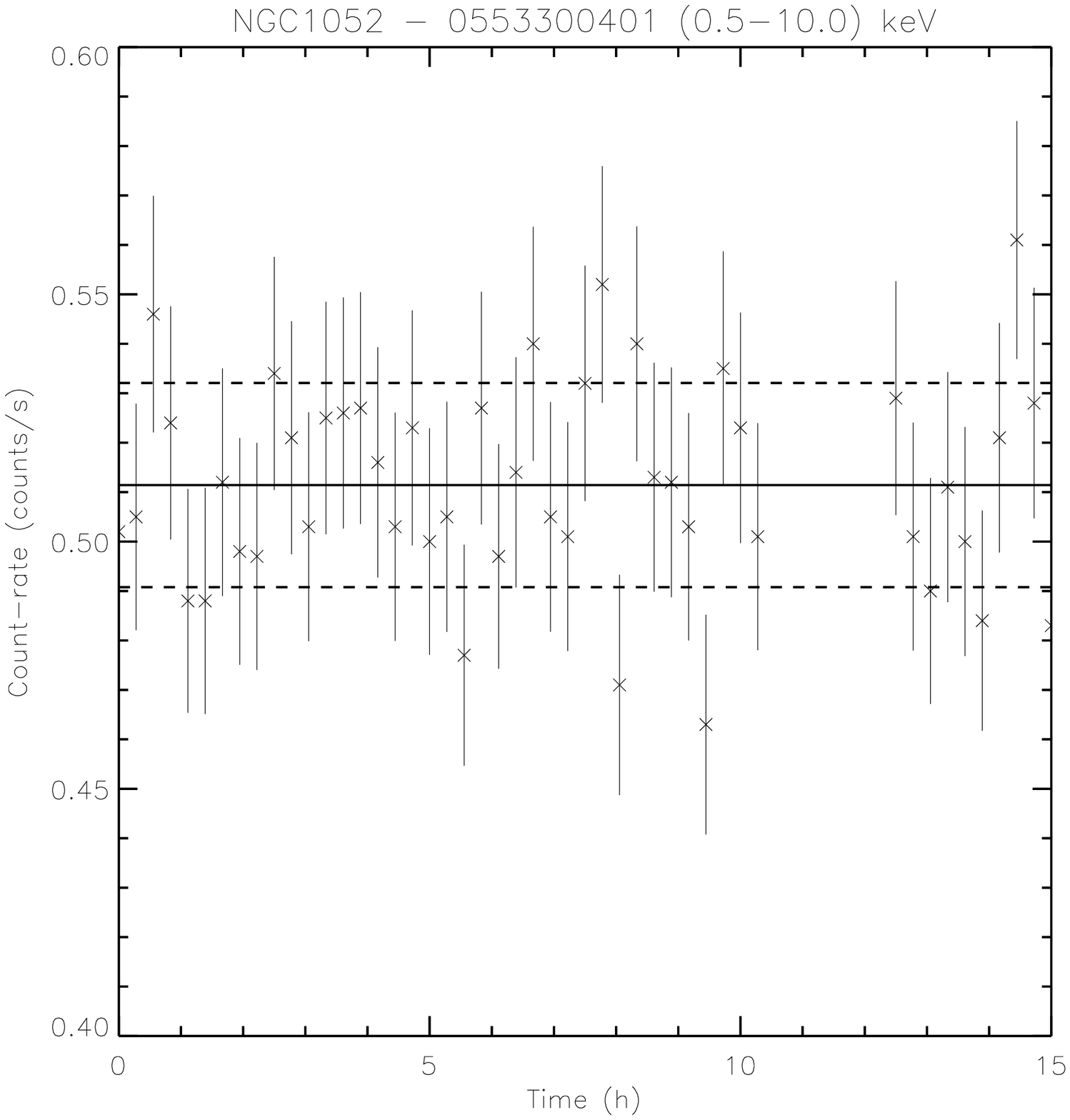}

\includegraphics[width=0.4\textwidth]{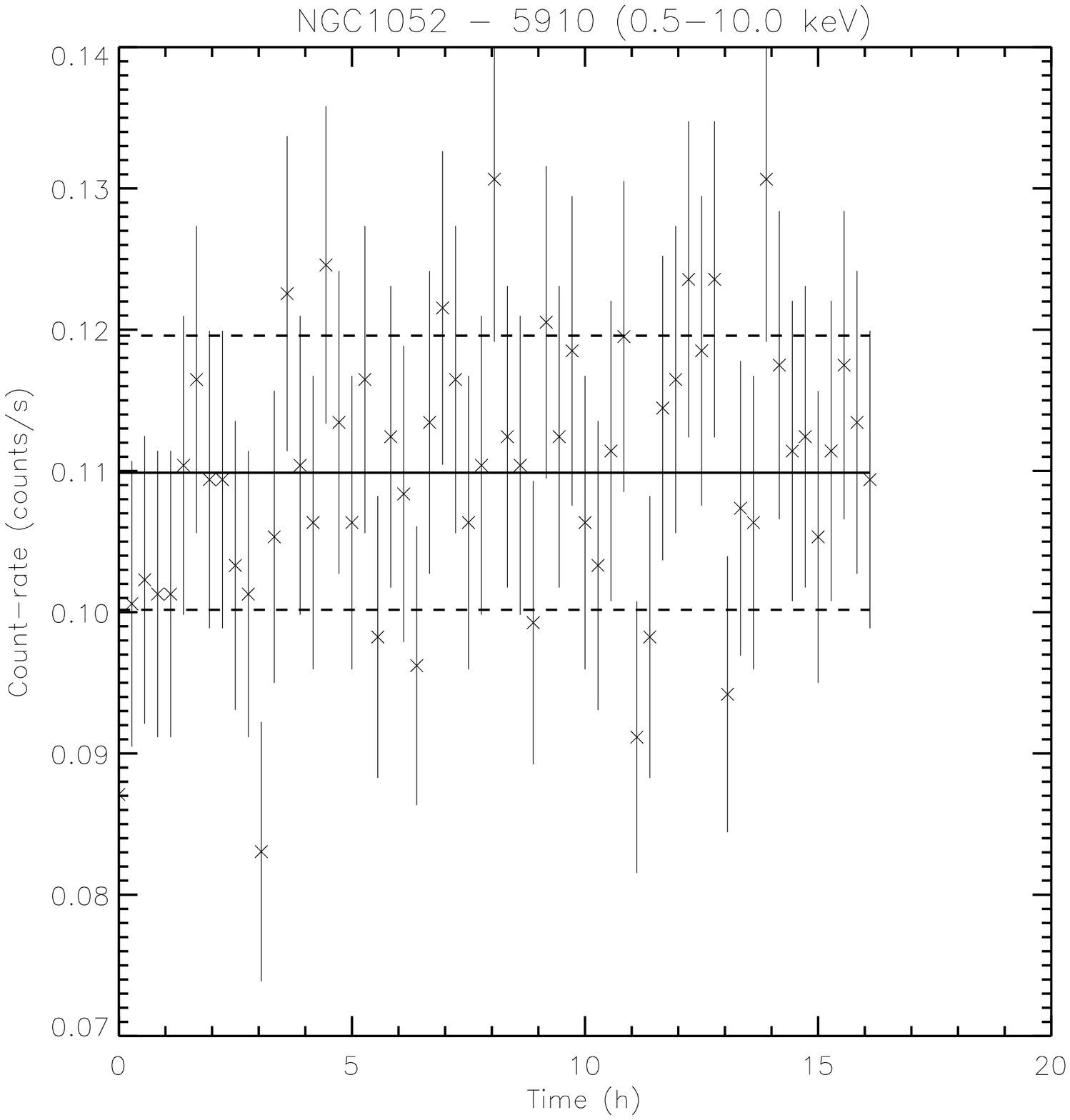}

\end{figure*}

\begin{figure*}[H]
\caption{ \label{lightcurves3627} Lightcurves for NGC\,3627.}
\centering
\includegraphics[width=0.4\textwidth]{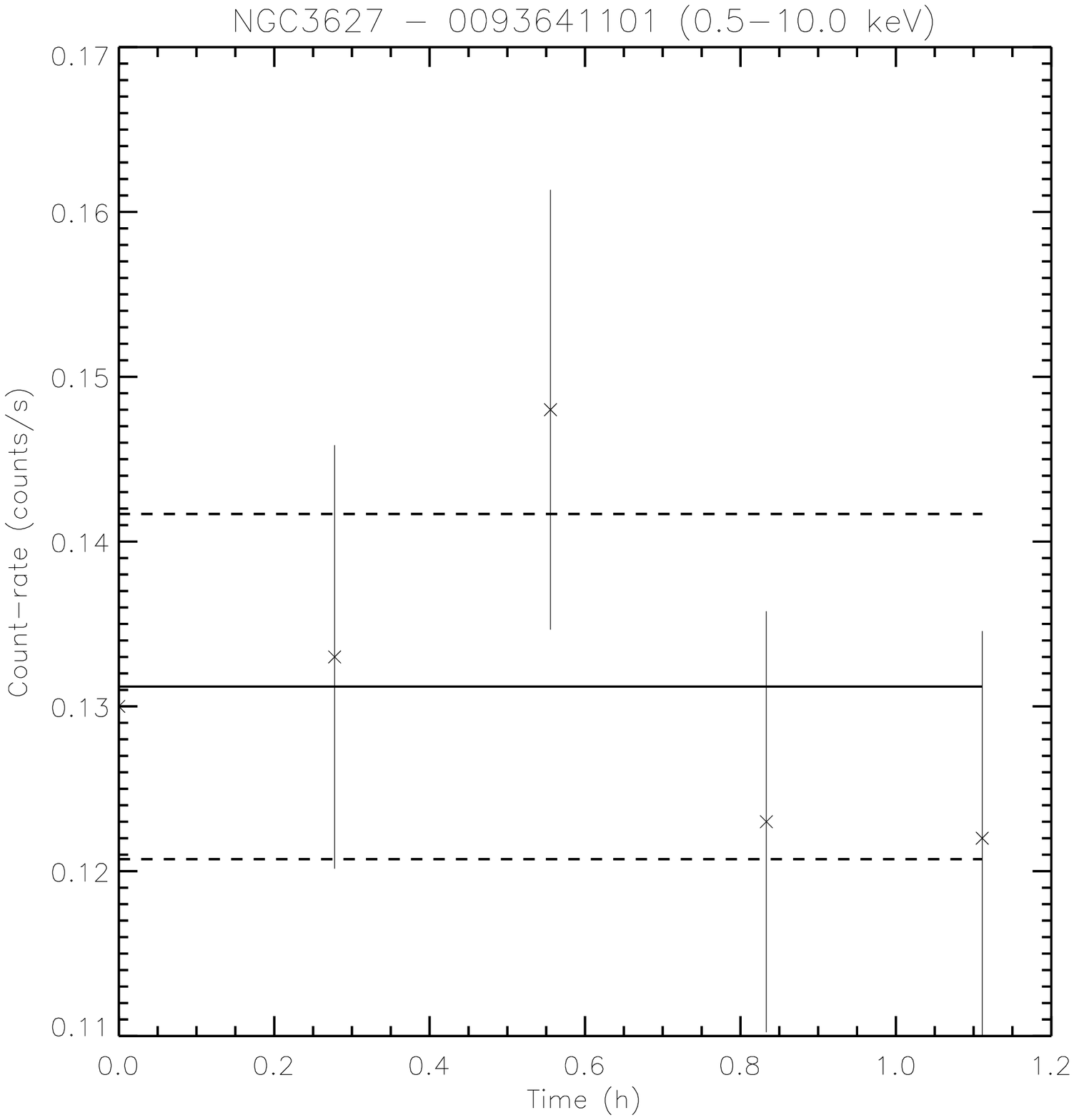} \includegraphics[width=0.4\textwidth]{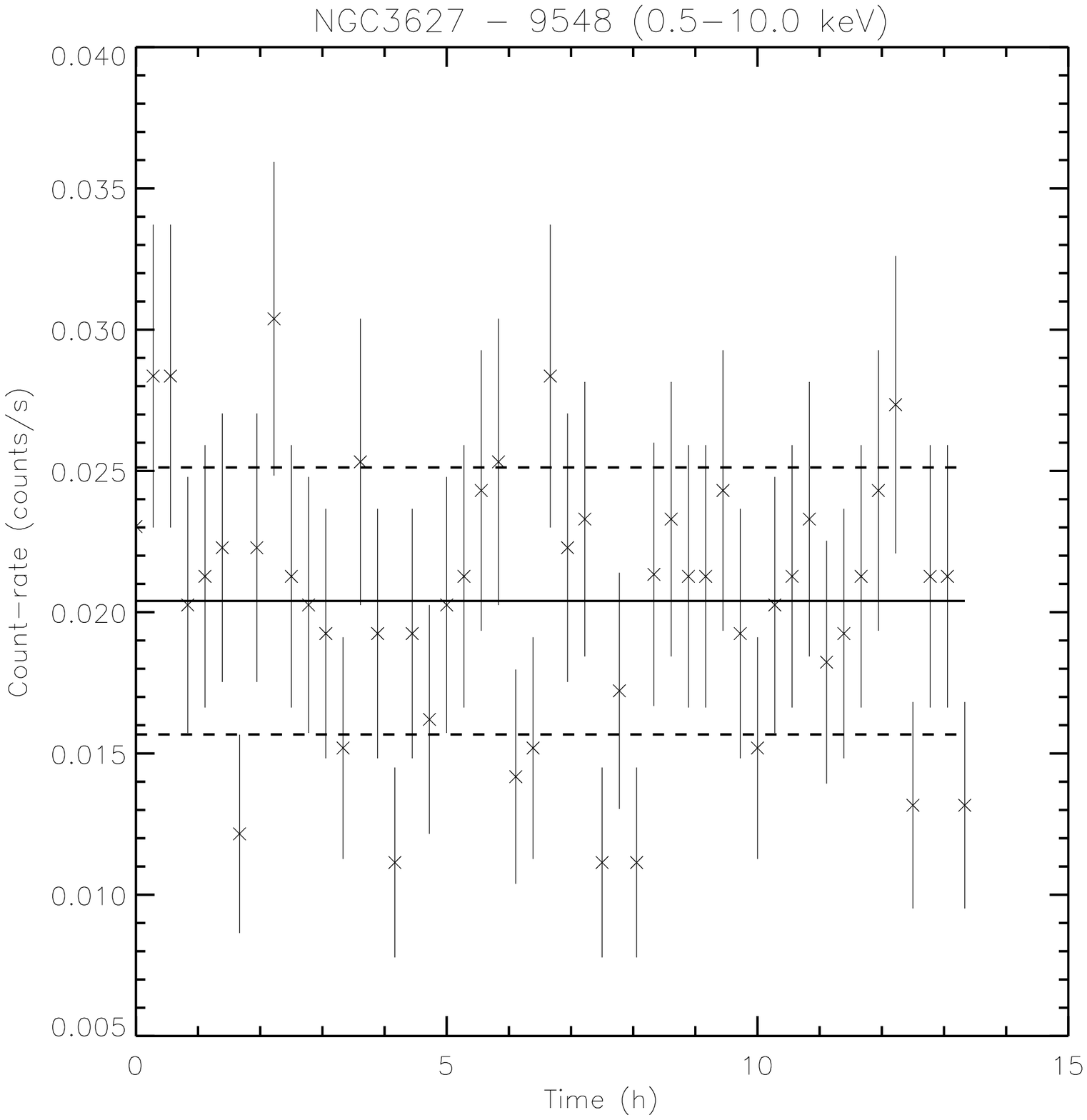}
\end{figure*}

\begin{figure*}[H]
\caption{ \label{lightcurves4261} Lightcurves for NGC\,4261.}
\centering
\includegraphics[width=0.4\textwidth]{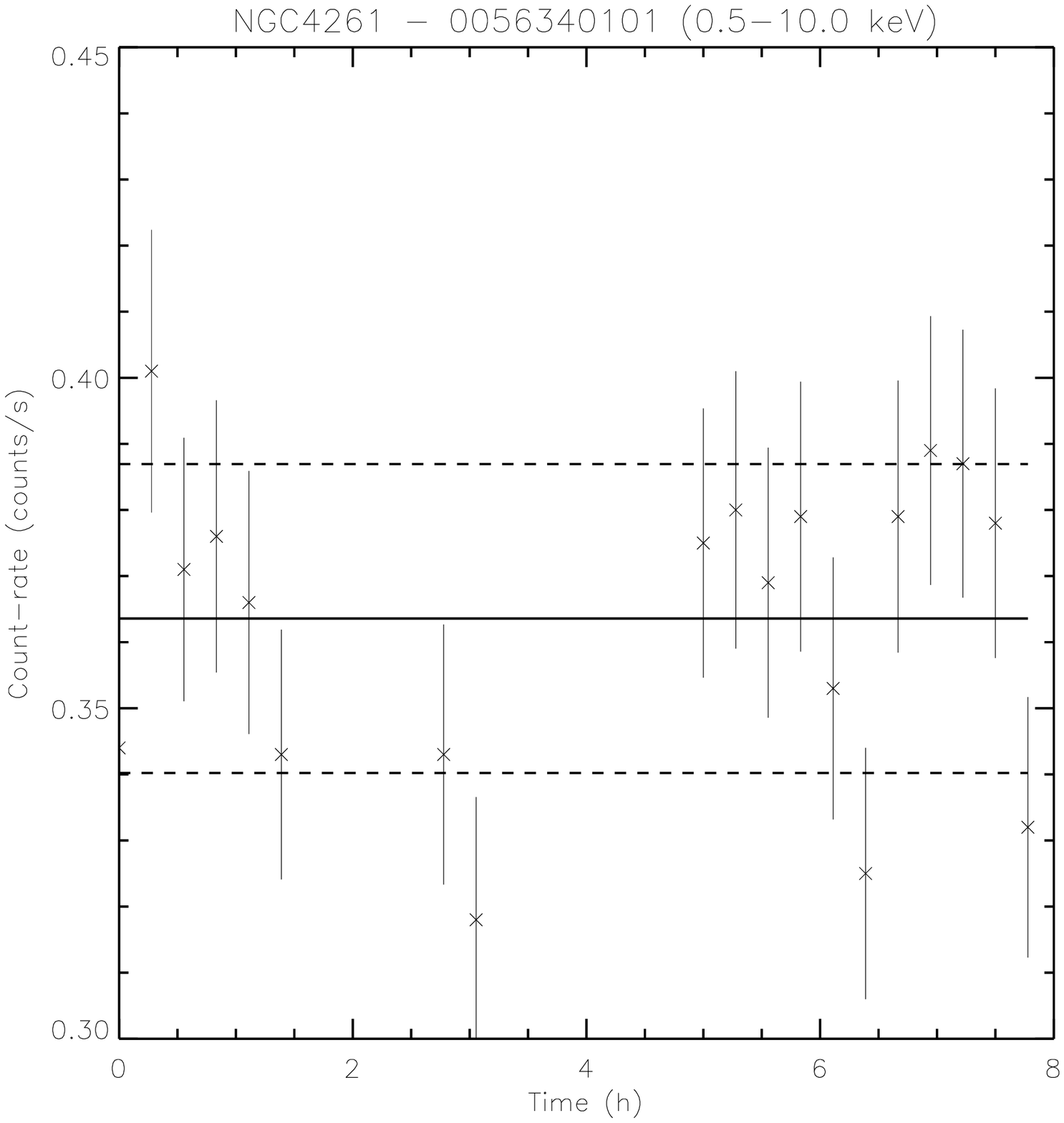} \includegraphics[width=0.4\textwidth]{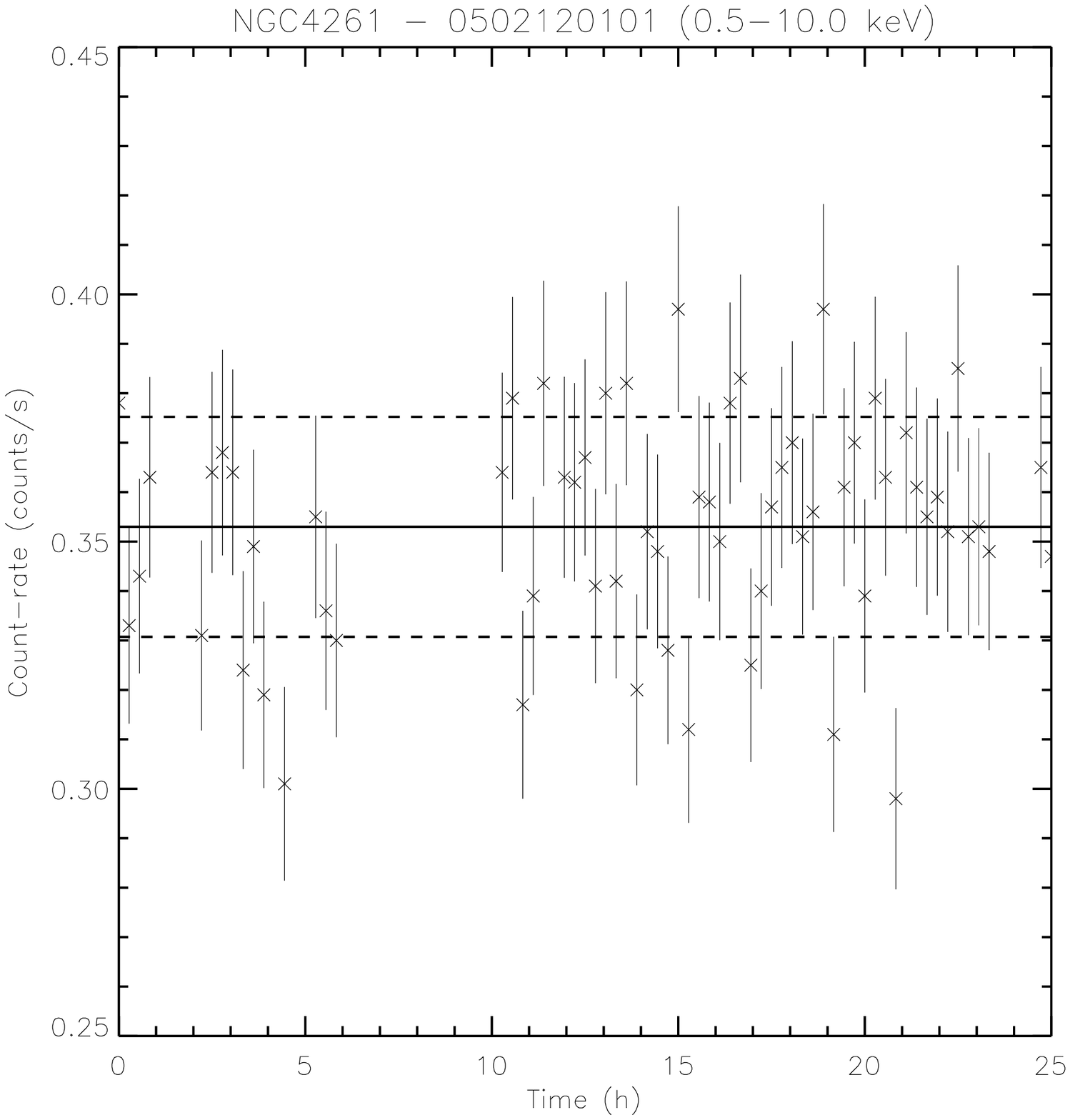}
\includegraphics[width=0.4\textwidth]{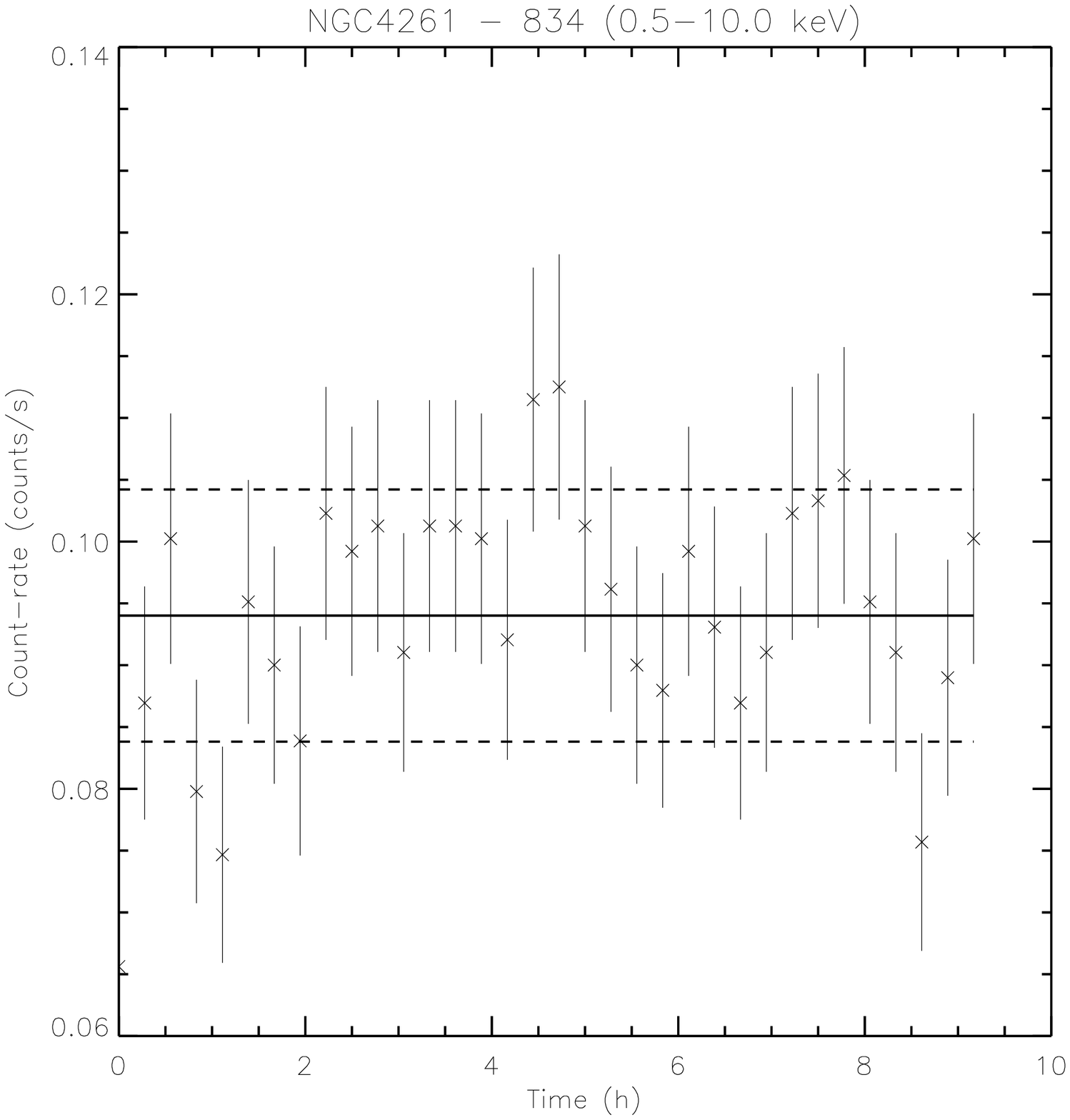} \includegraphics[width=0.4\textwidth]{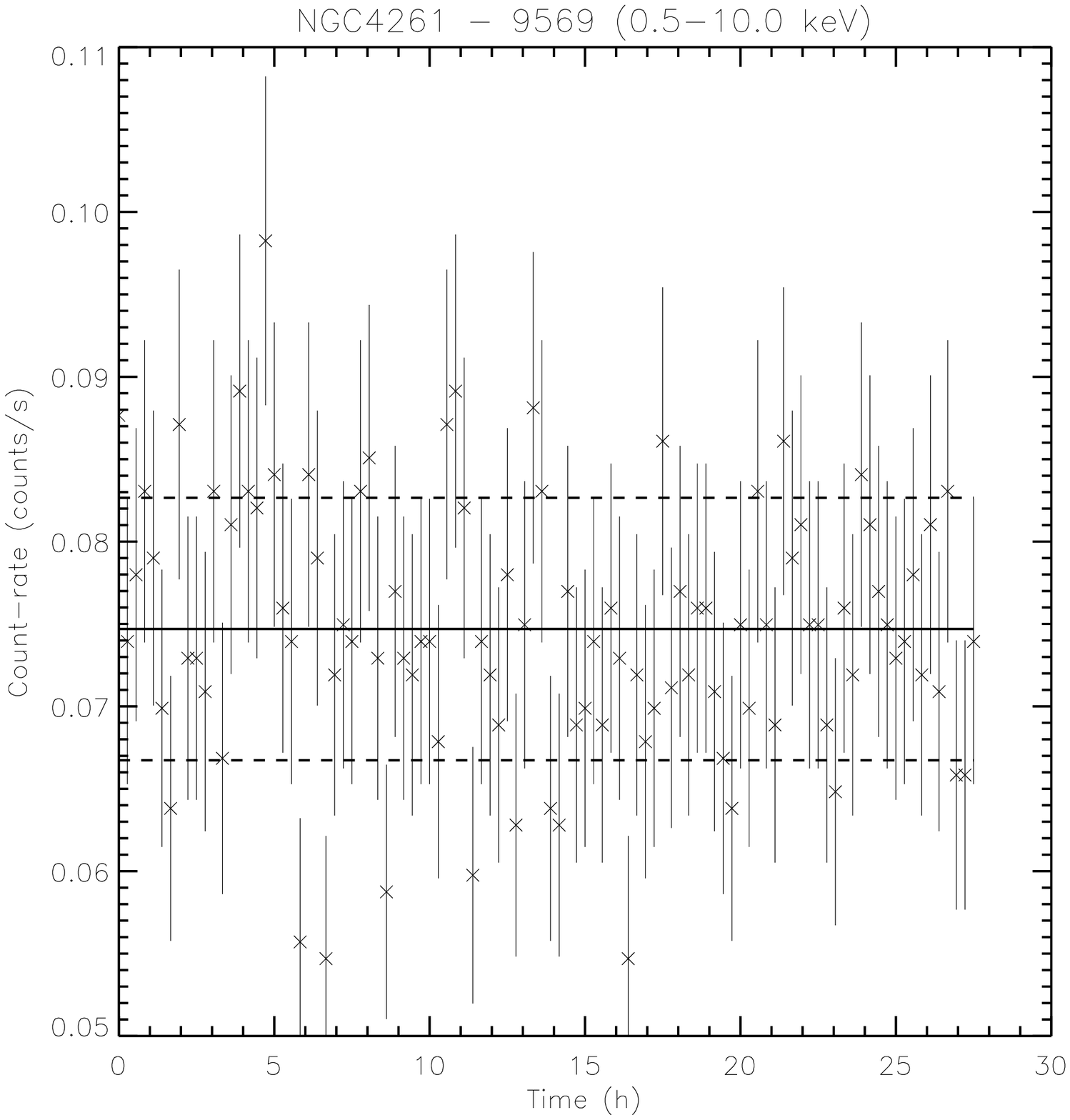}
\end{figure*}

\begin{figure*}[H]
\caption{ \label{lightcurves4278} Lightcurves for NGC\,4278.}
\centering
\includegraphics[width=0.4\textwidth]{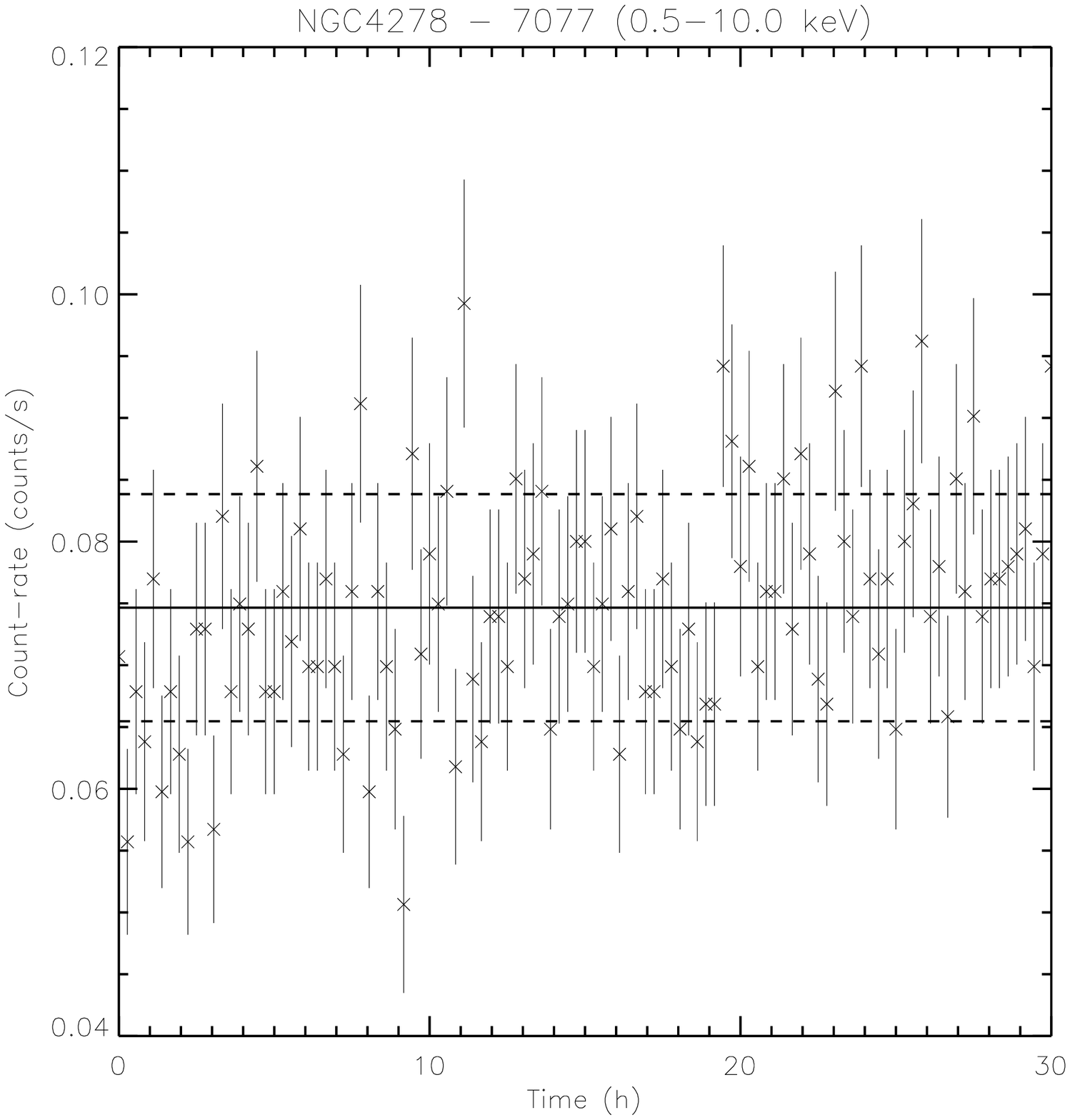} \includegraphics[width=0.4\textwidth]{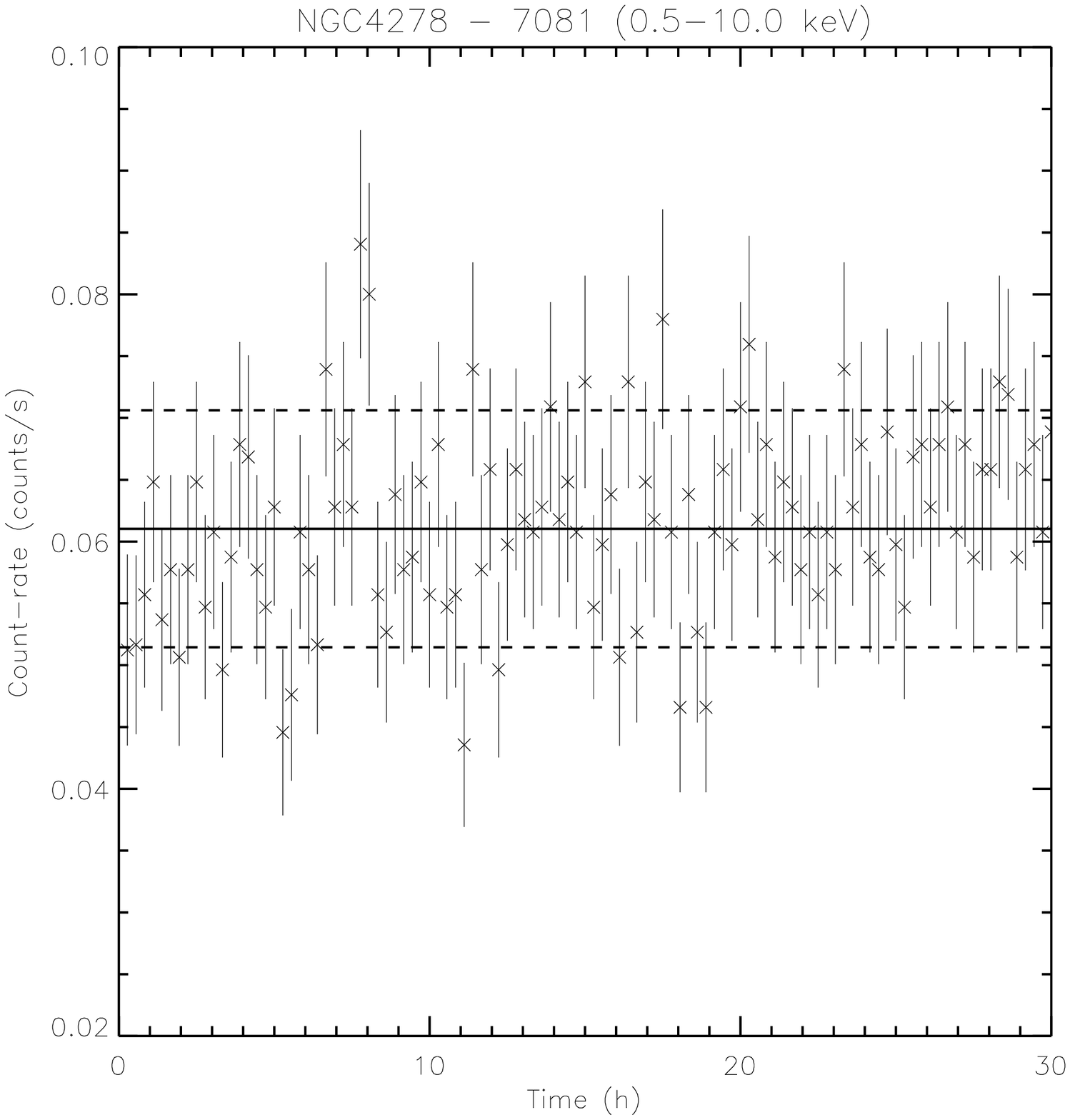}
\includegraphics[width=0.4\textwidth]{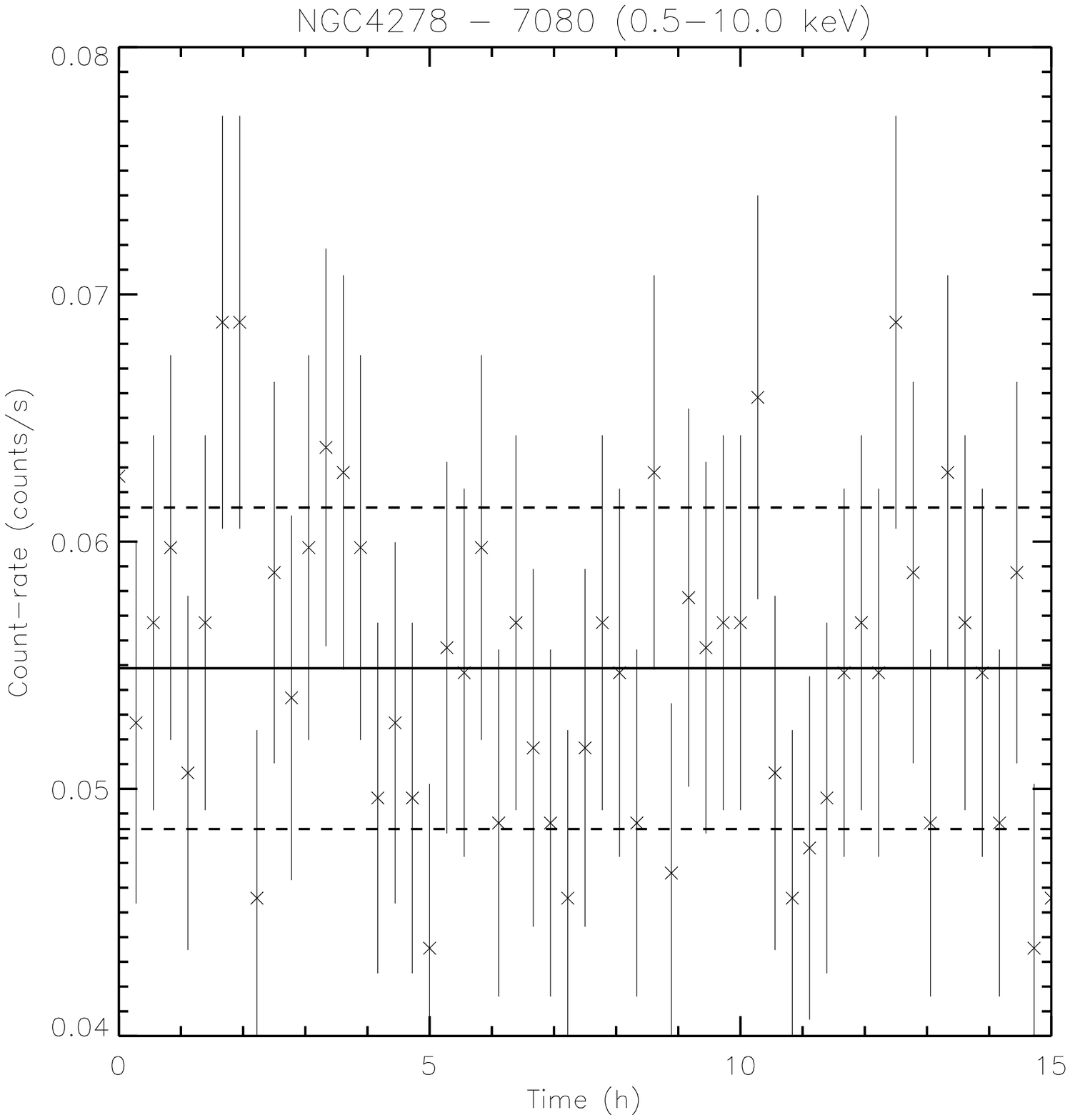} \includegraphics[width=0.4\textwidth]{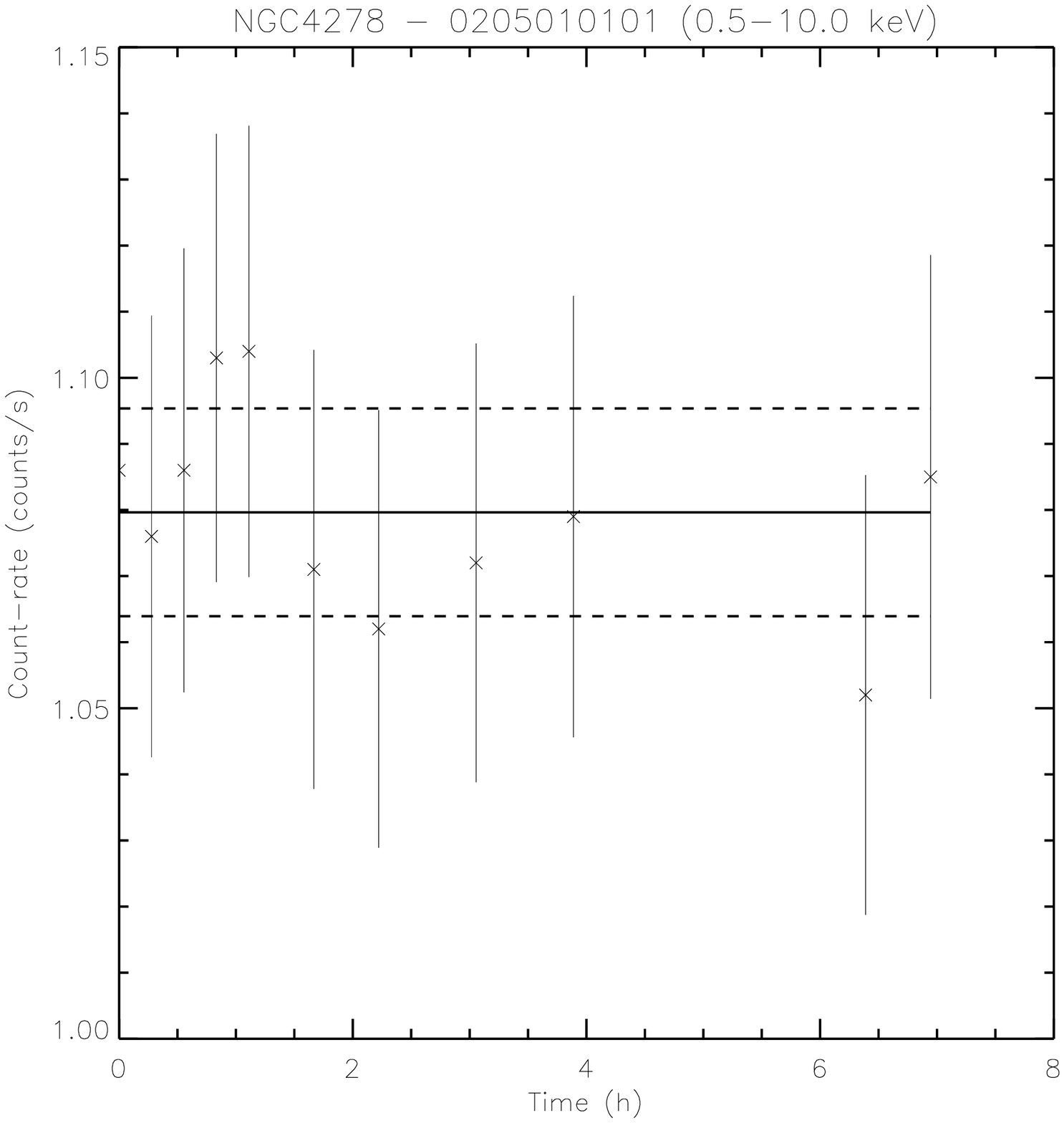}
\end{figure*}

\begin{figure*}[H]
\caption{ \label{lightcurves4552} Lightcurves for NGC\,4552.}
\centering
\includegraphics[width=0.4\textwidth]{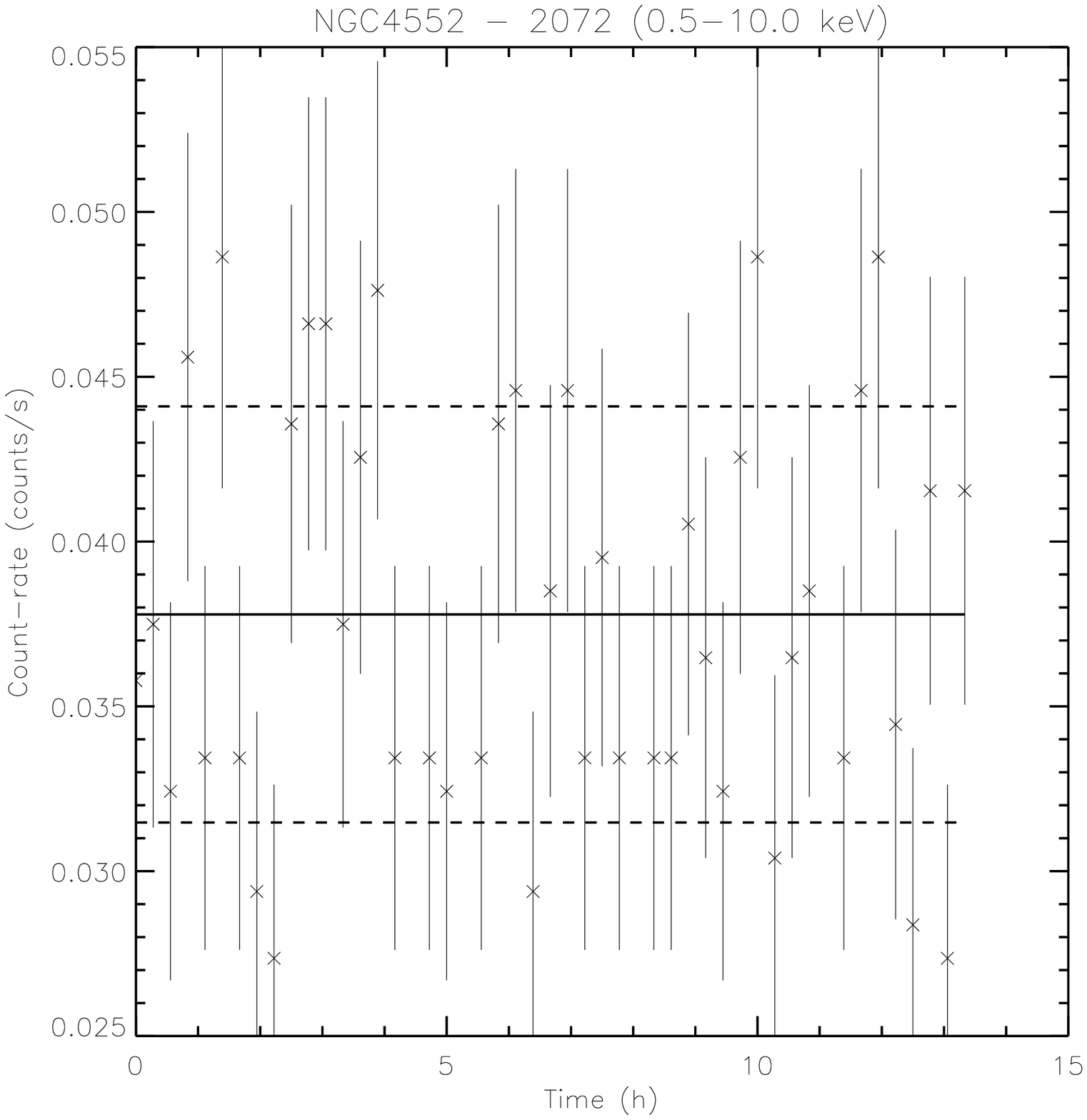}\includegraphics[width=0.4\textwidth]{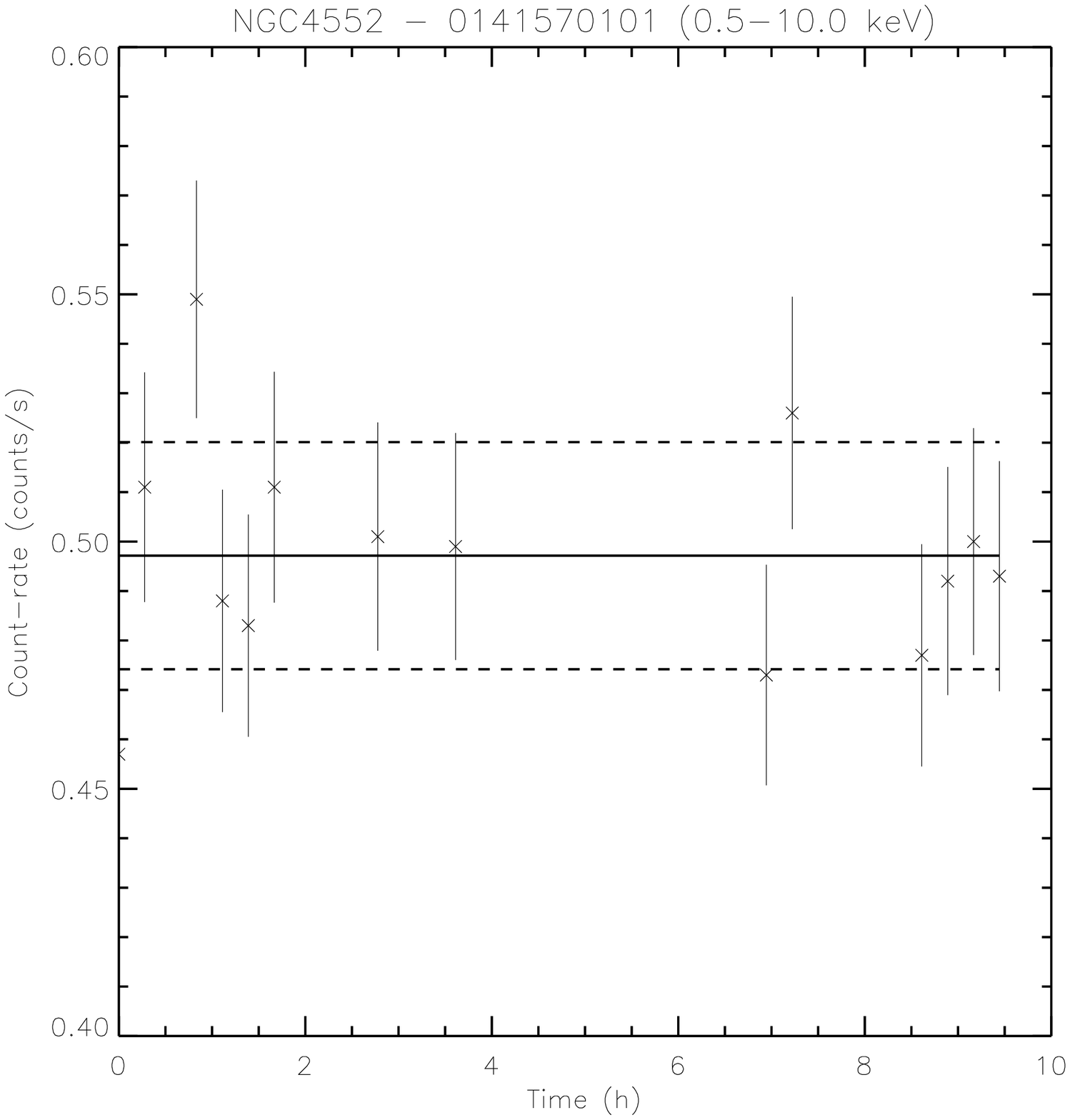} 
\end{figure*}

\begin{figure*}[H]
\caption{ \label{lightcurves5846} Lightcurves for NGC\,5846.}
\centering
\includegraphics[width=0.4\textwidth]{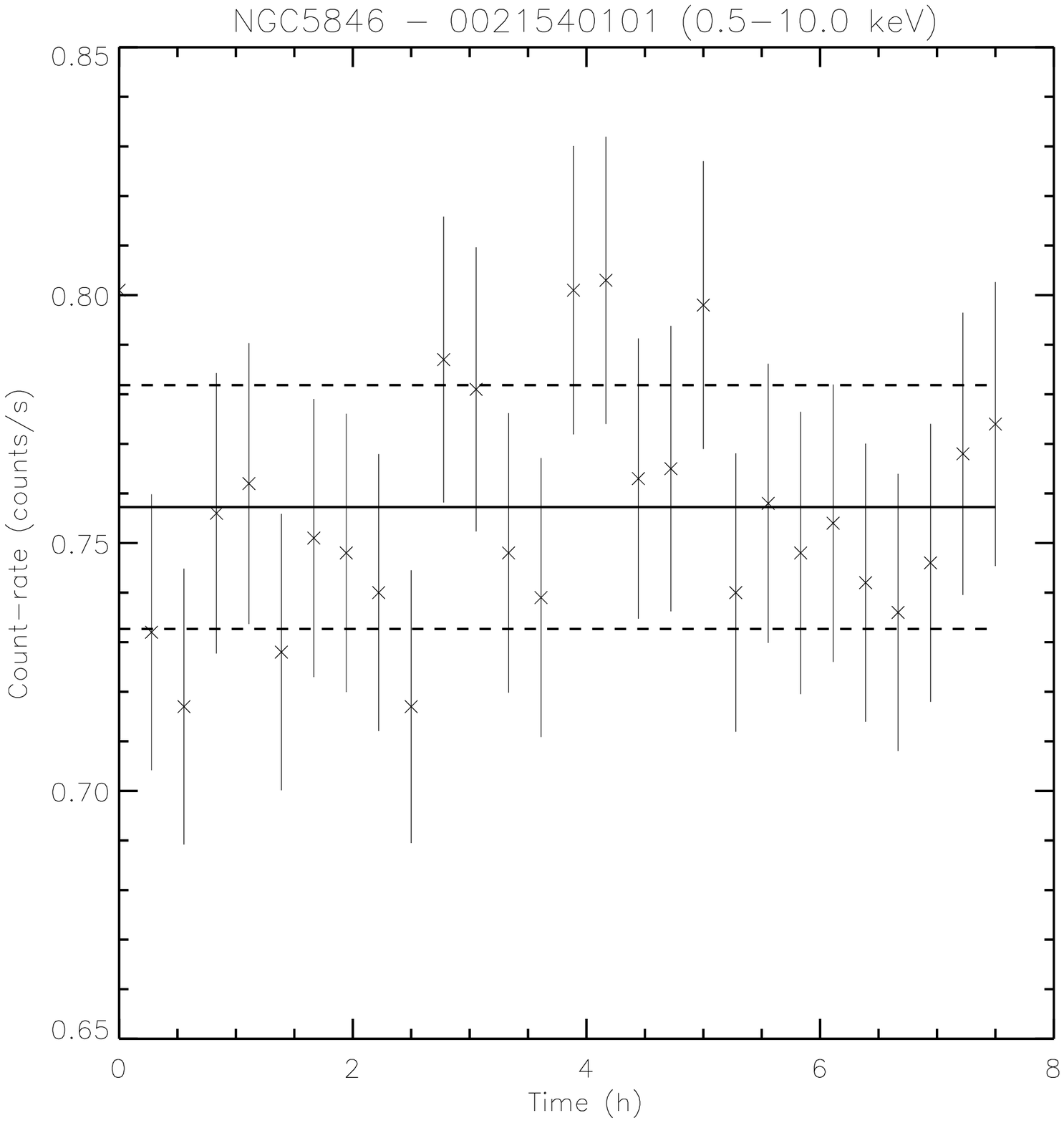} \includegraphics[width=0.4\textwidth]{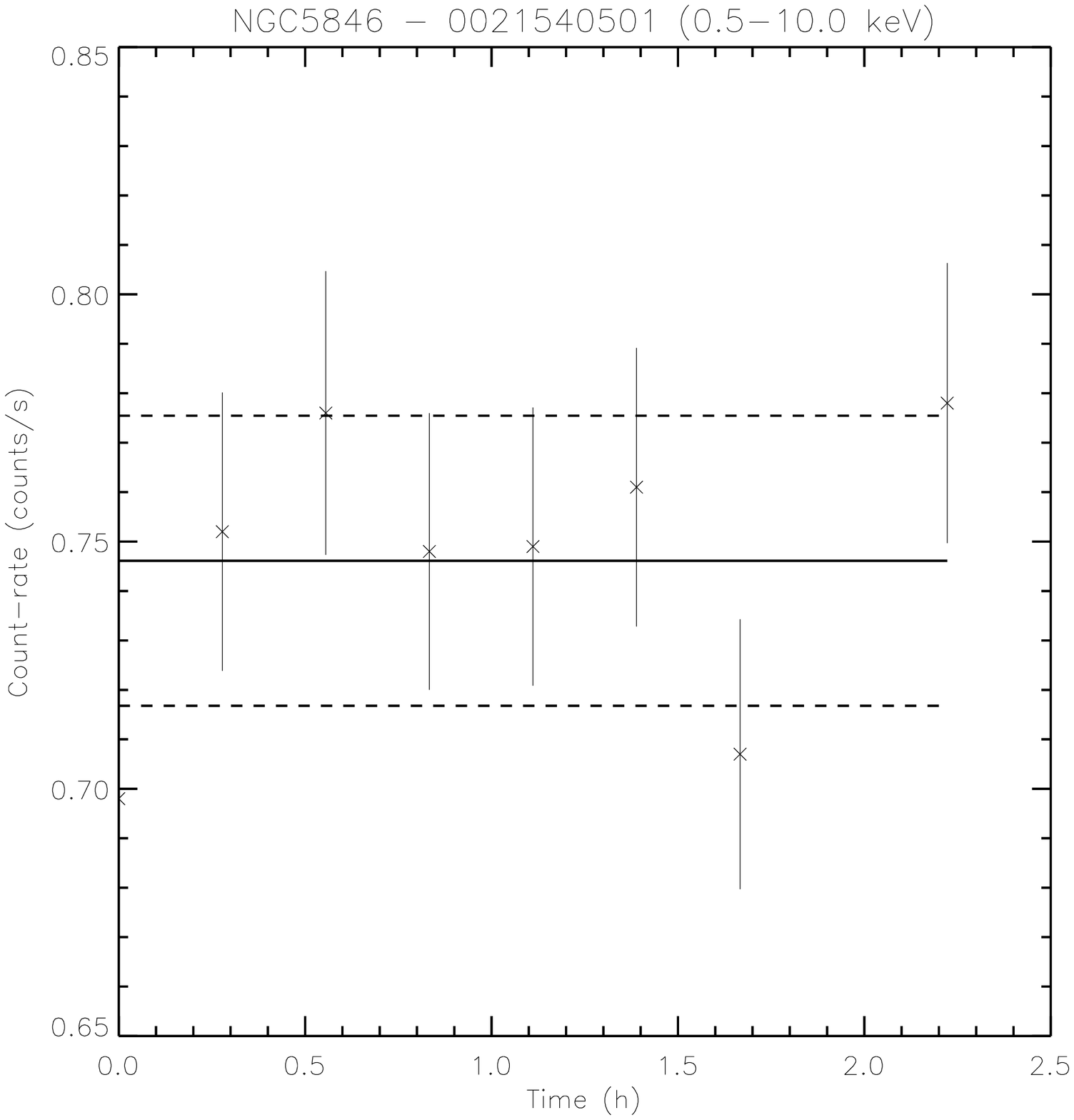}
\includegraphics[width=0.4\textwidth]{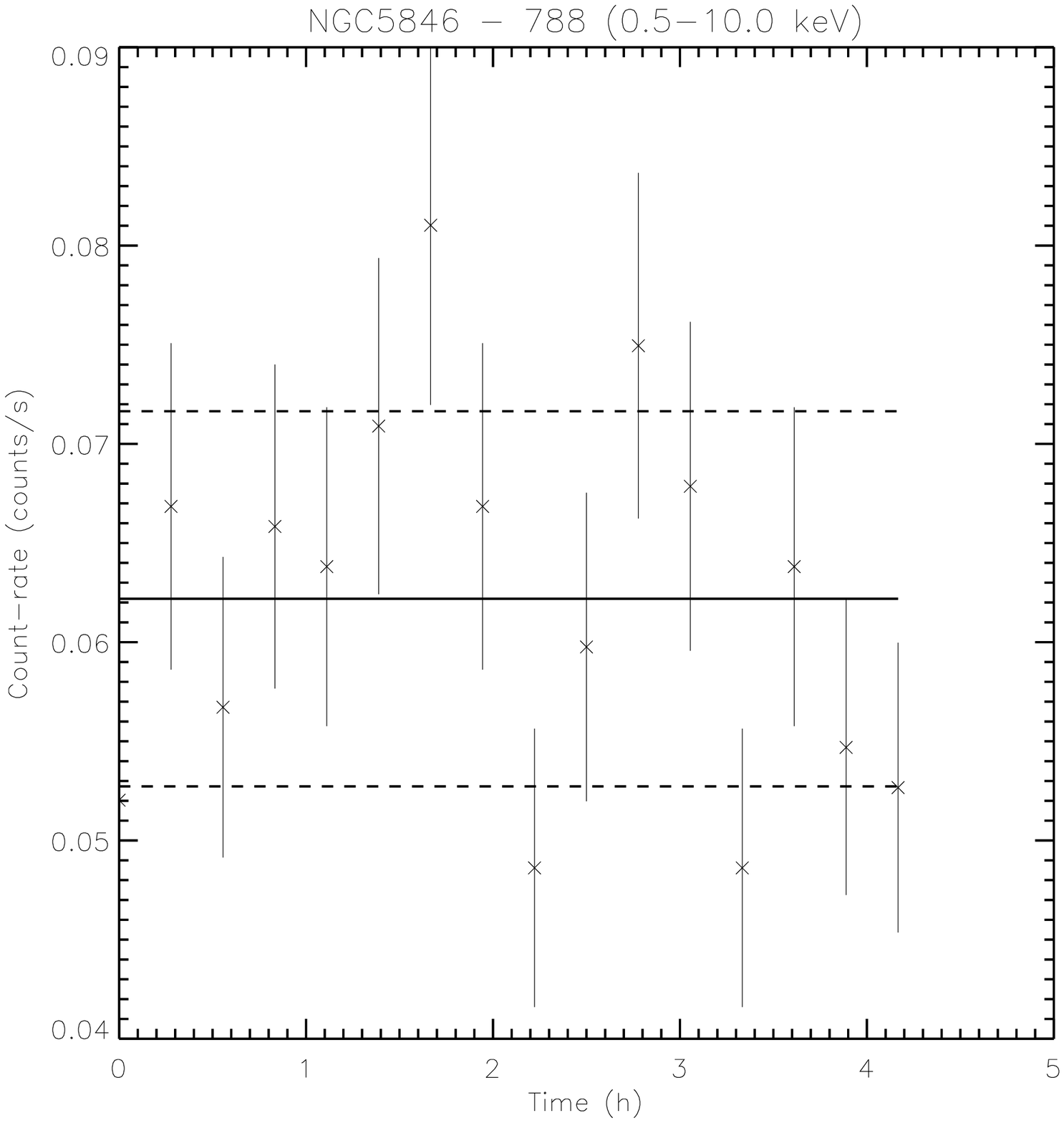} \includegraphics[width=0.4\textwidth]{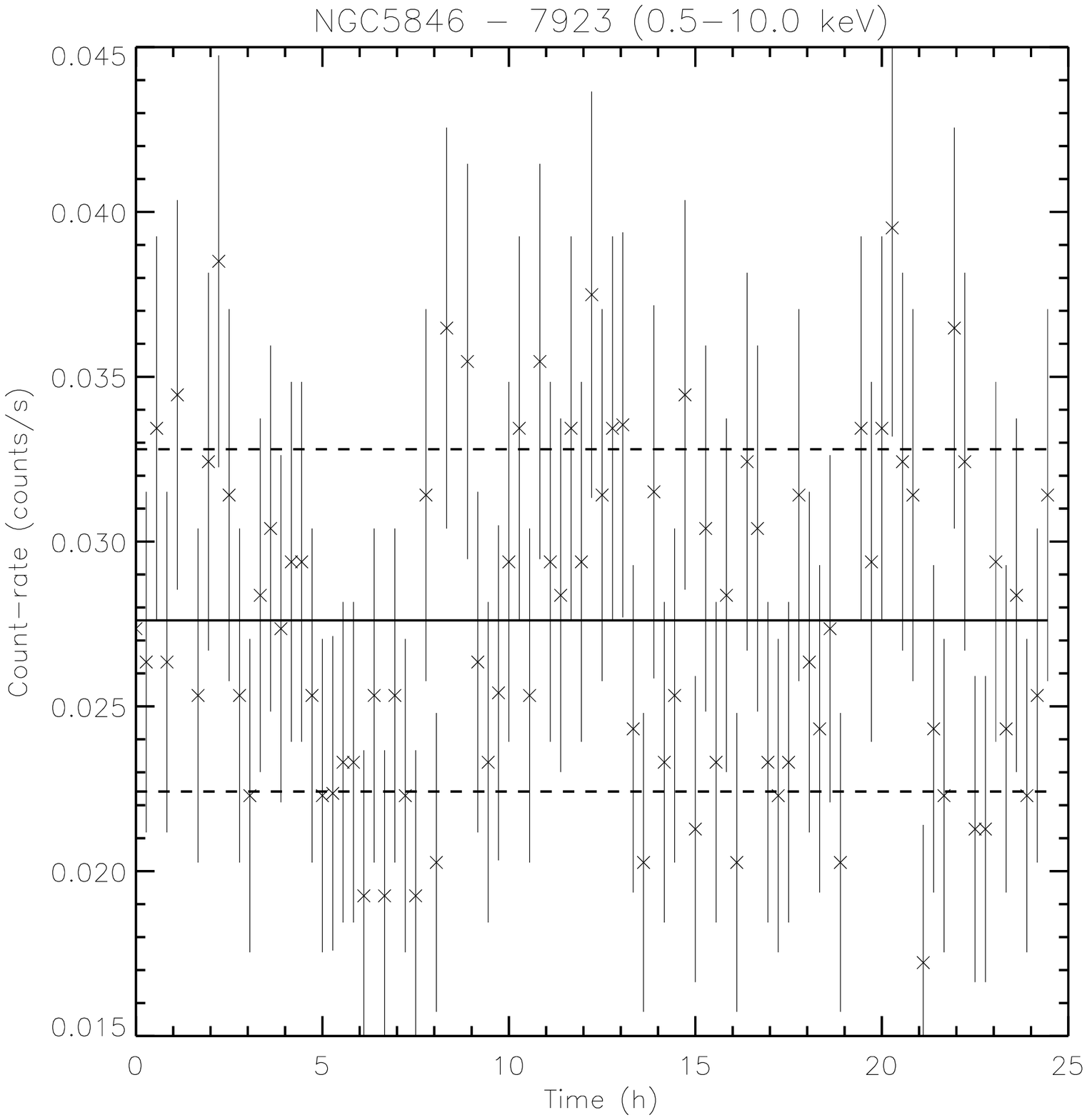}
\end{figure*}

Light curves in the 0.5-10 keV band for the source and background were extracted using the {\sc dmextract} task for \emph{XMM}-Newton and {\sc evselect} task for \emph{Chandra} with a 1000~s bin. The light curve from the source was manually screened for high background and flaring activity. This means that when the background light curve showed “flare”-like events and/or prominent decreasing/increasing trends, we did not use those intervals. After rejection of the respective time intervals, the total useful time for each observation was usually less than the original exposure time (see Table \ref{obs}). The light curves are shown in Figs. \ref{lightcurves1052}-\ref{lightcurves5846}, where the solid line represents the mean value of the count-rate and the dashed lines represent 1$\sigma$ standard deviation.

\section{\label{method}Methodology}

The spectral fitting process comprises two steps: (1) individual analysis of each observation to determine the best fit for each spectrum, and (2) simultaneous fitting of the set of spectra of the same object at different epochs. The spectral fitting was done using XSPEC\footnote{http://heasarc.gsfc.nasa.gov/xanadu/xspec/} version 12.7.0.

\subsection{\label{individual} Individual spectral analysis}

We have performed an individual study of the best fit model for each observation. For this, we follow the method used by \cite{omaira2009a}, where five different models were used:

\begin{enumerate}

\item ME: A pure thermal model (MEKAL in XSPEC). The thermal emission is responsible for the bulk of the X-ray energy distribution. The free parameters in this model are the column density, $N_H$, the temperature, $kT$, and the normalization, $Norm$.

\item PL: A single power law model, which corresponds to a non-thermal source representing an AGN. The column density, $N_H$, is added as a free parameter, to take the absorption by matter along our line of sight to the target into account. The free parameters in this model are the column density, $N_H$, the slope of the power law, $\Gamma$, and the normalization, $Norm$.

\item 2PL: A model containing two power laws with the same slope, $\Gamma$. Here the bulk of the hard X-rays is due to a primary continuum described by a power law, and the soft X-rays come from a scattering component described by the other power law. The free parameters in this model are the column densities, $N_{H1}$ and $N_{H2}$, the slope of the power-law, $\Gamma$, and the normalizations, $Norm_1$ and $Norm_2$.

\item MEPL: A composite of a thermal plus a single power law model. The AGN dominates the hard X-rays, but the soft X-rays require an additional thermal contribution. The free parameters in this model are the column densities, $N_{H1}$ and $N_{H2}$, the temperature, $kT$, the slope of the power law, $\Gamma$, and the normalizations, $Norm_1$ and $Norm_2$.

\item ME2PL: A composite of a thermal plus two power laws model. This model is like MEPL but including the contribution of the thermal emission at soft X-rays. This is the \emph{Compton-thin} Seyfert 2 \emph{baseline} model used by \cite{guainazzi2005a}. The free parameters in this model are the column densities, $N_{H1}$ and $N_{H2}$, the temperature, $kT$, the slope of the power law, $\Gamma$, and the normalizations, $Norm_1$ and $Norm_2$.

\end{enumerate}

For models 2PL, MEPL, and ME2PL we used two absorbers, $N_{H1}$  and $N_{H2}$. These absorbers are included in the models as abs($N_{H1}$)*PL + abs($N_{H2}$)*PL, abs($N_{H1}$)*Mekal + abs($N_{H2}$)*PL, and abs($N_{H1}$)*(PL) + Mekal + abs($N_{H2}$)*PL. The Galactic absoption was included in each model and was fixed to the predicted value (Col. 7 in Table \ref{general}) using the {\sc nh} tool within {\sc ftools} \citep{dickeylockman1990, kalberla2005}. We searched for the presence of the neutral iron fluorescence emission line, $FeK\alpha$, adding a narrow Gaussian with centroid energy fixed at the observed energy corresponding to a rest frame at 6.4 keV. Two Gaussians were also included to model recombination lines, FeXXV at 6.7 keV and FeXXVI at 6.95 keV.

We impose the following conditions to the resulting best-fit parameters to ensure a physical meaning to the best-fit solutions: $\Gamma > 0.5$, $N_{H1} > N_{Gal}$ and $N_{H2} > N_{H1}$.

We select as the best fit the spectral fitting with $\chi^2/d.o.f$ closer to unity and the F-test probability less than $10^{-5}$ when comparing with a simpler model \citep{ftest}. Thus, the best fit model is the simplest model which represents the data.

\subsection{ \label{simultaneous} Simultaneous spectral analysis}

The aim of this analysis is to detect variability and study the physical parameters governing its pattern for these sources. For that we simultaneously fitted the spectra for each object to the same model. The baseline model for this simultaneous fit was the best fit obtained for the individual fitting of the observations. When the best fit for individual sources did not match for all the observations, we used the most complex one. For each galaxy, the initial values for the parameters were set to those obtained for the spectrum with the largest number of counts.

The simultaneous fit was made in three steps. First, every spectrum was fitted with the same model and all the parameters linked to the same value (hereinafter SMF0). If this model was able to fit all the spectra, then the source is not variable. SMF0 was used as the baseline model for the next step otherwise. Secondly, we let the parameters: $N_{H1}$, $N_{H2}$, $\Gamma$, $Norm1$, $Norm2$, and $kT$ vary individually (hereinafter SMF1). Among these we chose the best fit as that with $\chi_{r}^2=\chi^2/d.o.f$ closest to the unity and that improved the SMF0 fit (using the F-test). The result of SMF1 was used as the baseline model for the next step. Finally, we also included the possibility that two parameters could vary together to explain the variability pattern of the sources. For that purpose we fitted each set of data, letting the parameter found as the best fit in SMF1 vary together with any of the other parameters of the fit (hereinafter SMF2). Again the $\chi_{r}^2$ and F-test were used to determine whether this further complexity of the spectral fitting results in a significant improvement of the final fit.

The final best fit could be (1) SMF0: the best simultaneous fit with each parameters tied together for all the observations (i.e. non variable source); (2) SMF1: in the best simultaneous fit only one parameter was allowed to vary among the observations, and (3) SMF2: the best simultaneous fit was that with two parameters allowed to vary during the observations.

Further complexity of the spectral fitting (e.g. three parameters allowed to vary) were not required for our set of data (see Section \ref{results}).

\subsubsection{ \label{apertures} Different appertures}

When data from \emph{Chandra} and \emph{XMM}-Newton were used together, an additional analysis was performed to ensure that the extra-nuclear emission is not producing the observed variability.

\begin{figure*}[H]
\caption{ \label{images} Images for \emph{Chandra} data (left) and \emph{XMM}-Newton data (right) for the sources in the 0.5-10 keV band. Big circles represent 25'' apertures. Small circles in the left figures  represent the nuclear extraction aperture used with \emph{Chandra} observations (see table \ref{obs}). }
\centering
\includegraphics[width=\textwidth]{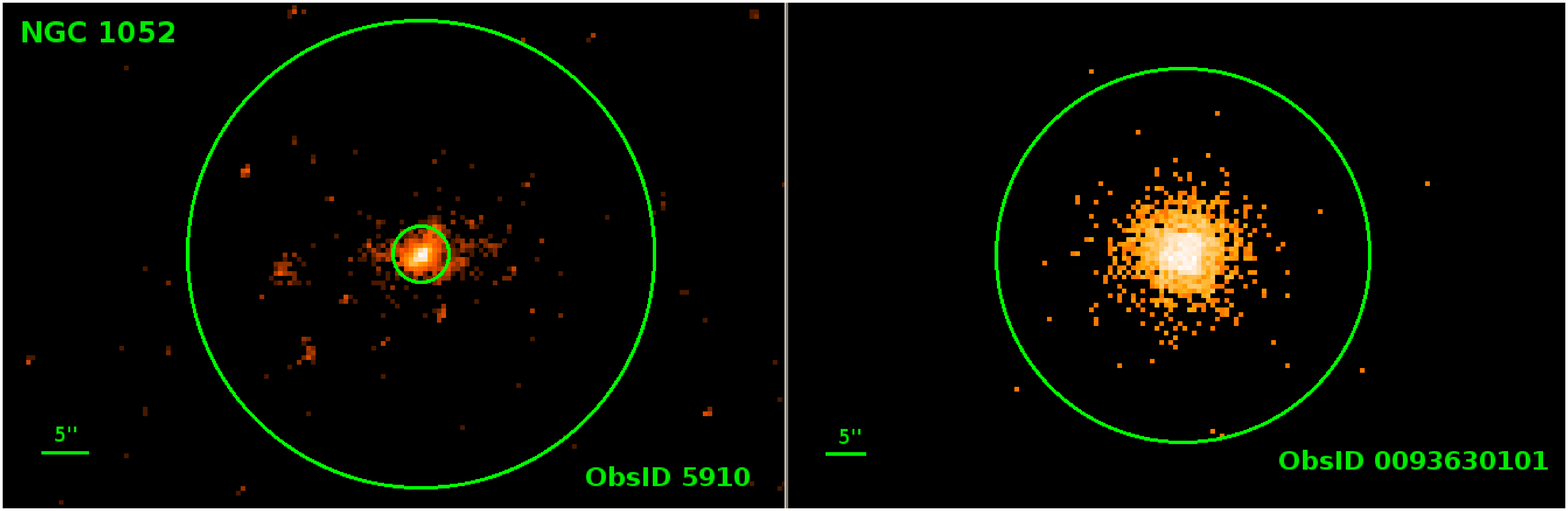}

\vspace*{0.5cm}
\includegraphics[width=\textwidth]{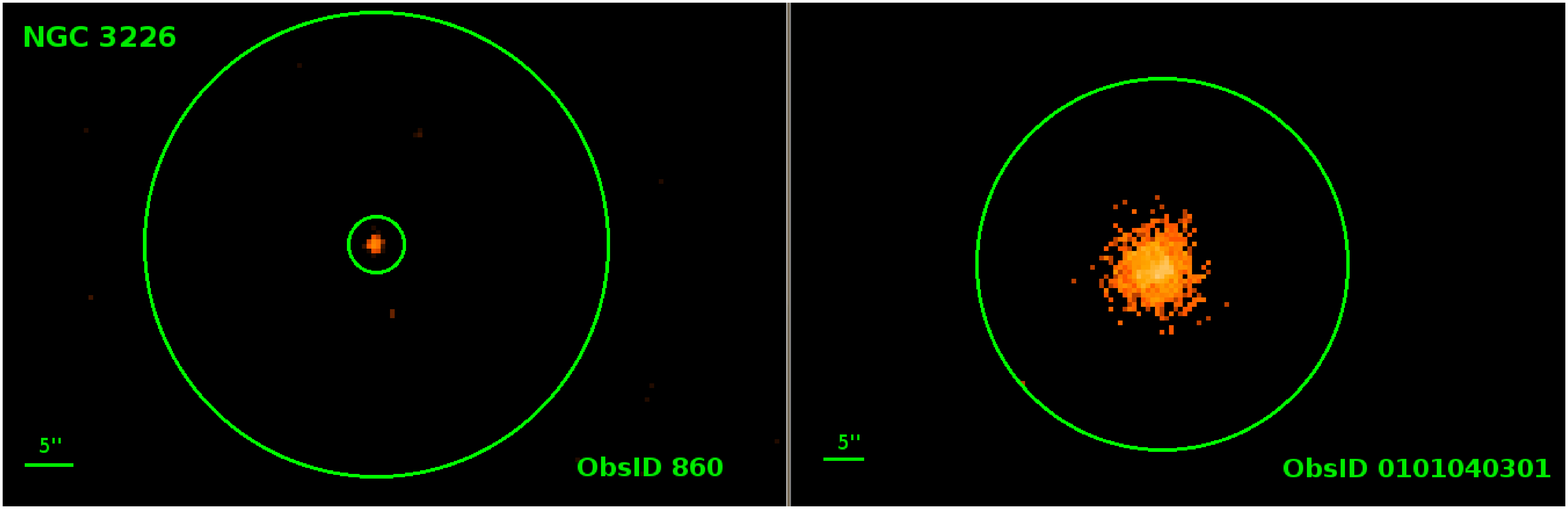}

\vspace*{0.5cm}
\includegraphics[width=\textwidth]{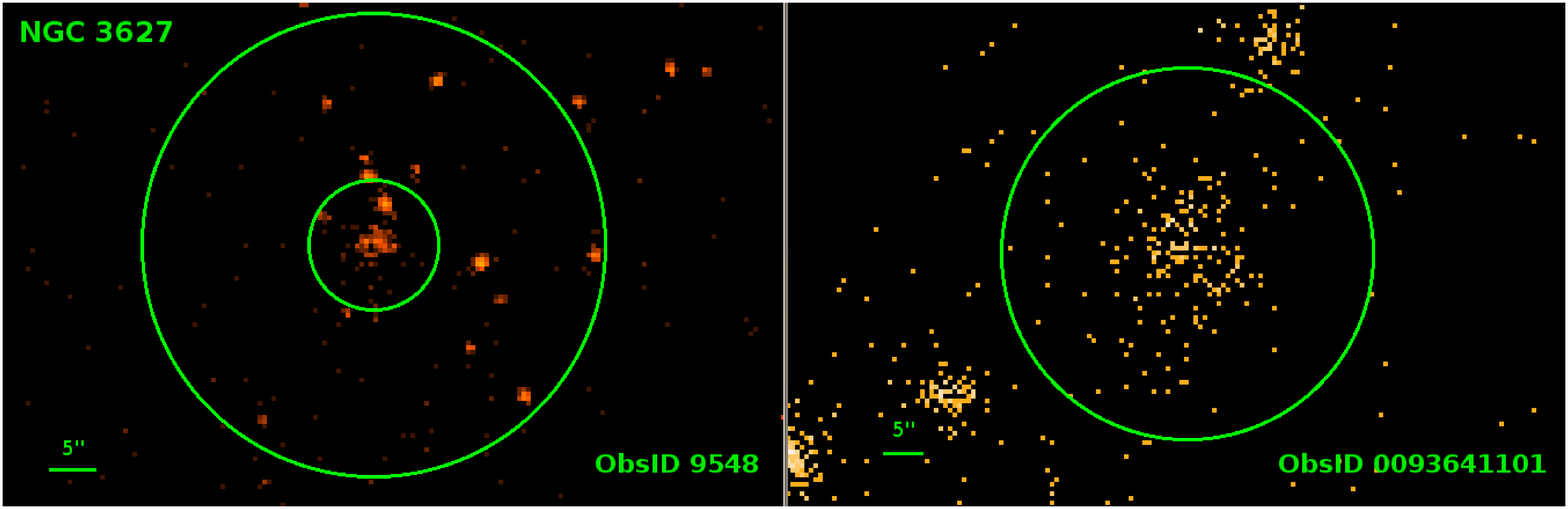}

\vspace*{0.5cm}
\includegraphics[width=\textwidth]{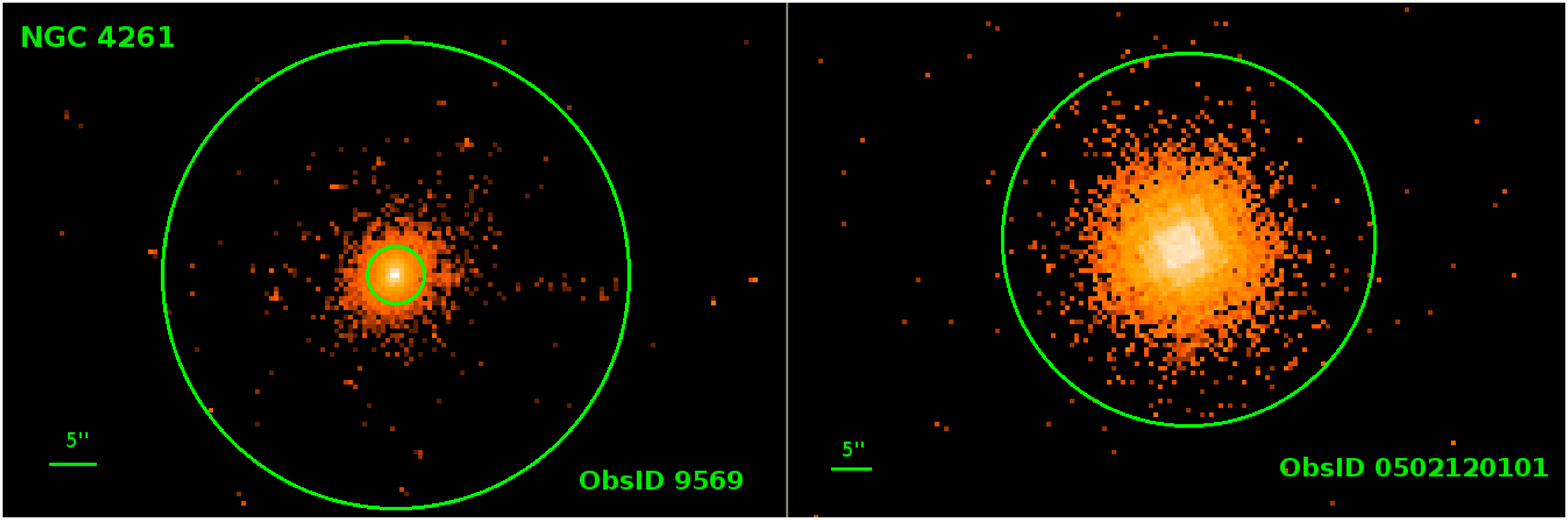}
\end{figure*}

Firstly, we extracted a spectrum from \emph{Chandra} data with an aperture radius of 25$\arcsec$. Secondly, a spectrum of an annular region was extracted from \emph{Chandra} data, whith $R_{ext}=25\arcsec$ and $R_{int}=R_{ \emph{Chandra}}$ (Col. 4 in Table \ref{obs}). When the contamination by the annular region to the 25$\arcsec$ \emph{Chandra} data emission was higher than 50\% in the 0.5-10.0 keV energy band (see Section \ref{results}), we did not consider the joint analysis since the accuracy of the derived parameters could be seriously affected. For lower contamination levels, we considered that \emph{Chandra} data could be used to estimate the contribution of the annular region to the \emph{XMM}-Newton spectrum. We extracted the \emph{Chandra} spectrum in that ring (from $R_{int}$ to $R_{ext}$) and fitted the five models explained in Section \ref{individual}. Then we included the resulting model, with its corresponding parameters frozen, in the fit of the \emph{XMM}-Newton nuclear spectrum (the circular region with $R_{ext}$), and extracted the parameters for the nuclear emission. Appendix \ref{images} shows the images corresponding to the data used for this analysis, where the different apertures are shown. This analysis was made for all the seven objects with \emph{Chandra} and \emph{XMM}-Newton spectra taken in similar dates. 

\subsection{Flux variability}

We computed X-ray luminosities for the individual and simultaneous fits. UV luminosities were also obtained when possible (see below).

\subsubsection{X-ray luminosities}

We computed the X-ray luminosities and sigma errors for the soft and hard bands using XSPEC. Note that distances of the sources were taken from NED. We assumed a given object as variable if the luminosity variation was larger than 3$\sigma$ and as non-variable when its variations were below 1$\sigma$.

\subsubsection{\label{uvlum} UV luminosities}

Simultaneous \emph{XMM}-Newton Optical Monitor (OM)\footnote{http://xmm.esac.esa.int/external/xmm\_user\_support/ documentation\-/technical/OM/} data were used to compute the UV luminosities. This monitor has three different filters in the UV range, UVW2 centred at 1894$\AA$ (1805-2454) $\AA$, UVM2 centred at 2205$\AA$ (1970-2675) $\AA$, and UVW1 at 2675$\AA$ (2410-3565) $\AA$. We used these three filters whenever possible.

We used the OM observation FITS source lists\footnote{ftp://xmm2.esac.esa.int/pub/odf/data/docs/XMM-SOC-GEN- ICD-0024.pdf} to get the photometry. We checked that the photometry using IRAF and the Vega magnitude system and calculating the luminosities by using SAS to estimate the count rate gave similar results. When OM data were not available we searched for UV information in the literature (see Appendix \ref{previous}).

When simultaneous observations from X-rays and UV with \emph{XMM}-Newton were available, we computed the X-ray to UV flux ratio defined as:

\begin{equation} \alpha_{ox} = \frac{log(L_x/L_{UV})}{log(\nu_x/\nu_{UV})} \end{equation}

where $L_{UV}$ were computed from UVM2 and UVW1 filters and $Lx$ was computed using the integral:

\begin{equation} F_x (0.5-2.0 keV) = \int_{0.5keV}^{2.0keV} F_{\nu}(2 keV) \Big(\frac{\nu}{\nu_{2keV}}\Big)^{(1-\Gamma)} d\nu \end{equation}

leading to:

\begin{equation} L(2 keV) = \frac{L(0.5-2.0 keV)}{\nu_{2 keV}} \frac{2-\Gamma}{1-0.25^{2-\Gamma}} \end{equation}

\subsection{Short time scale variability}

We analyzed the light curves for each observation to check for variability on short time scales in the sample.

Assuming a constant count rate for the whole observation, we calculated the $\rm{\chi^2/d.o.f}$ test as a first approximation to the variations. We considered the source to be variable if the count rate differed from the average above 3$\rm{\sigma}$ (or 99.7\% probability).

To check the variability amplitude of the light curves, we calculated the normalized excess variance, $\rm{\sigma_{NXS}^2}$. We followed prescriptions given by \cite{vaughan2003} to estimate $\rm{\sigma_{NXS}^2}$ and its error, $\rm{err(\sigma_{NXS}^2)}$  (see also \cite{omaira2011a}):

\begin{equation} \sigma_{NXS}^2 = \frac{S^2 - <\sigma_{err}^2>}{<x>^2} \end{equation}

\begin{equation} err(\sigma_{NXS}^2) = \sqrt{\frac{2}{N} \Big(\frac{<\sigma_{err}^2>}{<x>^2}\Big)^2 + \frac{<\sigma_{err}^2>}{N} \frac{4\sigma_{NXS}^2}{<x>^2} } \end{equation}

where $x$, $\sigma_{err}$ and $N$ are the count rate, its error and the number of points in the light curve, respectively, and $S^2$ is the variance of the light curve:

\begin{equation} S^2 = \frac{1}{N-1} \sum_{i=1}^N (x_i - <x>)^2 \end{equation}

%
\section{\label{results}Results}

\subsection{Individual objects}

Here we present the results of the variability of the seven sources individually. General results are given in Section~\ref{specvar}. Each subsection describes the following: the observations used in the analysis (Table \ref{obs}), variations of the hardness ratio (from Col. 8 in Table \ref{obs}), individual and simultaneous best fit and the parameters varying in the model (see Tables \ref{ftestcol1}, \ref{ftestcol2}, \ref{results} and Figure \ref{bestfig}), X-ray flux variations (see Table \ref{lumincorr} and Figure \ref{luminfig}), the analysis of the annular region when data of \emph{Chandra} and \emph{XMM}-Newton were used together (Table \ref{annulus} and Appendix \ref{images}), and the simultaneous fittings of these observations (Table \ref{simultanillo}), short term variability from the analysis of the light curves (see Table \ref{estcurvas} and Appendix \ref{lightcurves1052}-\ref{lightcurves5846}), and UV luminosities when simultaneous data from the OM monitor was available (Table \ref{luminUV}, Figure \ref{luminfig}). Moreover, a summary of the variability is given in Table \ref{variability}. Notes and comparisons with previous works for individual objects are included in Appendix \ref{previous}.

\subsubsection{NGC\,1052}

We used one \emph{Chandra} and four \emph{XMM}-Newton observations. These four \emph{XMM}-Newton observations were taken from August 2001 to August 2009, and the \emph{Chandra} observation was taken in August 2000 (see Table \ref{obs}).

Variations of 33\% (10\%) in HR were obtained between the first and the last \emph{XMM}-Newton (\emph{Chandra}) observations (see Column 8 in Table \ref{obs}).

The individual fits gave ME2PL as the best fit. In this case SMF2 was used, being the best representation of the observed differences (see Fig. \ref{bestfig}) when varying $Norm_2$ and $N_{H2}$ (see Tables \ref{ftestcol1}, \ref{ftestcol2} and \ref{bestfit}). Variations were 49\% ($Norm_2$) and 31\% ($N_{H2}$) between the first and the last observations.

In Fig. \ref{luminfig}, variations of the soft and hard intrinsic luminosities of the simultaneous fitting are presented. At soft energies we found variations at 8.3$\sigma$ (20\%), and at hard energies at 7.5$\sigma$ (20\%), in a period of 8 years. The largest variation for $Norm_2$ was found between the second and the third observations (see Table \ref{bestfit}), with an interval of 3 years, where both soft and hard luminosities varied 12\% and 29\% respectively (see Table \ref{lumincorr}). The strongest variation for $N_{H2}$ was obtained between the first and the second observations (see Table \ref{bestfit}), with a 32\% change in 4 years.

We compared the \emph{Chandra} observation from 2000 with the \emph{XMM}-Newton observation from 2001 (see Figure \ref{images}), following the prescriptions given in Sect.~ \ref{apertures}. The spectral analysis of \emph{Chandra} data was included in Tables \ref{bestfit} and \ref{lumincorr}, which gave ME2PL as the best fit. The annular region represented a 10\% of the 25$\arcsec$ \emph{Chandra} aperture luminosity in the 0.5-10 keV band. The \emph{Chandra} data for the 25$\arcsec$ radius circular region provided intrinsic luminosities representing 22\% (75\%) for the soft (hard) energy of the emission from \emph{XMM}-Newton data (see Table \ref{lumincorr}). After taking into account the contribution from the annular region, the analysis indicated no changes in one year period (see Tables \ref{annulus} and \ref{simultanillo}) in the nuclear emission. 

According to the values of $\chi^2_r$ and $\sigma^2$ showed in Table \ref{estcurvas}, the analysis from the light curves did not show short time scale variations (see Figure \ref{lightcurves1052}), neither \emph{XMM}-Newton nor in \emph{Chandra} data, since variations were below 3$\sigma$.

In Fig. \ref{luminfig} UV luminosities (Table \ref{luminUV}) are represented for the UVW2 and UVM2 filters. Variations at 7.3$\sigma$ (or 16-25 \%) and 2$\sigma$ (or 2-23\%) were obtained, respectively.

\subsubsection{ \label{res3226} NGC\,3226}

We used two \emph{XMM}-Newton observations in November 2000 and December 2006 and one \emph{Chandra} observation from December 1999. 

Variations in the HR of 100\% were obtained both in \emph{Chandra} and in \emph{XMM}-Newton data (see Column 8 in Table \ref{obs}). However, ObsID 0400270101 from \emph{XMM}-Newton was not used for further discussion (see below). 

The observation from 2000 gave as best fit the 2PL model, while that from 2006 was best fitted with a PL. The simultaneous spectral fitting was better represented by a 2PL model varying $N_{H2}$ (i.e. SMF1, Figure \ref{bestfig}), with a 74\% amplitude variation. The resulting parameters (see Tables \ref{bestfit} and \ref{lumincorr}) indicate that X-ray soft and hard intrinsic luminosity variations were below 1$\sigma$ over a period of 6 years.

We compared the \emph{Chandra} observation from 1999 with the \emph{XMM}-Newton observation from 2000. The spectral analysis of \emph{Chandra} data was included in Tables \ref{bestfit} and \ref{lumincorr}. The contribution from the annular region to the 25$\arcsec$ aperture \emph{Chandra} data was 20\% in the 0.5-10.0 keV band. The \emph{Chandra} spectrum extracted with 25$\arcsec$ aperture represented 62\% (70\%) of the \emph{XMM}-Newton soft (hard) emission (Table \ref{annulus}). When the contribution from the annular region was taken into account, the simultaneous fit resulted in variations of $N_{H2}$ (93\%) and $Norm_2$ (57\%), with 37\% (81\%) variations in the soft (hard) energies in one year period (Table \ref{simultanillo}).  

We analyzed short time scale variability from individual light curves, by calculating $\chi^2$ and $\sigma^2$ (see Table \ref{estcurvas} and Figure \ref{lightcurves3226} (up)). Data from 2006 showed $\chi^2_r = 4.7$ and a normalized excess variance $\rm{\sigma_{NXS}=}$1.8 $\rm{\pm}$ 0.1 $\rm{\times 10^{-2}}$ indicating variability (see Figure \ref{lightcurves3226}, upper-right). However, NGC\,3227 is a strong variable Seyfert 1 located 2$\arcmin$ from NGC\,3226. Thus, we analyzed the possibility that NGC\,3226 was contaminated by its emission. We extracted a light curve from NGC\,3227 and from a circular region between both galaxies close to NGC\,3226 (background), and the same pattern of variability was found. In Figure \ref{lightcurves3226} we present the light curves for the background (middle-left) and NGC\,3227 (middle-right), and the \emph{XMM}-Newton image (down). The normalized excess variance for the background light curve was $\rm{\sigma_{NXS}=}$1.5 $\rm{\pm}$ 0.5 $\rm{\times 10^{-2}}$. Therefore, we conclude that NGC\,3226 might be contaminated by emission from NGC~3227, so its short time scale variability cannot be assessed.
%
Only UVW1 observations are available from OM (see Table \ref{luminUV} and Figure \ref{luminfig}), with a 10-12 \% variation (7.4$\sigma$).

\subsubsection{\label{res3627} NGC\,3627}

We used one \emph{XMM}-Newton observation in May 2001 and another \emph{Chandra} observation in March 2008 (see Table \ref{obs}). We recall that different apertures (8$\rm{\arcsec}$ for \emph{Chandra} and 25$\rm{\arcsec}$ for \emph{XMM}-Newton) were used for the extraction of the nuclear spectrum.

Since observations were obtained with different instruments, comparisons of HR were avoided.

According to $\chi^2_r$, both spectra were individually best-fitted with a MEPL model. The best simultaneous fit implied variations in $N_{H2}$ (Figure \ref{bestfig}), from no absorption to $N_{H2}=1.28 \times 10^{22} cm^{-2}$ in a 7 year period (i.e. SMF1, Table \ref{bestfit}). Even if observed fluxes vary, when computing intrinsic luminosities (see Table \ref{lumincorr}) for this model we get variations below 1$\sigma$ in the soft and the hard energies, indicating no variations in 7 years (Figure \ref{luminfig}). 

To compare \emph{Chandra} and \emph{XMM}-Newton data, we carried out the analysis explained in Sect.~\ref{apertures}. 
The contribution of the annular region to the 25$\arcsec$ aperture \emph{Chandra} spectrum is $\sim$92\% in the 0.5-10 keV band (see Table \ref{annulus}). Thus we assumed that \emph{XMM}-Newton data were strongly contaminated by emission surrounding the nucleus, which avoided the joint use of Chandra and \emph{XMM}-Newton data for this object (see Section \ref{apertures}).

The analysis of the light curves did not show short time scale variability (see Table \ref{estcurvas} and Figure \ref{lightcurves3627}), since all measures were below 2$\sigma$ from the average.

\subsubsection{NGC\,4261}

We used two \emph{Chandra} (May 2000 and February 2008) and two \emph{XMM}-Newton observations (December 2001 and December 2007) for this object. Given the different resolutions in both sets of observations we firstly performed the analysis separately (see Table \ref{obs}).

Variations of 19\% and 0\% in HR were obtained for \emph{Chandra} and \emph{XMM}-Newton, respectively (from Column 8 in Table \ref{obs}).

The best fit for the \emph{Chandra} spectra is a ME2PL model from the individual analysis. The simultaneous fit without allowing to vary any parameter (i.e. SMF0) resulted in a good fit ($\chi^2/d.o.f = 1.27$). Varying one parameter did not give an improvement in the final fit. Therefore, the source seemed to be non-variable (see Table \ref{bestfit} and Figure \ref{bestfig}). X-ray luminosity variations were below 1$\sigma$ for the soft and the hard bands over a period of 8 years (Table \ref{lumincorr} and Figure \ref{luminfig}).

The individual analysis of the two observations with \emph{XMM}-Newton again gave as a best fit the ME2PL model, and variations in the parameters did not improve the fit (i.e. the best fit was SMF0). Thus, we obtained a non-variable source (Table \ref{bestfit} and Figure \ref{bestfig}). X-ray luminosity variations were below 1$\sigma$ in a 6 year period in this case (Table \ref{lumincorr} and Figure \ref{luminfig}).

To compare data from \emph{Chandra} and \emph{XMM}-Newton, the procedure explained in section \ref{apertures} was applied to this object. We compared obsID 9569 from \emph{Chandra} and obsID 0502120101 from \emph{XMM}-Newton data (Figure \ref{images}) since they are the closest in time. The contribution of the emission from the annular region was 37\% in the 0.5-10 keV band emission in the 25$\rm{\arcsec}$ aperture \emph{Chandra} data. Intrinsic luminosities of the 25$\rm{\arcsec}$ aperture \emph{Chandra} spectrum represented 37\% (66\%) of the soft (hard) emission from \emph{XMM}-Newton data. The simultaneous fit between these data taking into account the annular contribution resulted in a non-variable object (Table \ref{simultanillo}).

To check for short time scale variability we analyzed the light curves for each observation (see Table \ref{estcurvas} and Figure \ref{lightcurves4261}). No short time variability was detected for this object, since all measurements were below 2$\sigma$ from the average.

Considering the UV range (Table \ref{luminUV}), the variations amounted to 9-11 \% (10.3$\sigma$) in the UVW1 filter and 28-39 \% (9.3$\sigma$) in the UVM2 filter (see Figure \ref{luminfig}).

\subsubsection{NGC\,4278}

We only used three of the nine observations taken by \emph{Chandra}, in March 2006, February 2007 and April 2007, and the \emph{XMM}-Newton observation in May 2004 (Table \ref{obs}).

HR variations amounted to 4\% for the runs with useful spectroscopic \emph{Chandra} data. Taking into account all the observations from \emph{Chandra}, 40\% HR variations were found between 2000 and 2010 (from Column 8 in Table \ref{obs}).

Our best fit for \emph{Chandra} data was a MEPL model with $Norm_2$ varying (i.e. SMF1, see Figure \ref{bestfig}). This parameter varied 30\% between the first and the last observation ($\approx$ 1 year apart) (Table \ref{bestfit}). X-ray intrinsic luminosity variations (Table \ref{lumincorr}) were within 11.2$\sigma$ (9.6$\sigma$) for the soft (hard) energies. This corresponded to variation amplitudes of 26\% (29\%, see Figure \ref{luminfig}). It is remarkable that flux variations were 11\% (13\%) between the second and the third observations (two months apart), with a 13\% variation in $Norm_2$ for the same period (Tables \ref{bestfit} and \ref{lumincorr}).

We compared the \emph{XMM}-Newton observation in 2004 with the \emph{Chandra} observation in 2006, which is the closest in time (see Figue \ref{images}). We applied the procedure explained in section \ref{apertures}. The spectral analysis of \emph{XMM}-Newton spectrum, included in Tables \ref{bestfit} and \ref{lumincorr}, gave a PL as the best fit model. The contribution of the annular region is 38\% in the 0.5-10.0 keV band to the emission in the 25$\rm{\arcsec}$ aperture \emph{Chandra} data. The \emph{Chandra} spectrum extracted with 25$\rm{\arcsec}$ aperture represented 25\% (20\%) of the soft (hard) \emph{XMM}-Newton emission. When the contribution of the annular region was taken into account, the resulting $\Gamma$ was in agreement with that from \emph{XMM}-Newton data (see Table \ref{annulus}). The simultaneous fit shows 15\% of variations in the normalization of the PL along two years (Table \ref{simultanillo}).

The analysis of the light curves (see Table \ref{estcurvas} and Figure \ref{lightcurves4278}) did not show short time scale variability, either in \emph{Chandra} or in \emph{XMM}-Newton data.

\subsubsection{NGC\,4552}

We used a \emph{Chandra} observation taken in April 2001 and a \emph{XMM}-Newton observation taken in July 2003. We recall that different apertures were used (3$\rm{\arcsec}$ for \emph{Chandra} and 25$\rm{\arcsec}$ for \emph{XMM}-Newton data).

Since observations were obtained with different instruments, comparisons of HR were avoided.

Both observations needed MEPL for the individual best fit. When varying parameters, the best fit was obtained when $Norm_1$ and $Norm_2$ varied (i.e. SMF2, $\chi^2_r $=1.21, see also Figure \ref{bestfig}), with 93\% and 78\% amplitude variations (see Table \ref{bestfit}). We found intrinsic luminosity variations at 21.5$\sigma$ (14.1$\sigma$) in the soft (hard) energies, i.e., 87\% (79\%) amplitude variations in a period of two years (Table \ref{lumincorr}).

The \emph{Chandra} image of this object revealed many X-ray sources surrounding the nucleus (see Figure \ref{images}). The contribution of the annular region to the 0.5-10.0 keV band emission in the 25$\rm{\arcsec}$ aperture \emph{Chandra} data was 23\%. Extraction from \emph{Chandra} data with 25$\rm{\arcsec}$ aperture resulted in a 72\% (60\%) emission of the soft (hard) \emph{XMM}-Newton data. The simultaneous fit resulted in a variable object, where $Norm_1$ (21\%) and $Norm_2$ (37\%) varied, with  29\% (37\%) flux variations in the soft (hard) energies (see Figure \ref{luminfig} and Table \ref{simultanillo}).

Short time scale variations were not found (see Table \ref{estcurvas} and Figure \ref{lightcurves4552}).

\subsubsection{NGC\,5846}

We used two observations from \emph{XMM}-Newton (January and August 2001) and other two observations from \emph{Chandra} (May 2000 and June 2007). The analysis was firstly made separately due to different apertures (see Table \ref{obs}).

HR Variations of 1\% and 6\% were obtained for \emph{XMM}-Newton and \emph{Chandra}data, respectively (from Column 8 in Table \ref{obs}).

In the case of \emph{Chandra} observations, we only made the analysis up to 3 keV, due to low count-rate at hard energies. In this case SMF0 was used, resulting in a non-variable source when fitting a MEPL model (Table \ref{bestfit} and Figure \ref{bestfig}) with flux variations below 1$\sigma$ in a period of 7 years (Table \ref{lumincorr} and Figure \ref{luminfig}). The \emph{XMM}-Newton data did not show variability (the best fit model is MEPL, see Table \ref{bestfit} and Figure \ref{bestfig}), and flux variations below 1$\sigma$ on the soft and hard energies for a period of 7 months (Table \ref{lumincorr}).

The contribution to the 0.5-10.0 keV band emission from the 25$\rm{\arcsec}$ aperture \emph{Chandra} data was 73\% from the annular region, 
what avoided the joint use of Chandra and \emph{XMM}-Newton data for this object. No short time scale variability was found for this object 
(see Table \ref{estcurvas} and Figure \ref{lightcurves5846}). The availability of UV data for a single epoch precludes any attempt to get information on its variability.

\subsection{\label{specvar}Spectral variability}

A rough description of the spectral shape is provided by HR. Consequently, a first approximation of the spectral variations can be obtained from the variation of HR. The use of HR allows the inclusion of observations for which the number of counts are not enough for a proper spectral fitting. Note that, since the calculation of HR is based on number counts, we only used data coming from the same instrument for comparisons. We considered that variations greater than 20\% (2$\sigma$ error) in HR may correspond to variable objects (NGC\,1052, NGC\,3226, and NGC\,4278). Note that variations in HR smaller than 20\%  can be found in variable 
objects, since the variations in soft and hard energies may have 
different signs and somewhat compensate in the final calculation of HR. 

The individual fitting of each observation revealed that composite models (2PL, MEPL or ME2PL) were needed in all cases. A thermal component was used in six objects, all of them with kT$\approx$0.60 keV.

We fitted all available data, for the same object and model, varying different parameters to get information on the variability pattern. Fig. \ref{bestfig} shows the best fit (top panel) with the residuals of the individual observations (bottom panels). In order to analyze the data from \emph{Chandra} and \emph{XMM}-Newton data jointly, we estimated the influence of extra-nuclear emission on the results (see Section \ref{apertures}). For two objects (NGC\,3627 and NGC\,5846) the contamination by emission surrounding the nucleus in \emph{XMM}-Newton data was so high (up to 50\%) that we avoided the comparisons in these cases. For the remaining five sources (NGC\,1052, NGC\,3226, NGC\,4261, NGC\,4278, and NGC\,4552), the joint analysis was attempted.

Two objects are compatible with being non-variable sources, namely NGC\,4261 and NGC\,5846. For these two objects the same conclusion is reached if we isolate the analysis of \emph{Chandra} and \emph{XMM}-Newton data, both with similar spectral parameters (see Table \ref{bestfit}). Moreover, NGC\,3627 was not longer used for further discussion, since contamination prevents of any variability analysis. Four objects are variable, namely NGC\,1052, NGC\,3226, NGC\,4278, and NGC\,4552. NGC\,1052 and NGC\,3226 showed variations in $N_{H2}$ (31\% and 93\%) and $Norm_2$ (49\% and 57\%) (in eight and one years), while NGC\,4278 showed changes in $Norm_2$ (30\%) (in three years). NGC\,4552 showed variations in $Norm_1$ (21\%) and $Norm_2$ (37\%) in a two years period. All the variations occur at hard energies, and these variations are related to the absorber and/or to the nuclear power. Even if the small number of sources precludes any statistical characterization, there does not seem to exist any clear relation with either the LINER type (1 or 2) or to the Eddington ratios (see Table \ref{variability}). A larger sample of LINERs is needed to search for any variability pattern to be eventually more frequently observed than others.

\subsection{Flux variability}

X-ray soft and hard luminosities (see Table \ref{lumincorr}) are shown in Figure \ref{luminfig}. Two objects are compatible with no variations of the central engine (NGC\,4261, and NGC\,5846). Four sources are compatible with being variable, NGC\,1052 with 20\% variations in both bands, NGC\,3226 with 37\% (81\%) variations in the soft (hard) band, NGC\,4278 with 26\% (29\%) variations in the soft (hard), and NGC\,4552 with 29\% (37\%) variations in the soft (hard) band. Sources showing flux variability in the soft and hard bands also showed spectral variability (see Sec. \ref{specvar}).

We also studied the UV variability for the sources, by studying UV and X-ray data obtained simultaneously with \emph{XMM}-Newton (available for three galaxies). All of them are variable at UV frequencies (see Table \ref{luminUV} and Fig. \ref{luminfig}). \emph{XMM}-OM provided UV fluxes at three epochs for NGC\,1052; $\sim$ 20\% variations were obtained with the filters UVW2 and UVM2. NGC\,3226 was observed with the filter UVW1, showing 11\% variation. NGC\,4261 was observed with the filters UVW1 and UVM2, with 10\% and 33\% variations, respectively. Two objects, namely NGC\,1052 and NGC\,3226, showed UV and spectral variability, while another one, NGC\,4261, showed UV variability but not spectral variability.

\subsection{Light curves}

Table \ref{estcurvas} provides the values for $\rm{\chi^2_r}$ (and the probability of variability) and the $\rm{\sigma^2_{NXS}}$. Objects are considered to be variable when the count rates are different 3$\rm{\sigma}$ from the average. {No short term variability (from hours to days) was found in our sample \footnote{Note that we do not take into account obsID 0400270101 from NGC\,3226 (see Section \ref{res3226}).}.}


\section{\label{discusion}Discussion}

We have performed an X-ray spectral analysis to search for variability in seven LINER nuclei, three type 1.9 (NGC\,1052, NGC\,3226, and NGC\,4278) and four type 2 (NGC\,3627, NGC\,4261, NGC\,4552, and NGC\,5846). We used data from \emph{Chandra} and \emph{XMM}-Newton satellites with observations at different epochs. Whenever possible, we made the analysis separately for each instrument to avoid corrections due to different apertures. The results obtained for the long term variability of NGC\,3627 will not be used for further discussion (see Section \ref{res3627}).

Our main results are:

\begin{itemize}

\item Short term variability: No variations were found on time scales from hours to days.

\item Long term variability: Four out of the six objects (NGC\,1052, NGC\,3226, NGC\,4278, and NGC\,4552) were variable in X-rays on time scales from months to years. The shortest variation is found for NGC\,4278 in time scales of two months. Simultaneous observations in the UV for three objects (NGC\,1052, NGC\,3226, and NGC\,4261) revealed variations on time scales of years.

\item Main driver for the variability: Among the variable sources, NGC\,4278 presented variations in $\rm{Norm_2}$, NGC\,4552 in $\rm{Norm_1}$ and $\rm{Norm_2}$, and NGC\,1052 and NGC\,3226 in $\rm{N_{H2}}$ and $\rm{Norm_2}$. In all the variable LINERs variations occur at hard energies.
%
\end{itemize}

\subsection{Short and long time scale variability}

In our sample of LINERs we analyzed variability from hours to days (short term) from the analysis of the light curves for each observation (see Col. 6 in Table \ref{obs}), and from months to years (long term) from the simultaneous fitting of the different observations (see Col. 10 in Table \ref{variability}).

Concerning short time-scales, \cite{pian2010} studied four type 1 LINERs and found variations of 30\% in a half a day in  NGC\,3998, and 30\% variations in the hard (1-10 keV) X-rays in $\sim$ 3 hours in M~81. The other two sources in their sample (NGC\,4203 and NGC\,4579) did not show short time scale variability. The power specral density (PSD) profiles of the 14 LINERs included in the sample of 104 AGN in \cite{omairavaughan2012}, did not show short-term variability except for two objects (3C218 and NGC\,3031). All these studies suggest that about 20\% of LINERs show short time-scale variability. This percentage goes to zero in our own study, since none of our seven sources showed short time scale variability, according to the $\rm{\chi^2_r}$ and the normalised excess variance, $\rm{\sigma_{NXS}^2}$, (see Table \ref{estcurvas}). \cite{younes2010} found variability in $\sim$ 1.5 hours for NGC\,4278 on the \emph{XMM}-Newton observation, where the flux increased 10\%. Using the same observation we found a 3\% flux increase in the same period, and a null probability of being a variable source. The difference is most probably due to the different appertures used in the analysis. Adding up all the studied LINERs from this and previous papers, the percentage of variable LINERs at short time-scales is 16\%.

Long term spectral variability is clearly found for four objects in our sample.
NGC\,1052 needed variations in $\rm{N_{H2}}$ (49\%) and $\rm{Norm_2}$ (31\%) in a period of eight years, NGC\,3226 also varied $\rm{N_{H2}}$ (93\%) and $\rm{Norm_2}$ (57\%) in a period of one year, NGC\,4278 varied $\rm{Norm_2}$ (30\%) in a period of one year, and NGC\,4552 varied $\rm{Norm_1}$ (21\%) and $\rm{Norm_2}$ (37\%) in two years.

Long term variability is common among LINERs. \cite{younes2011} studied a sample of type 1 LINERs, where seven out of nine sources showed long term variability (i.e. months and/or years). Two of their objects are in common with our sample, namely NGC\,3226 and NGC\,4278. We found similar spectral characteristics and the same parameters varying for both objects. In the case of NGC\,3226, they found that the $\rm{N_{H}}$ varied 72\% and $\rm{Norm}$ varied 48\% when fitting a simple power law. This is similar to our results, although we used two absorbers instead of one. They found flux variations of 49\% (46\%) in the soft (hard) band between the \emph{Chandra} and \emph{XMM}-Newton observations, while we found 37\% (81\%), the differences most probably due to the different models used. In the case of NGC\,4278 they used seven public observations from the archive, while we only used three. Despite this, the same spectral variation was found ($\rm{Norm_2}$). \cite{younes2010} found for NGC\,4278 a 31\% (20\%) variation at soft (hard) energies, where we found 26\% (29\%) variation for the same observations. \cite{pian2010} found long term variability in two of their four type 1 LINERs using \emph{Swift} data. Therefore, long term variability if found in $\sim$65\% of the LINERs, significantly larger than that of LINERs at short time scales.

Three kinds of long term X-ray variability patterns have been found in our sample: (1) variations in the soft excess (NGC\,4552); (2) variations of the obscuring matter (NGC\,1052 and NGC\,3226); and (3) variations of the intrinsic source (NGC\,1052, NGC\,3226, NGC\,4278, and NGC\,4552). As in the case of NGC\,4552, variations at soft energies has already been reported in the literature for the type 2 LINER NGC\,4102 \citep{omaira2011b}. This behaviour is seen in type 1 Seyferts, where absorption variations are related to a partially ionized, optically thin material along the line of sight to the central source, the so called warm absorber \citep{reynolds1997, petrucci2013}. The variations due to the $\rm{N_H}$ of the X-ray absorbing gas that we see in two sources are well established for many famous type 2 Seyferts. These variations are thought to be related to the motion of clouds perpendicular to the line of sight of the observer to the AGN. These clouds produce partial eclipses of the AGN over time. In some cases, the fast movement of the clouds places them at the distance of the BLR, although in other cases the clouds seem to be located at further distances (few parsecs) from the AGN \citep[e.g. NGC1365][]{risaliti2002, risaliti2007, risaliti2010, braito2013}. The time scale of the $\rm{N_H}$ variations in our sample are consistent with this latter scenario. The most common pattern of variability among the LINERs in our sample (four cases) is the change on the intrinsic continuum of the source. \cite{mchardy2006} found that the time scale of the intrinsic variability increases for larger masses of the black hole and/or lower bolometric luminosities for objects where variability is related to the nuclear power\footnote{Although the dependence on the bolometric luminosity does not seem to be so strong according to \citet{omairavaughan2012}}. According to the revised relation between the BH mass, bolometric luminosities and time scales of variations, \cite{omairavaughan2012} predicted that LINERs (with $\rm{M_{BH} \sim 10^{8-9} M_{\odot}}$) do not vary in time scales lower than tens of days. Applying their formula to our four objects, predicted values for NGC\,1052, NGC\,3226, NGC\,4278, and NGC\,4552 were 13.3, 28.2, 118.0 and 449.9 days (see Table \ref{variability} for the values of $\rm{M_{BH}}$ and $\rm{L_{bol}}$), that is, time scales of days, months and years were expected for these objects. This is in agreement with our results for NGC\,4278 and NGC\,4552. Unfortunately, we do not have observations within days for NGC\,1052 and NGC\,3226, we were only able to search for variations on time scales from months to years. Moreover, in the cases of NGC\,1052, NGC\,3226, and NGC\,4552 coupled variations \footnote{Variations are needed for more than one parameter in the spectral fitting} were obtained. A study of a larger sample of LINERs will be required to constrain the time scale of the intrinsic variability for these sources and be able to understand whether LINERs match in the same scenario than more powerful AGN.

A first approximation of the variations can be obtained by the analysis of the hardness ratios. We have considered differences in HR larger than 20\% as a measure of spectral variation. For
 NGC\,1052, HR varied 33\% between the first and last \emph{XMM}-Newton
 observations and no variation is found for HR over the 5 year period 
 observed by
 \emph{Chandra}. These results are compatible with flux variations obtained
 when analyzing \emph{XMM}-Newton data. NGC\,3226 presents HR differences
 over 100\% in \emph{Chandra} data. For NGC\,4278 total HR variations
 amount to 40\% when using all the \emph{Chandra} available data, although
 the results from the \emph{Chandra} data used for spectral analysis
 appear to be compatible with no variations. NGC\,4261 and NGC\,5846
 are compatible with being non-variable objects, both with
 \emph{XMM}-Newton and \emph{Chandra} data, and with both analyses, spectroscopic and HR. NGC\,4261 shows HR differences of 19\%, which
 seem to be compatible with the flux variations obtained
 through the individual analysis of the source.

Variability among LINERs is not restricted to X-rays. The work done by \cite{maoz2005} was the first to show variability at UV frequencies in LINER galaxies, where all but three objects in their sample of 17 LINERs 1 and 2 were variable. From the literature, we found different studies for the LINERs in our sample using \emph{HST} data (see Appendix \ref{previous} for details). \cite{cappellari1999} studied FOC data for NGC\,4552, and found a factor 4.5 brightening between 1991 and 1993 (filter F342W), followed by a factor $\sim$2 dimming between 1993 and 1996 (filters F175W, F275W and F342W). \cite{maoz2005} studied both NGC\,1052 and NGC\,4552, concluding that both of them were variable on time scales of years. NGC\,4278 was studied by \cite{cardullo2008}, who found a luminosity increase of a factor 1.6 over 6 months. UV data were not available for the remaining two objects in our sample (NGC\,3627 and NGC\,5846). Thus, from the seven LINERs in our sample five seem to be variable at UV frequencies. Simultaneous X-rays and UV data were obtained from \emph{XMM}-Newton data for three objects (NGC\,1052, NGC\,3226, and NGC\,4261), all showing variability, whereas intrinsic variations in X-rays are not found for NGC\,4261. A possible explanation for this source to the non-simultaneous X-ray and UV variation could be the existence of time lags in both frequencies. Time lag explanations has already been reported for the NLSy1 galaxy NGC\,4051 by \cite{alston2013}. Repeated, simultaneous observations at X-rays and UV frequencies would be required for verifying this model. 

\subsection{Accretion mechanism}

\begin{figure*}[H]
\caption{ \label{gammaredd} Anticorrelation between the spectral index, $\Gamma$ from individual fits, vs. the Eddington ratio, $log(L_{bol}/L_{Edd})$, for our sample galaxies. The solid line represents the relation given by \cite{younes2011}, while the dashed line represents that by \citep{gucao2009}, both shifted to the same bolometric correction (see text).}
\centering
\includegraphics[width=\textwidth]{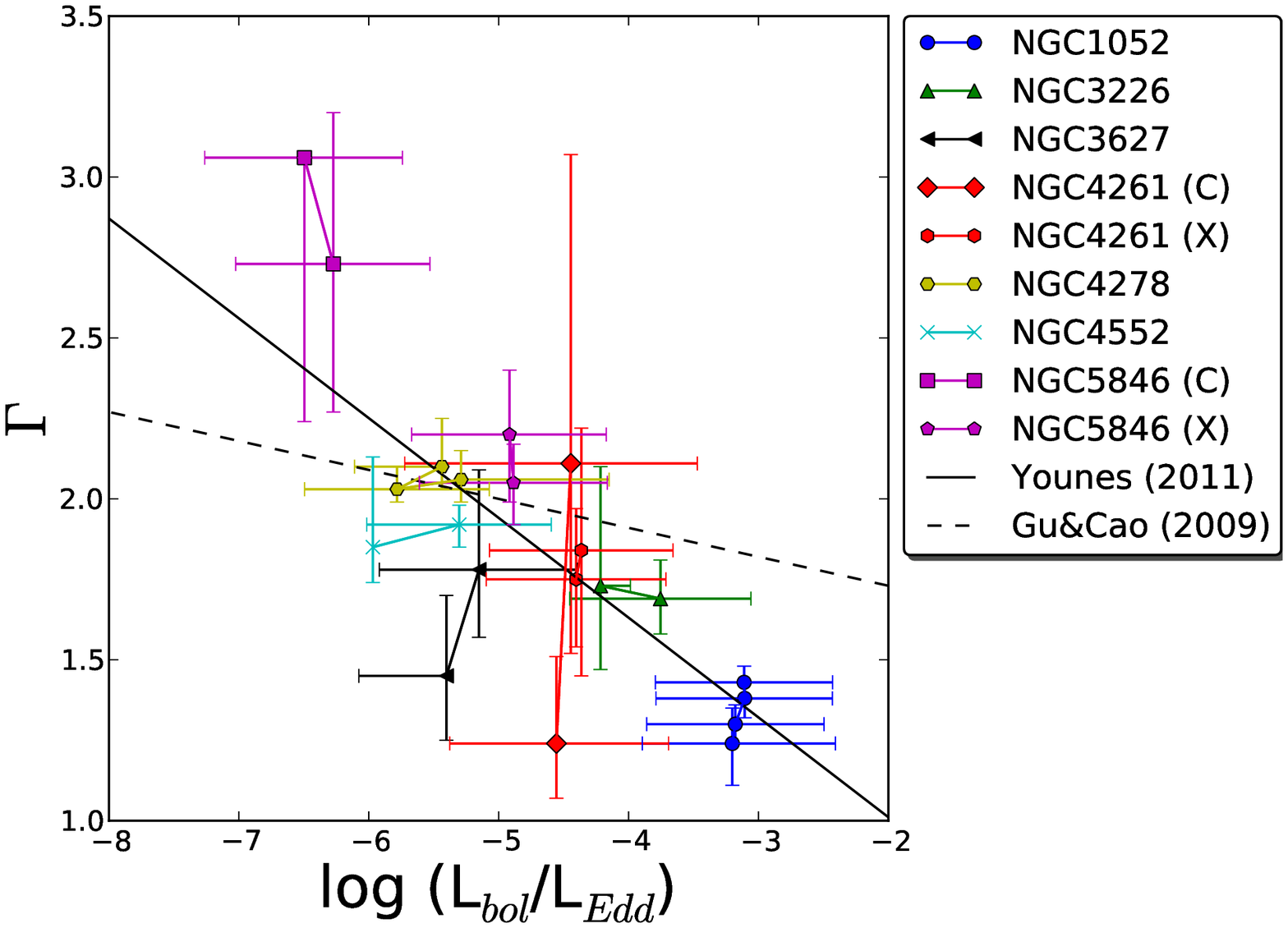} 
\end{figure*}

It has been suggested in the literature that the accretion mechanism in low luminosity active galactic nuclei (LLAGN) is different from that in more powerful AGN (e.g. Seyferts), and more similar to that in X-Ray Binaries (XRB) in their low/hard state \citep{yamaoka2005, yuan2007, gucao2009, younes2011, xu2011}. The X-ray emission is supposed to originate from the Comptonization process in advection dominated accretion flow (ADAF), where accretion is inneficient for $\rm{L_{bol}/L_{Edd} < 10^{-3}}$. At low accretion rates, the infalling material may never cool sufficiently to collapse into a thin disk (as is the case for efficient radiation), and an advection-dominated flow from the outermost radius down to the black hole could be formed \citep{narayanyi1994}. In powerful AGN a positive correlation between the hard X-ray photon index, $\rm{\Gamma}$, and the Eddington ratio, $\rm{L_{bol}/L_{Edd}}$ was found by \cite{shemmer2006}, who argued that the hard X-ray photon index depends primarily on the accretion rate. On the contrary, according to the results provided by \cite{mahadevan1997}, the lower the accretion rate, the less efficient is the cooling by comptonization and the X-ray region of the spectrum becomes softer and reaches lower luminosities. In this case a negative correlation between these magnitudes has been found for LLAGN \citep{gucao2009, younes2011}, and also for XRB. We present these parameters for each individual observation in our sample of LINERs in Figure \ref{gammaredd}, where the negative correlation is shown. $\rm{Log(L_{bol}/L_{Edd})}$ were calculated following the formulation given in \cite{eracleous2010}, using $\rm{L_{bol}=33L_{2-10 keV}}$. We corrected the equation given by \cite{younes2011} by this factor (they used $\rm{L_{bol}=16L_{2-10 keV}}$) and plotted it as a solid line. Our results are consistent with the correlation given by \cite{younes2011} for their sample of type 1 LINERs. \cite{emmanoulopoulos2012} found for the first time a `harder when brighter' (i.e. higher luminosities for harder spectra) X-ray behaviour for the LLAGN NGC\,7213, where they found variations in $\rm{\Gamma}$. However, we did not find this behaviour for any of the sources in our sample, Figure \ref{gammaredd} showing the consistency of our simultaneous fittings with no variations in $\rm{\Gamma}$.

We also computed the X-ray to UV flux ratio, $\rm{\alpha_{ox}}$ (see Section \ref{uvlum}). We calculated these values for all sources with simultaneous observations at X-ray and UV (see Table \ref{luminUV}), obtaining values of $\rm{\alpha_{ox}}$ between [-0.81, -1.66], in good agreement with previous studies \citep{maoz2007, younes2012}. Despite being similar to the $\rm{\alpha_{ox}}$ given for powerful AGN, these values were slightly lower \citep{maoz2007}. This may indicate the lack or absence of the `big blue bump'.

In the cases where $\rm{\alpha_{ox}}$ could be calculated more than once (NGC\,1052 and NGC\,4261) no variations were found within the errors.

\begin{figure*}[H]
\caption{ \label{alpharedd} Correlation between the spectral index, $\alpha_{ox}$, vs. the Eddington ratio, $log(L_{bol}/L_{Edd})$. Starred symbols correspond to the sources in \cite{younes2012}.}
\centering
\includegraphics[width=\textwidth]{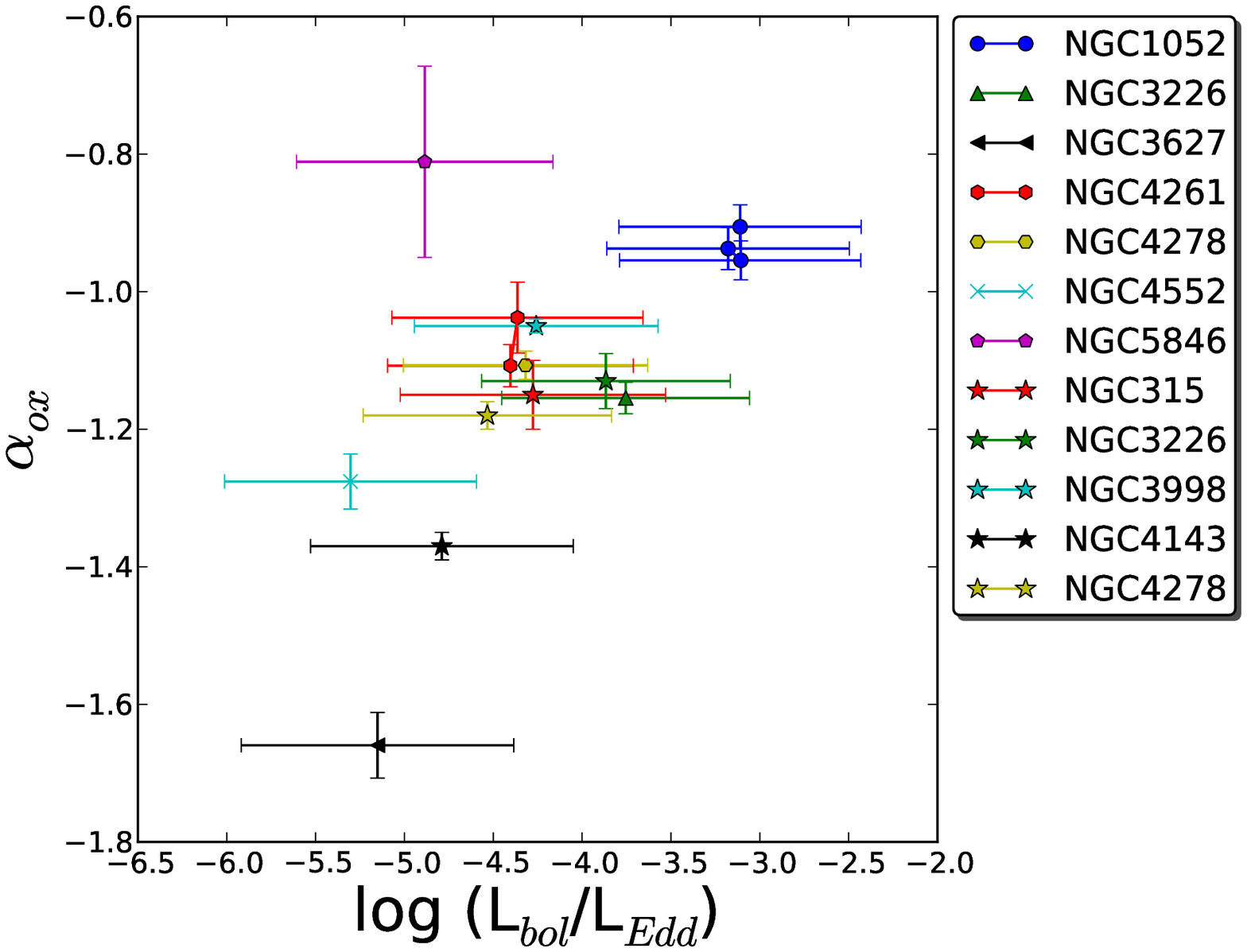} 
\end{figure*}

Another indication of a different emission process between powerful AGN and LINERs could be the positive correlation found by \cite{younes2012} between $\rm{\alpha_{ox}}$ and $\rm{log(L_{bol}/L_{Edd})}$, in contrast with the anticorrelation found for powerful AGN. In order to compare the results from \cite{younes2012} and ours, we recalculated $\rm{log(L_{bol}/L_{Edd})}$ for the sample in \cite{younes2012} following the relation given by \cite{eracleous2010} and using the data from \cite{younes2011}. In Fig. \ref{alpharedd} we plot this relation (see Table \ref{luminUV}), where the symbols used for the sources from \cite{younes2012} are stars. The results are in good agreement, indicating a correlation between $\rm{\alpha_{ox}}$ and $\rm{log(L_{bol}/L_{Edd})}$. \cite{younes2012} suggested that this behaviour can be understood within the framework of radiatively inefficient accretion flow models, such as ADAF.

%

\section{\label{conclusion}Conclusions}

A spectral variability analysis of seven LINER nuclei was performed using public data from \emph{Chandra} and \emph{XMM}-Newton. The main conclusions of this paper can be summarized as follows:

   \begin{enumerate}
   
      \item Variations greater than 20\% in the hardness ratio always correspond to objects showing spectral variability.

      \item Individual fits of each observation provided composite models as the best fit (2PL, MEPL, and ME2PL).

      \item No short time scale variability was found, in agreement with predictions.

      \item Spectral X-ray variability was found in four out of six objects. In all of them variations occurred at hard energies due to the absorber and/or the nuclear source, and variations in the soft energy were found only in NG\,4552. These variations occur on time scales of months and/or years. The shortest time scale was found for NGC\,4278, with variations of two months.

      \item We found an anticorrelation between the X-ray spectral index, $\Gamma$, and the Eddington ratio, $L_{bol}/L_{Edd}$. We have also found a correlation between the X-ray to UV flux ratio, $\alpha_{ox}$, and the Eddington ratio, $L_{bol}/L_{Edd}$. Both relations are compatible with inefficient flows being the origin of the accretion mechanism in these sources.

   \end{enumerate}

\begin{acknowledgements}

    We thank the anonymous referee for his/her helpful comments that helped to improve the paper, T. Mahoney for reviewing the English, J. Perea for help with the statistical analysis and the AGN group at the IAA for helpfull comments during this work. This work was financed by MINECO grant AYA 2010-15169, Junta de Andaluc\'{i}a TIC114 and Proyecto de Excelencia de la Junta de Andaluc\'{i}a P08-TIC-03531. LHG acknowledges financial support from the Ministerio de Econom\'{i}a y Competitividad through the Spanish grant FPI BES-2011-043319. OGM thanks Spanish MINECO through a Juan de la Cierva Fellowship. This research made use of data obtained from the \emph{Chandra} Data Archive provided by the \emph{Chandra} X-ray Center (CXC). This research made use of data obtained from the \emph{XMM}-Newton Data Archive provided by the \emph{XMM}-Newton Science Archive (XSA). This research made use of the NASA/IPAC extragalactic database (NED), which is operated by the Jet Propulsion Laboratory under contract with the National Aeronautics and Space Administration.

\end{acknowledgements}

\bibliographystyle{aa}
\bibliography{000referencias}

\begin{figure*}[H]
\caption{\label{bestfig}For each object, all X-ray spectra are plotted together in the first row. Best fits and their residuals are also shown, one row per observation from second row on. The legends contain the date (in the format yyyymmdd) and the obsID. Details given in Table \ref{obs}. }
\centering
\includegraphics[width=0.4\textwidth]{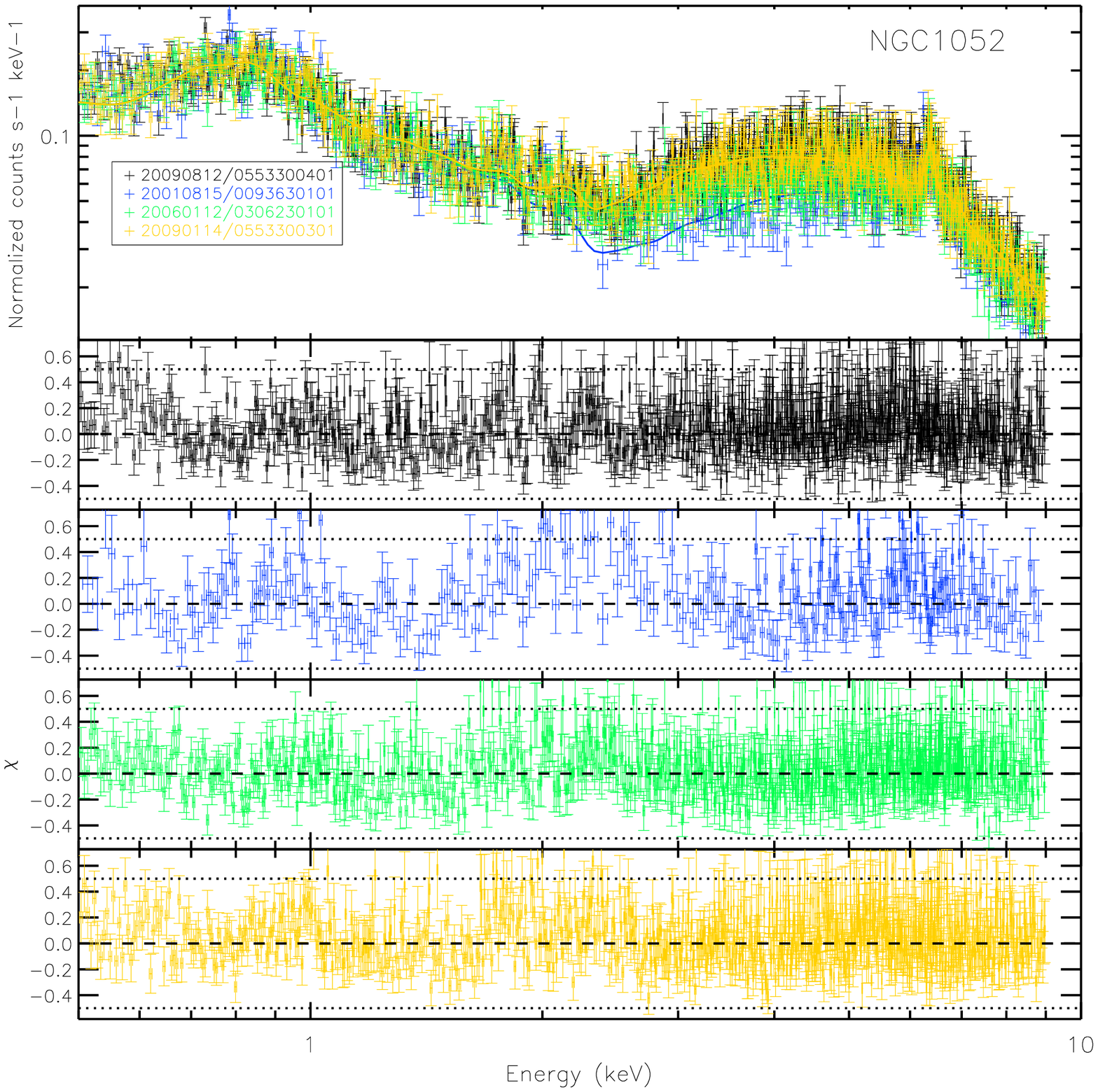} 
\includegraphics[width=0.4\textwidth]{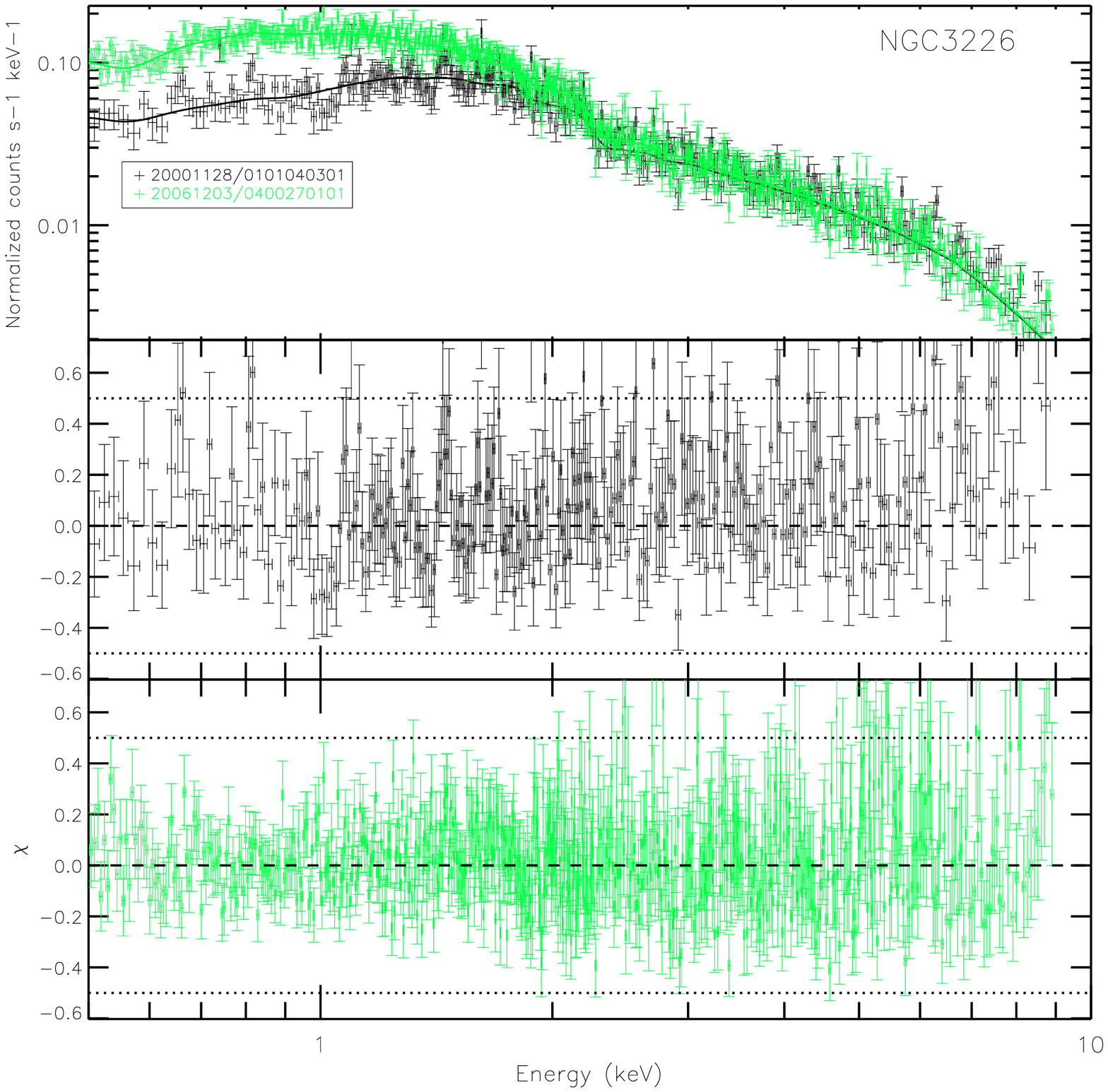}
\includegraphics[width=0.4\textwidth]{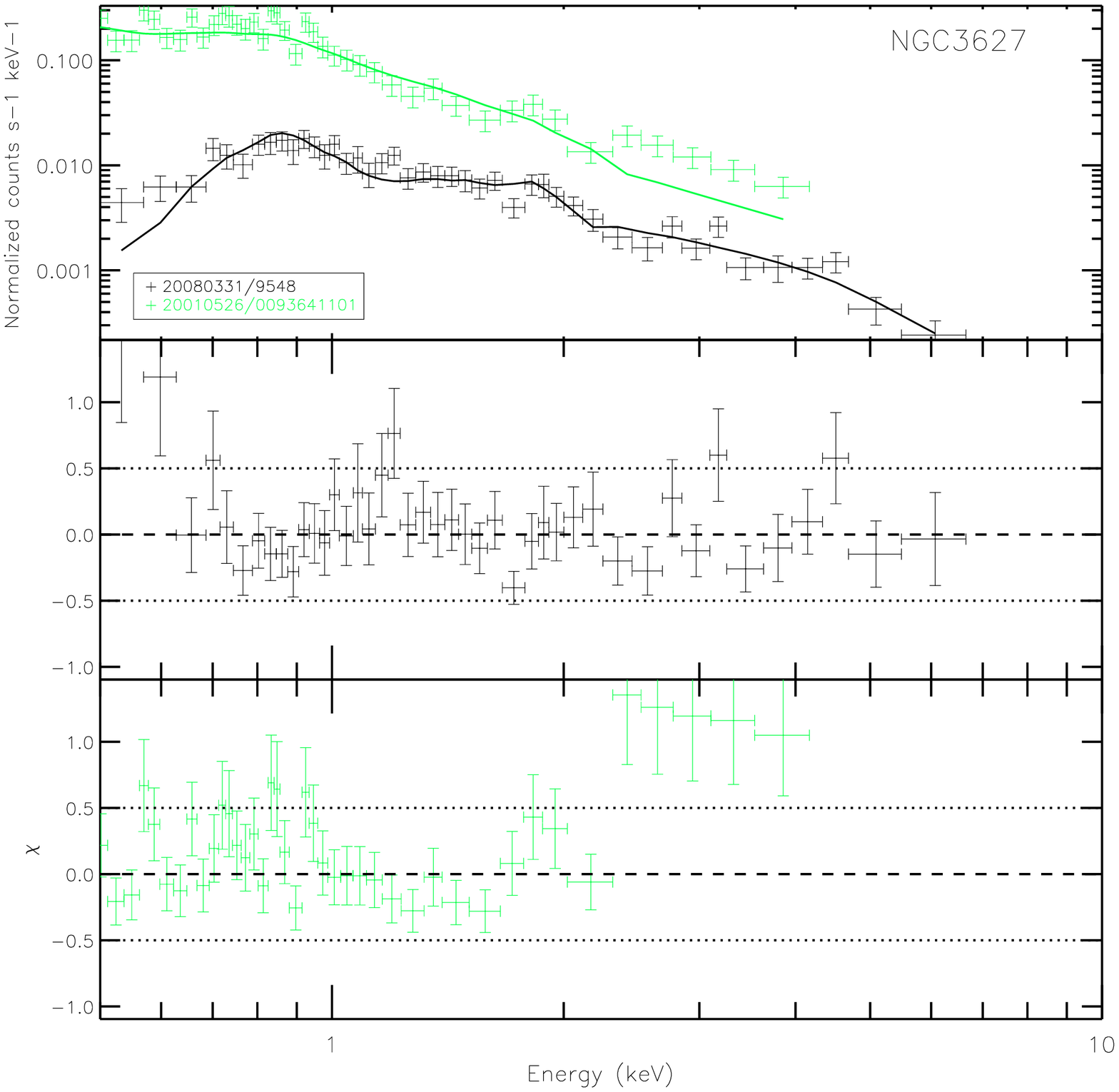} 
\includegraphics[width=0.4\textwidth]{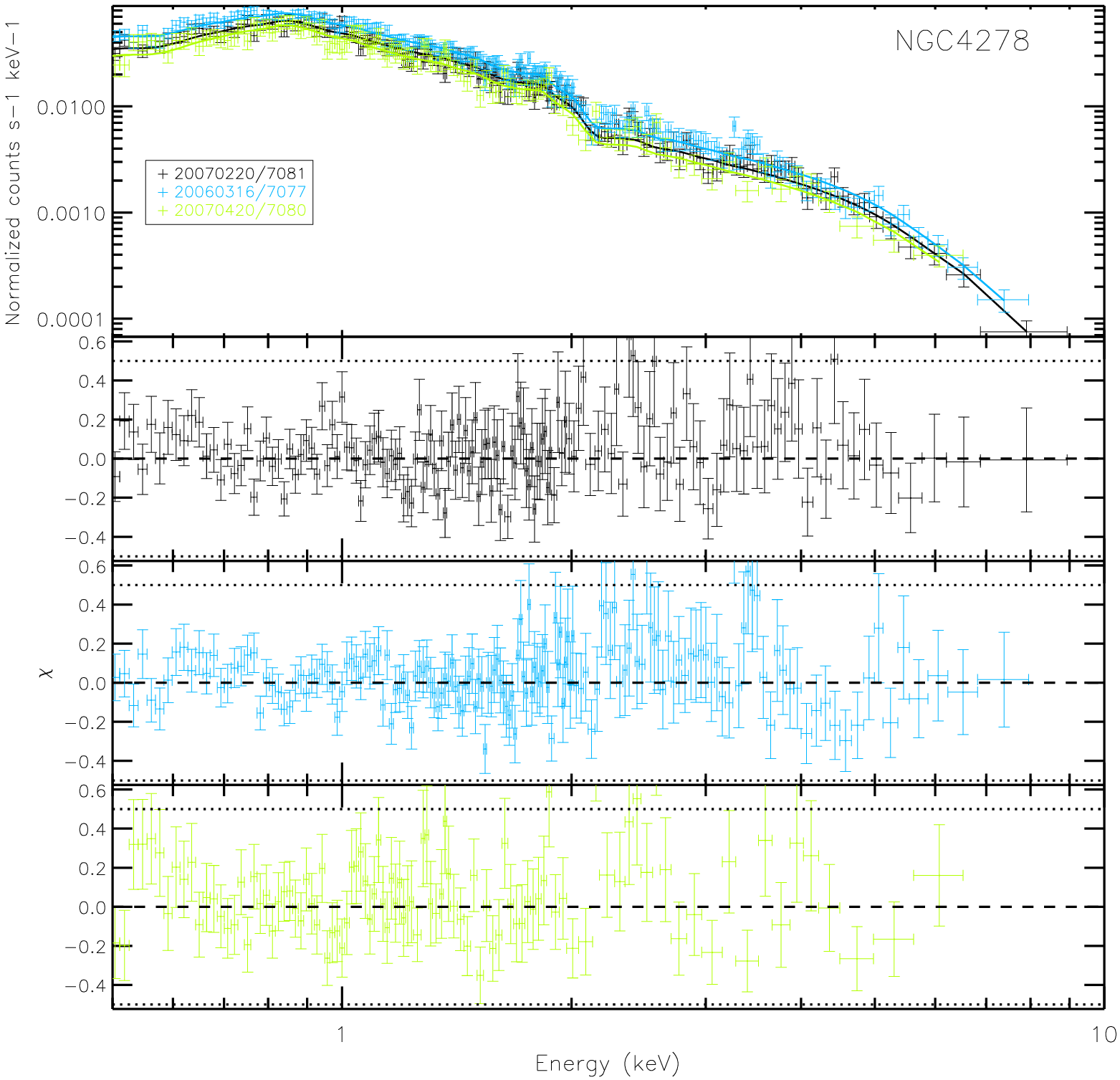}  
\includegraphics[width=0.4\textwidth]{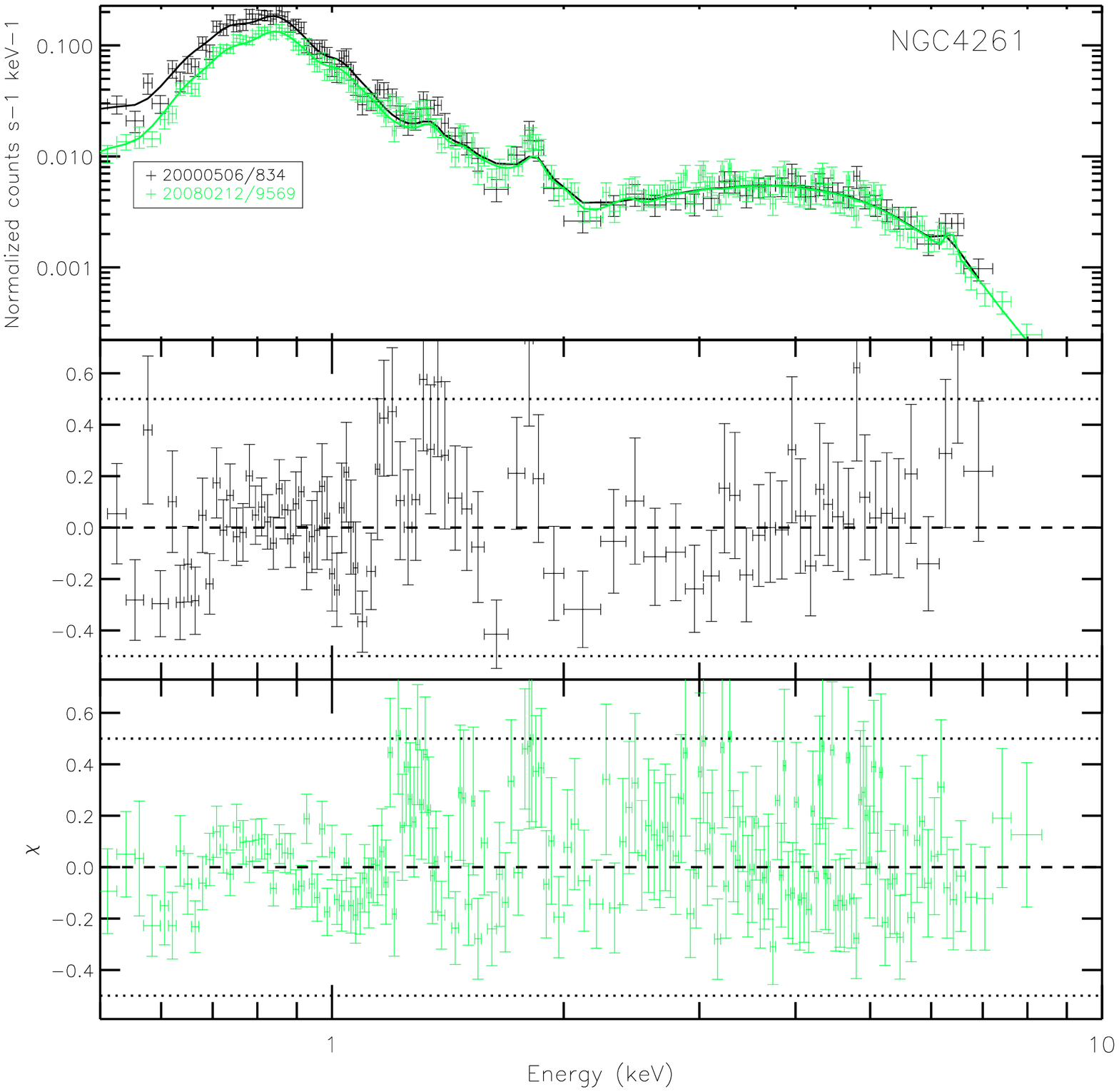} 
\includegraphics[width=0.4\textwidth]{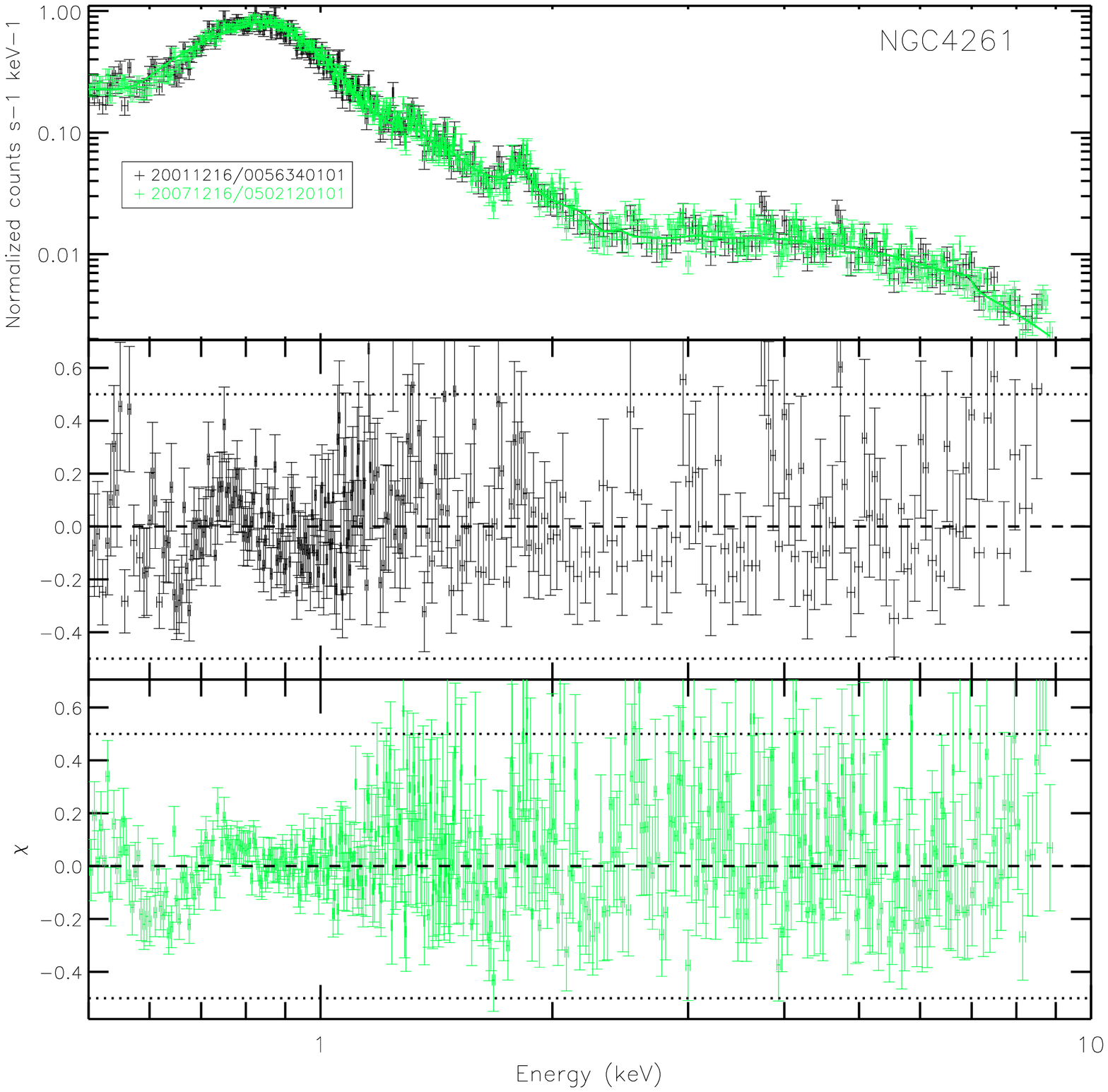}
\end{figure*}

\begin{figure*}[H]
\setcounter{figure}{0}
\caption{Cont.}
\centering
\includegraphics[width=0.4\textwidth]{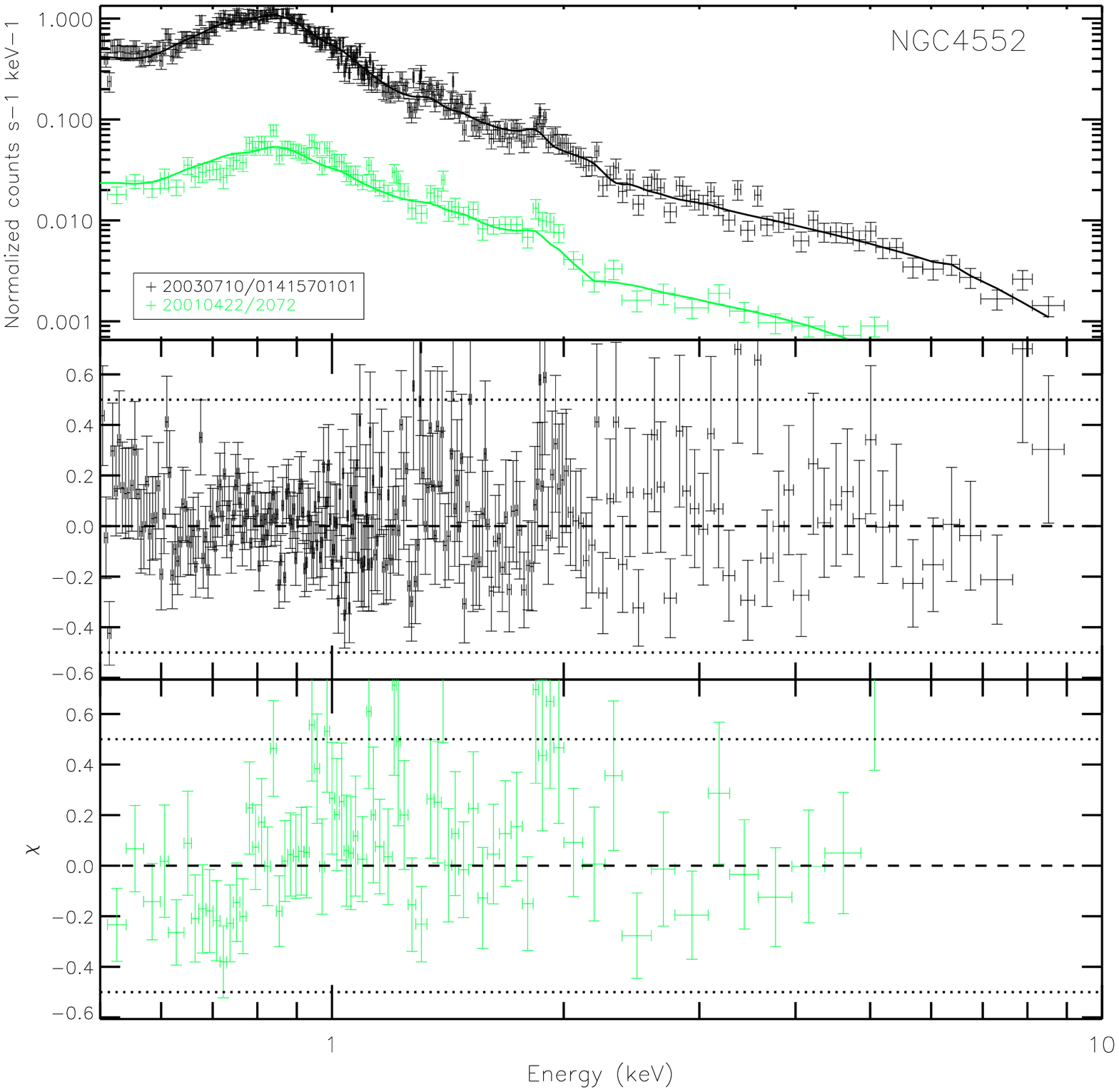}
\includegraphics[width=0.4\textwidth]{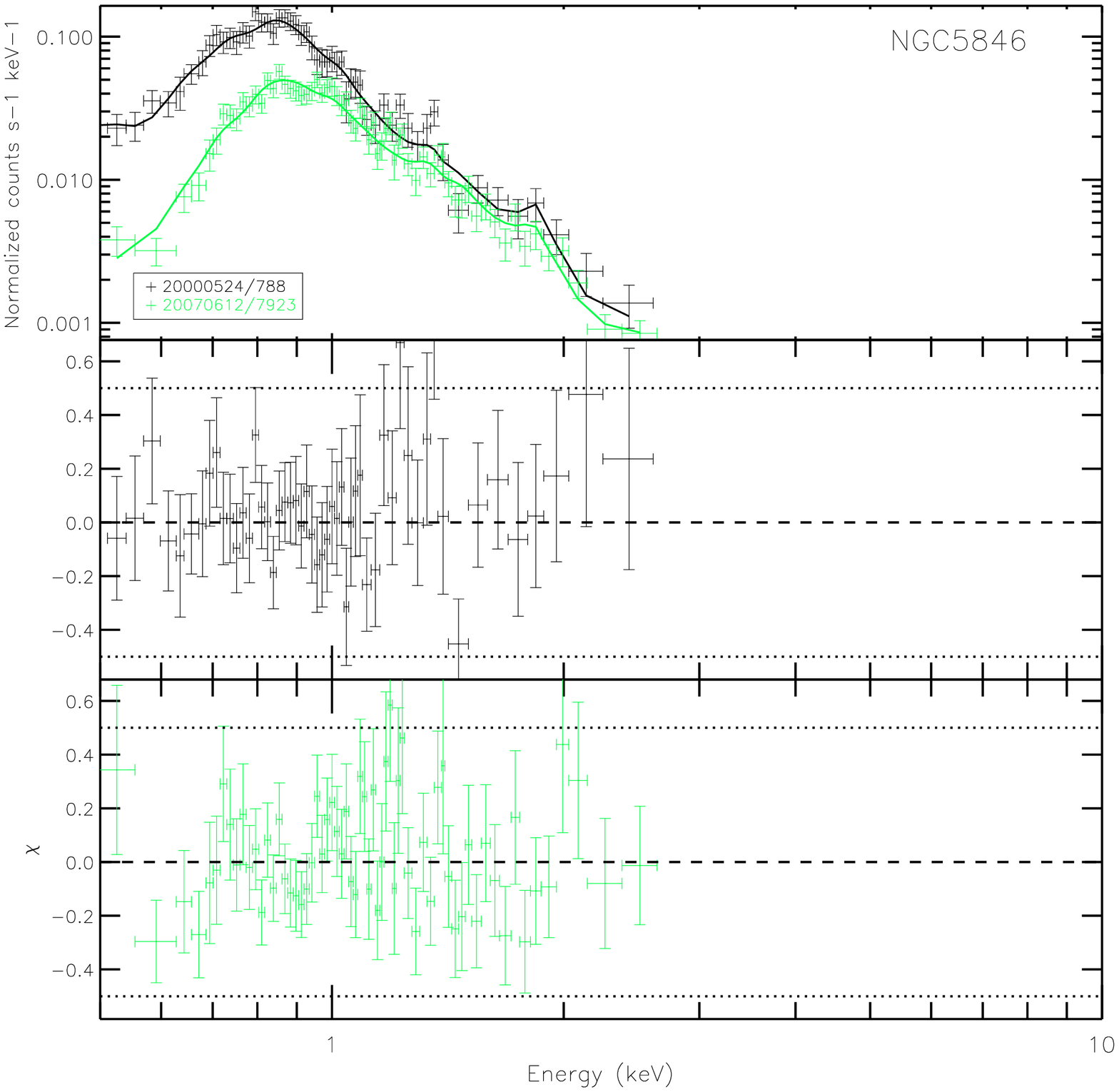} 
\includegraphics[width=0.4\textwidth]{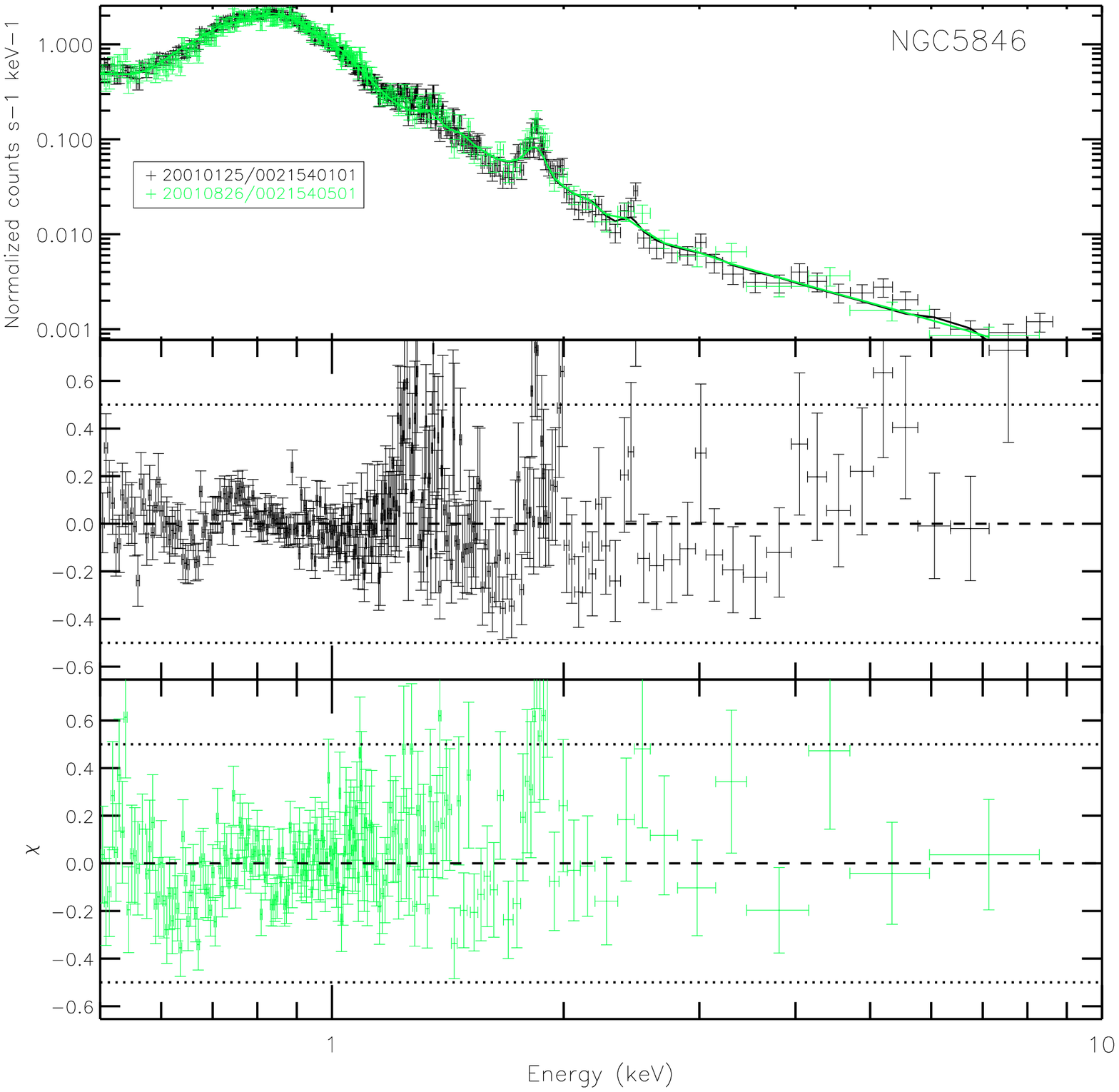}  
\end{figure*}

\begin{figure*}[H]
\centering
\caption{ \label{luminfig} Left: Intrinsic luminosities calculated for the soft (0.5-2.0 keV) and hard (2.0-10.0 keV) energies in the simultaneous fitting. Right: UV luminosities obtained from the data with the OM camera onboard \emph{XMM}-Newton, when available.}
\includegraphics[width=0.4\textwidth]{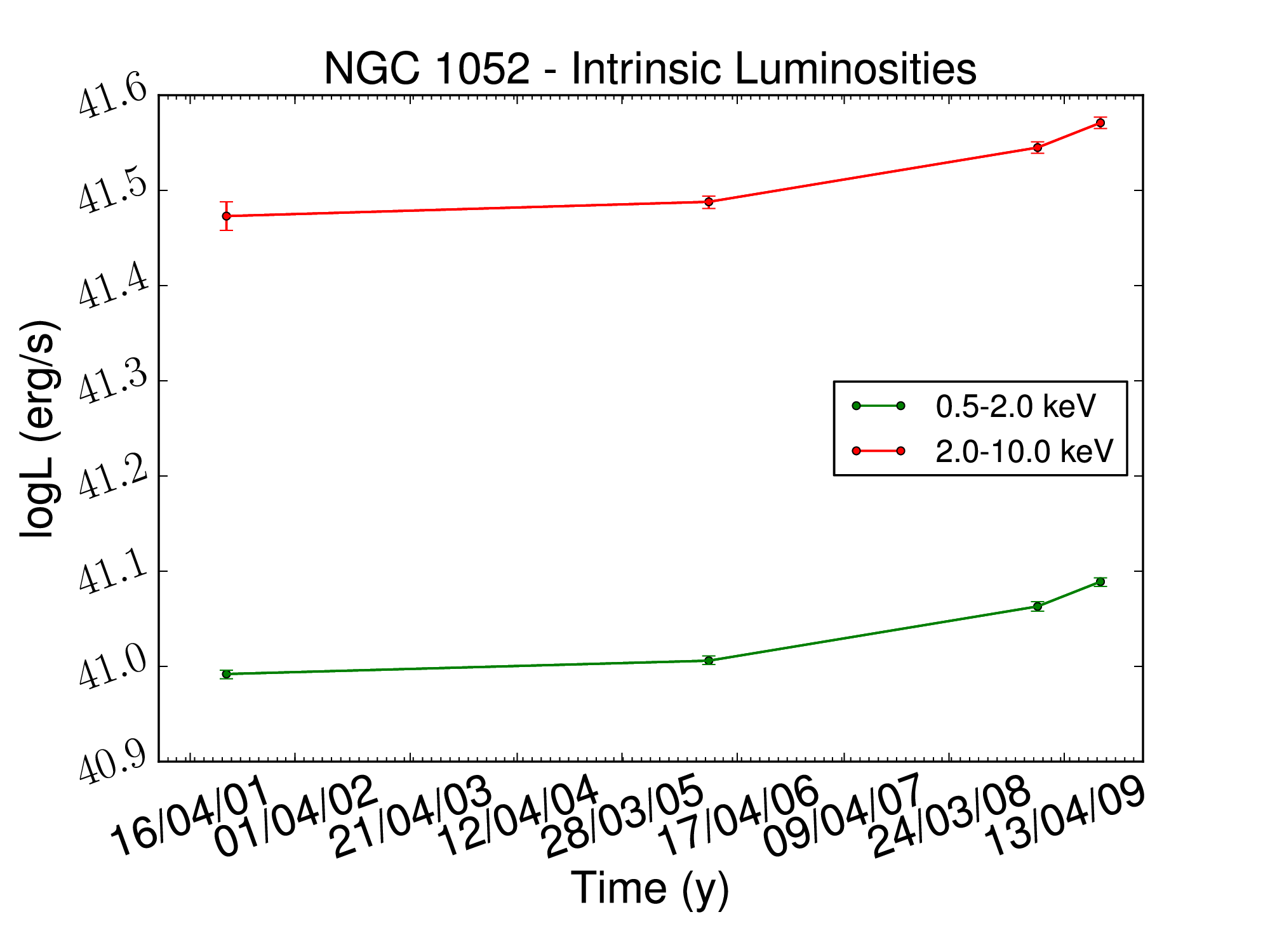} 
\includegraphics[width=0.4\textwidth]{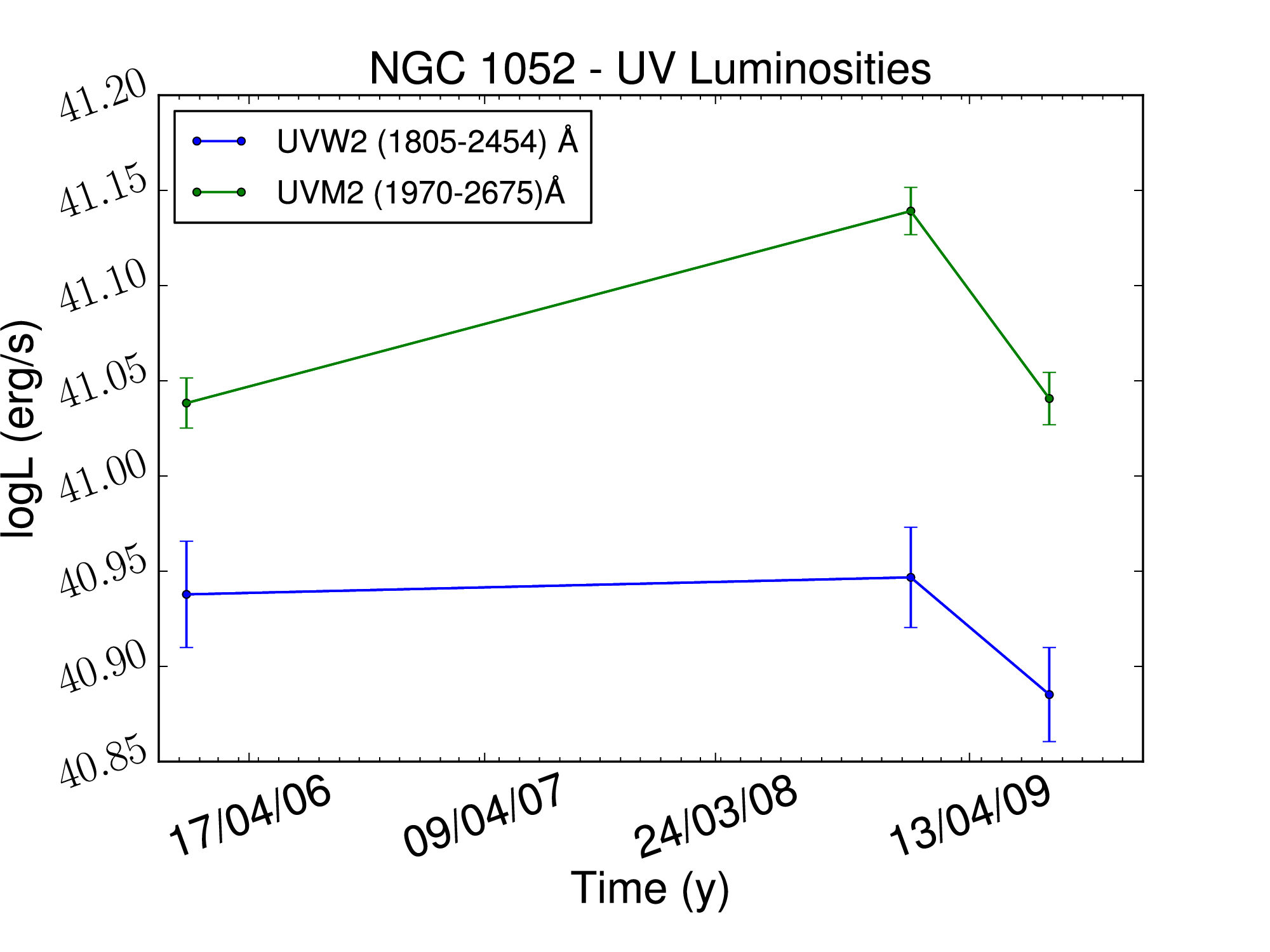}
\hspace*{7.5cm} \includegraphics[width=0.4\textwidth]{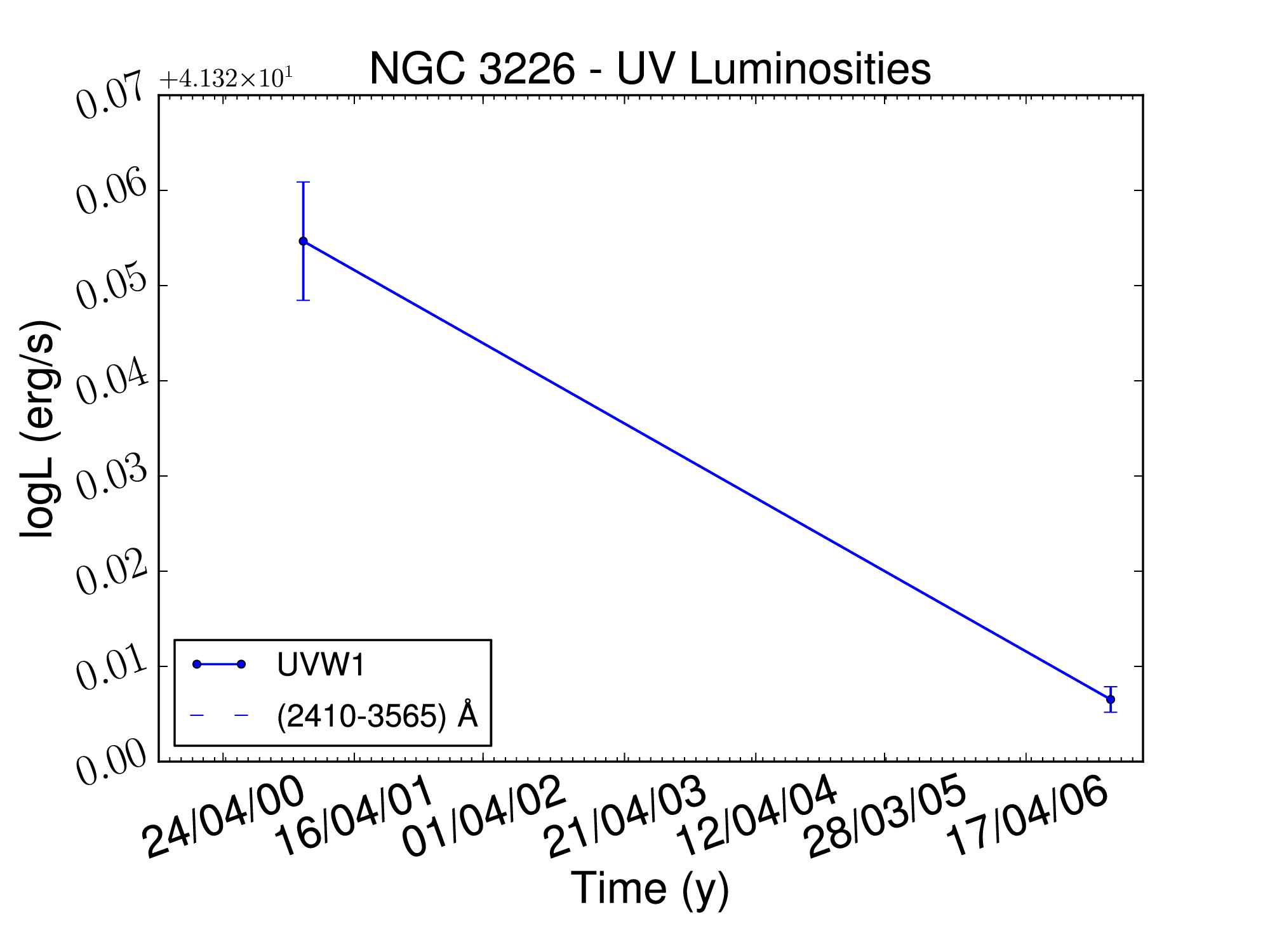}
\includegraphics[width=0.4\textwidth]{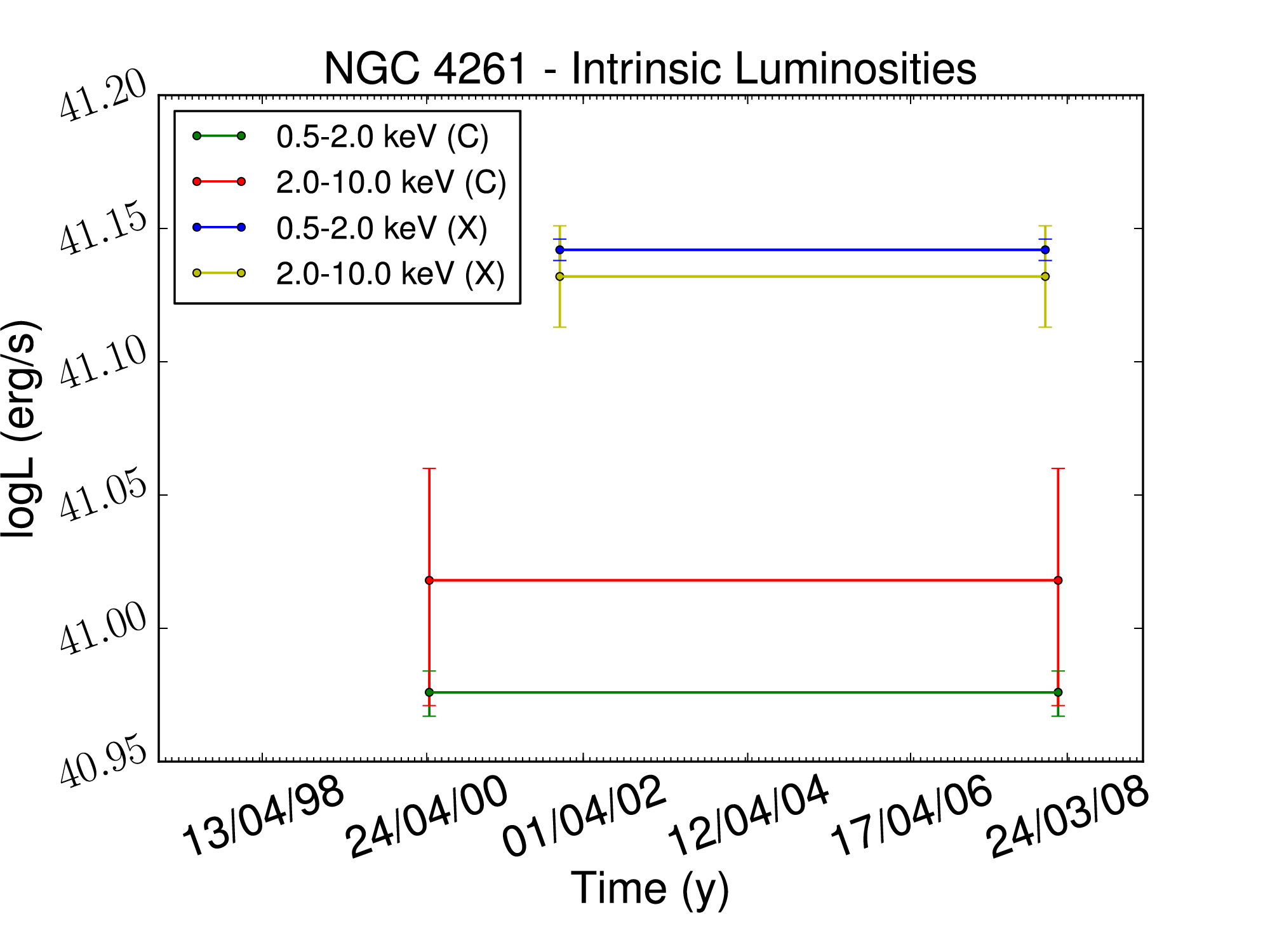} 
\includegraphics[width=0.4\textwidth]{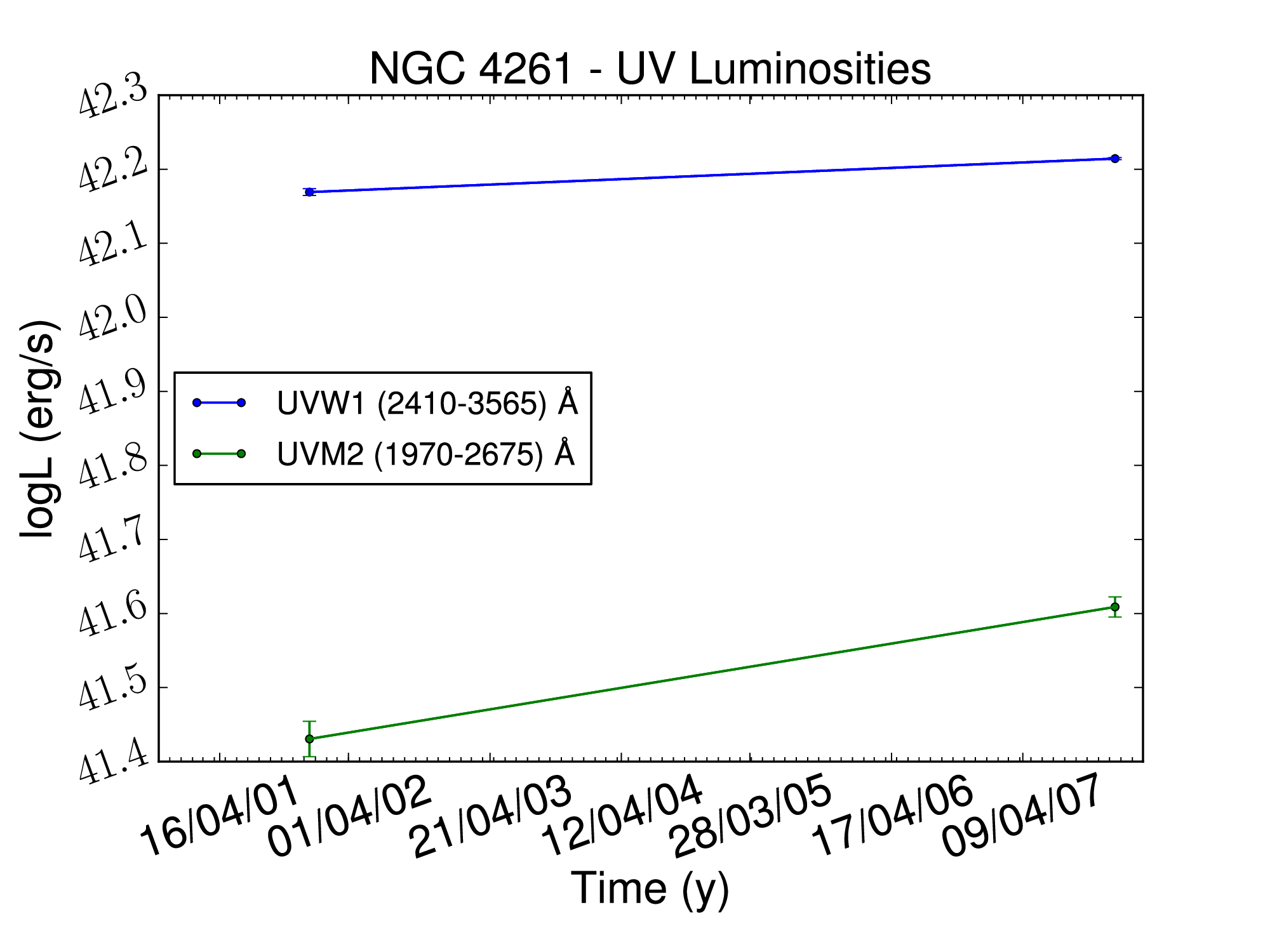}
\end{figure*}

\begin{figure*}[H]
\setcounter{figure}{1}
\caption{Cont.}
\hspace*{1.5cm} \includegraphics[width=0.4\textwidth]{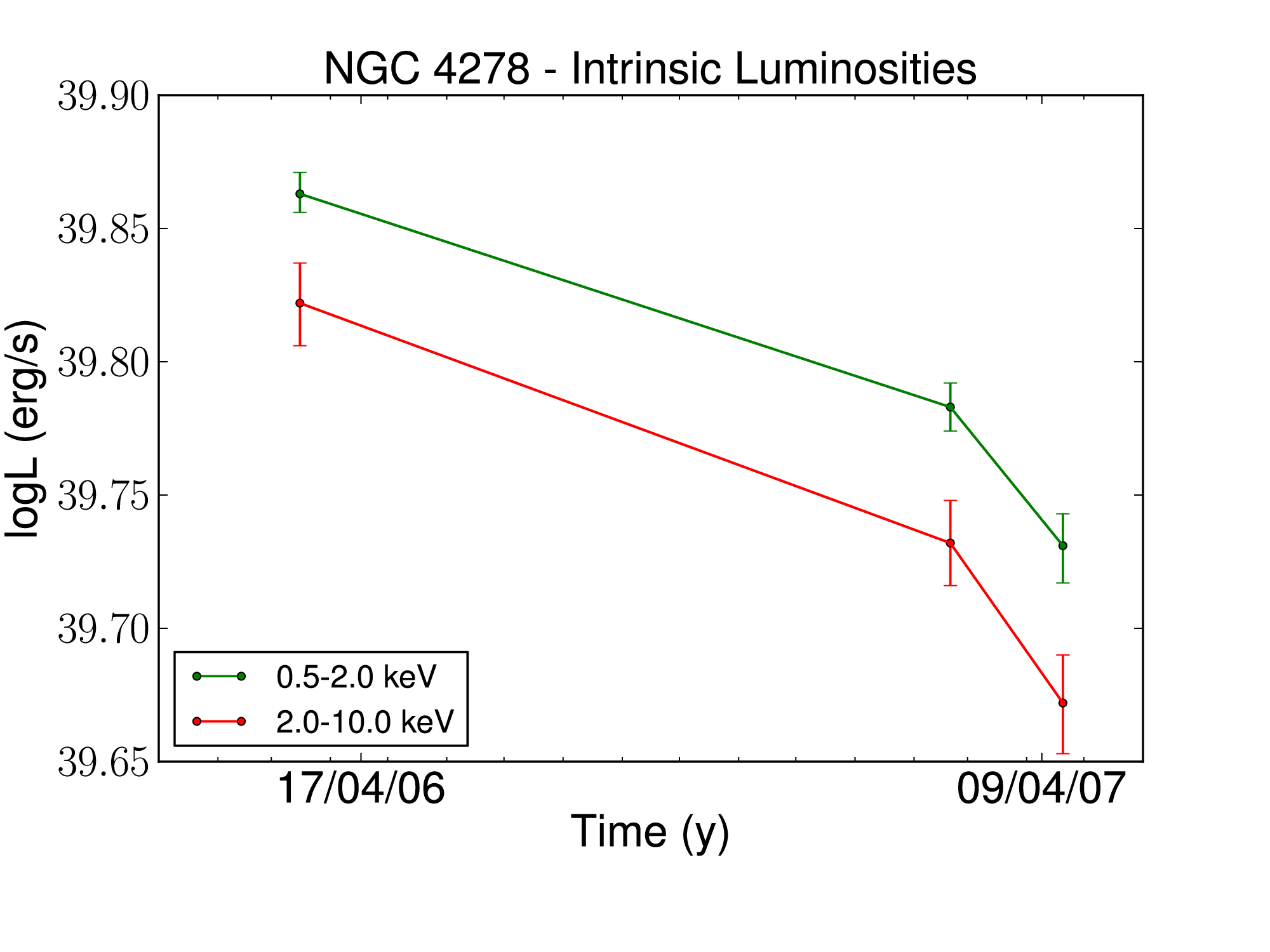} 
\hspace*{1.5cm} \includegraphics[width=0.4\textwidth]{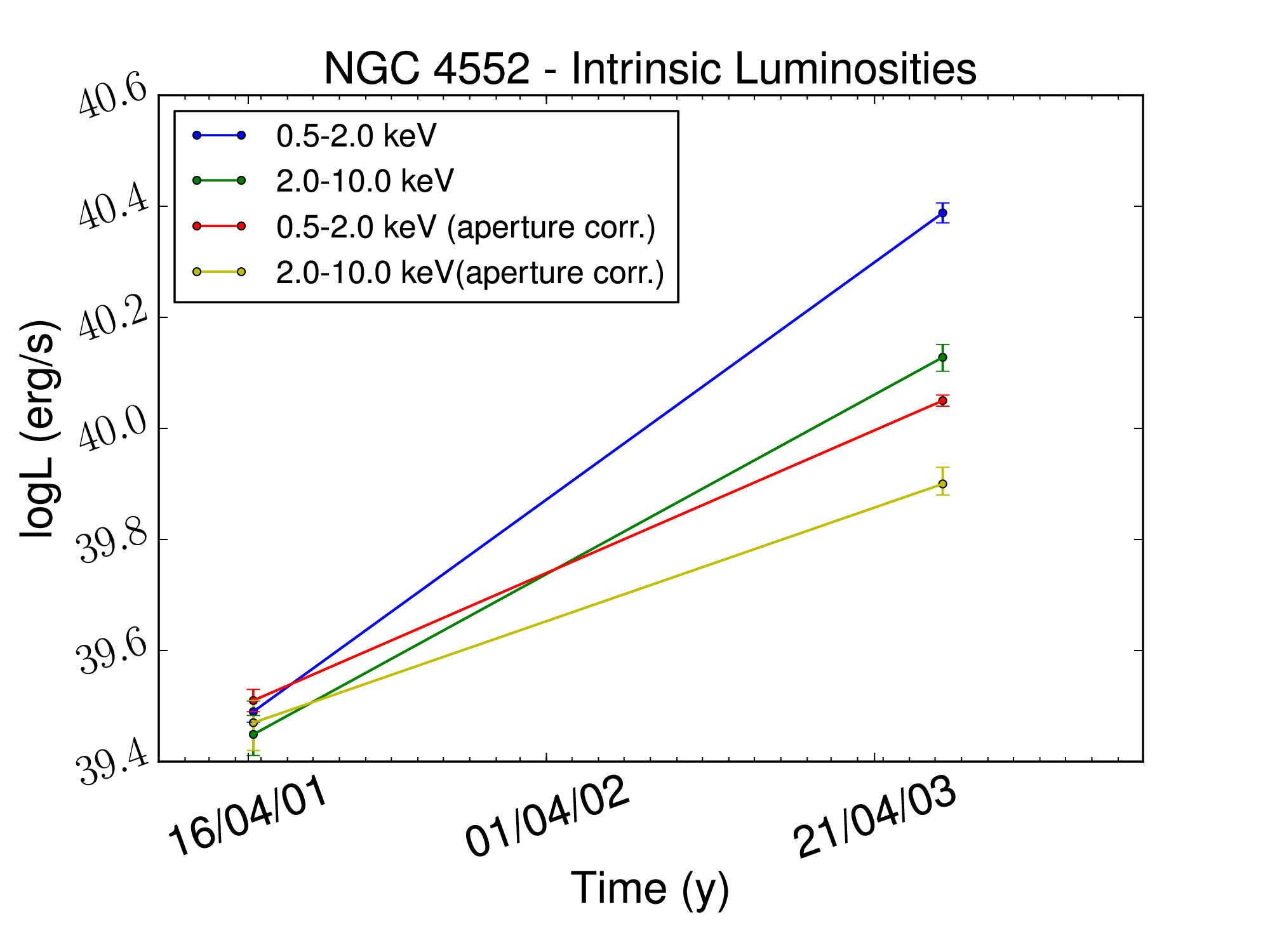}
\hspace*{1.5cm} \includegraphics[width=0.4\textwidth]{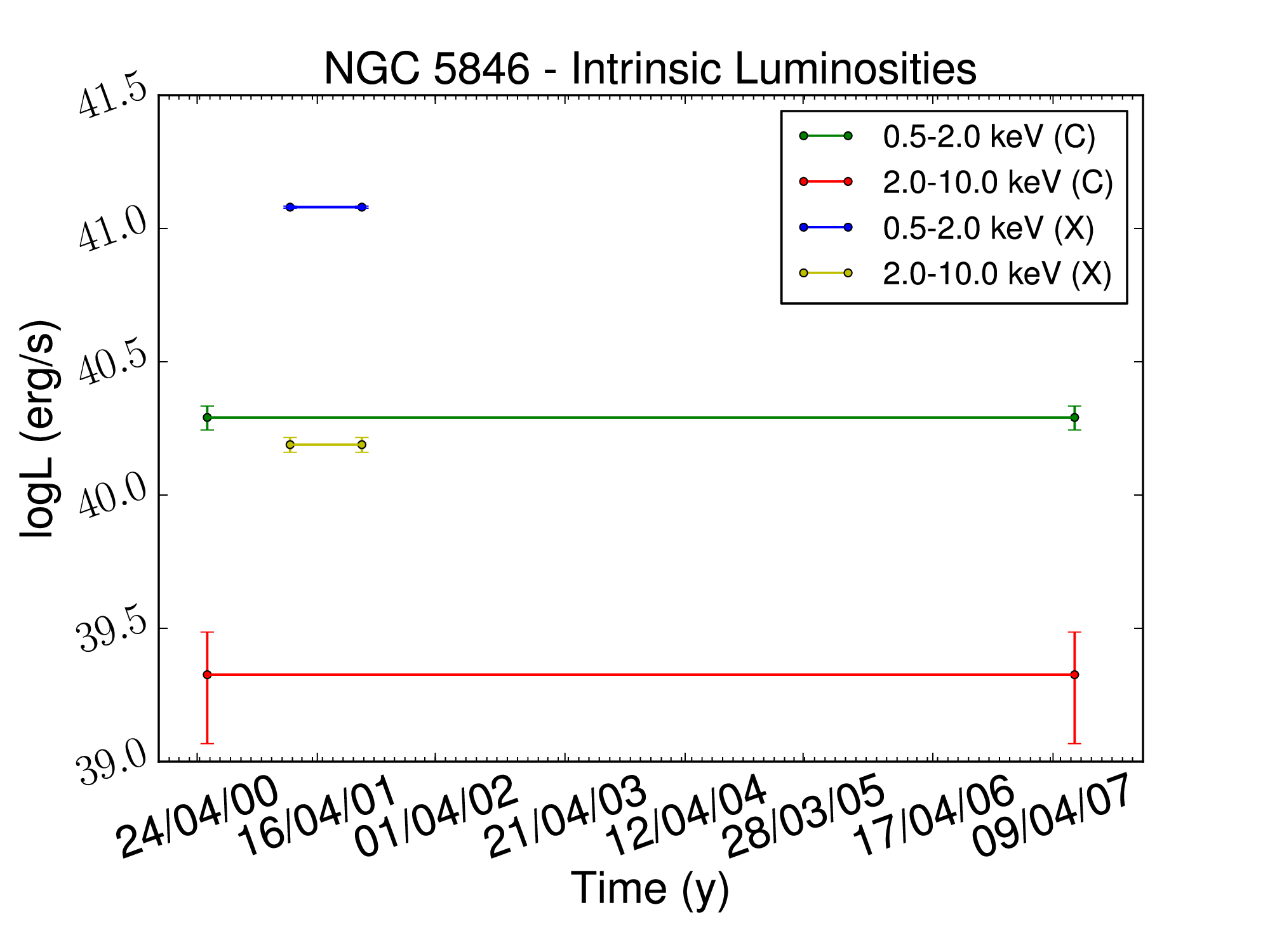}
\end{figure*}

\begin{table*}
\begin{center}
\caption{\label{general} General properties of the sample galaxies. Names (Col. 1 and 2), right ascension (Col. 3), declination (Col. 4), redshift (Col. 5), distance (Col. 6), galactic absorption (Col. 7), aparent magnitude in the Johnson filter B (Col. 8), reddening (Col. 9) and galaxy morphological type (Col. 10).  }
\begin{tabular}{lcccccccccc} \hline
\hline
Name & Other Name & RA & DEC & Redshift & Dist.  & NH(Gal) & $m_B$ & E(B-V) & Morph. Type \\
 & & & & & (Mpc) & (cm$^{-2}$) &  \\
(1) & (2) & (3) & (4) & (5) & (6) & (7) & (8) & (9) & (10)    \\  \hline
NGC~1052 & & 02 41 04.80 & +08 15 20.8 & 0.0049 & 19.41 & 0.0307  & 11.436 & 0.027 & E \\
NGC~3226 & UGC~5617 & 10 23 27.01 & +19 53 54.7 & 0.0059 & 23.55 & 0.0214  & 12.339 & 0.023 & E \\
NGC~3627 & M~66 & 11 20 15.03 & +12 59 29.6 & 0.00242 & 10.28 & 0.0243 & 9.735 & 0.033 & SABb(s) \\
NGC~4261 & UGC~7360 & 12 19 23.22 & +05 49 30.8 & 0.00737 & 31.62 & 0.0155 & 11.35 & 0.018 & E \\
NGC~4278 & M~98 & 12 20 06.83 & +29 16 50.7 & 0.00216 & 16.07 & 0.0177 & 11.042 & 0.029 & E \\
NGC~4552 & M~89 & 12 35 39.81 & +12 33 22.8 & 0.0038 & 15.35 & 0.0257 & 10.67 & 0.041 & E \\
NGC~5846 & UGC~9705 & 15 06 29.29 & +01 36 20.2 & 0.00622 & 24.89 & 0.0426 & 11.074 & 0.056 & E \\
\hline
\end{tabular}
\end{center}
\footnote*{All the distances from \cite{tonry2001} except that for that NGC~3627, which was taken from \cite{ferrarese2000}.}
\end{table*}

\begin{table*}
\begin{center}
\caption{\label{obs} Observational details. Instrument (Col. 1), obsID (Col. 2), date (Col. 3), aperture radius for the nuclear extraction (Col. 4), total exposure time (Col. 5), net exposure time (Col. 6), number of counts in the 0.5-10 keV band (Col. 7), hardness ratio (Col. 8)}, data from the optical monitor available (Col. 9). 
\begin{tabular}{lrccccccc} \hline \hline
& & & & & \\
\multicolumn{7}{c}{NGC 1052} \\ \hline
Satellite & OBSID & Date & Radius & Exptime & Net Exptime & Counts & HR & OM \\ 
 & & & ($\arcsec$) & (ksec) & (ksec) & (0.5-10 keV) & \\ 
  (1) & (2) & (3) & (4) & (5) & (6) & (7) & (8) & (9)    \\  \hline
\emph{XMM}-Newton$^c$ &  093630101 & 2001-08-15 & 25 & 16.3 & 11.2 & 5818 & 0.28 $^+_-$ 0.06 & No  \\
\emph{XMM}-Newton &  306230101 & 2006-01-12 & 25 & 54.9 & 44.8 & 25565 & 0.34 $^+_-$ 0.03 & Yes \\
\emph{XMM}-Newton &   553300301  & 2009-01-14 & 25 & 52.3 & 42.4 & 27367 & 0.39 $^+_-$ 0.02 & Yes \\
\emph{XMM}-Newton* &   553300501  & 2009-01-14 & 25 & 8.0 & 5.4 & 320 & - & No \\
\emph{XMM}-Newton & 553300401  & 2009-08-12 & 25 & 59.0 & 46.8 & 30643 & 0.42 $^+_-$ 0.02 & Yes \\
\emph{Chandra (ACIS-S)}* & 385 & 2005-09-18 & 3 & 2.4 & 2.3 & 270 & 0.36 $^+_-$ 0.24 & - \\
\emph{Chandra (ACIS-S)}*$^c$ & 5910 & 2000-08-29 & 3 & 60.0 & 59.2 & 6549 & 0.40 $^+_-$ 0.04 & - \\
 & & & & & \\
\multicolumn{7}{c}{NGC 3226}  \\ \hline
\emph{XMM}-Newton$^c$ & 0101040301 & 2000-11-28 & 15 & 40.1 & 30.2 & 6514 & 0.35 $^+_-$ 0.04 & Yes \\
\emph{XMM}-Newton & 0400270101 & 2006-12-03 & 15 & 107.9 & 93.1 & 28199 & -0.32 $^+_-$ 0.01 & Yes \\
\emph{Chandra (ACIS-S)}*$^c$ & 860 & 1999-12-30 & 3 & 47.0 & 46.6 & 476 & 0.53 $^+_-$ 0.12 & - \\
\emph{Chandra (ACIS-S)}* & 1616 & 2001-03-23 & 3 & 2.5 & 2.2 & 193 & -0.20 $^+_-$ 0.31 & - \\
 & & & & & \\
\multicolumn{7}{c}{NGC 3627}  \\ \hline 
\emph{XMM}-Newton$^c$ & 0093641101 & 2001-05-26 & 25 & 11.2 & 5.1 & 1181 & -0.63 $^+_-$ 0.02 & Yes \\
\emph{Chandra (ACIS-S)}* & 394 & 1999-11-03 & 8 & 1.8 & 1.8 & 7 & - & - \\
\emph{Chandra (ACIS-S)}$^c$ & 9548 & 2008-03-31 & 8 & 50.2 & 49.5 & 964 & -0.40 $^+_-$ 0.05 & - \\
\multicolumn{7}{c}{NGC 4261}  \\ \hline
\emph{Chandra (ACIS-S)} & 834  & 2000-05-06 & 3 & 35.2 & 34.4 & 3465 & -0.58 $^+_-$ 0.01 & - \\
\emph{Chandra (ACIS-S)}$^c$ & 9569 & 2008-02-12 & 3 & 102.2 & 100.9 & 7757 & -0.47 $^+_-$ 0.01 & - \\
\emph{XMM}-Newton & 0056340101 & 2001-12-16 & 25 & 33.2 & 21.3 & 10730 & -0.68 $^+_-$ 0.01 & Yes \\
\emph{XMM}-Newton$^c$ & 0502120101 & 2007-12-16 & 25 & 127.0 & 63.0 & 32156 & -0.68 $^+_-$ 0.01 & Yes \\
 & & & & & \\
\multicolumn{7}{c}{NGC 4278}  \\ \hline
\emph{Chandra (ACIS-S)}* & 398 & 2000-04-20 & 3 & 1.4 & 1.4 & 303 & -0.55 $^+_-$ 0.05 & - \\
\emph{Chandra (ACIS-S)}* & 4741 & 2005-02-03 & 3 & 37.9 & 37.5 & 20\% pileup & - & - \\
\emph{Chandra (ACIS-S)}$^c$ & 7077 &  2006-03-16 & 3 & 111.7 & 110.3 & 9182 & -0.66 $^+_-$ 0.01 & - \\
\emph{Chandra (ACIS-S)}* & 7078 & 2006-07-25 & 3 & 52.1 & 51.4 & 18\% pileup & - & - \\
\emph{Chandra (ACIS-S)}* & 7079 & 2006-10-24 & 3 & 106.4 & 105.1 & 16\% pileup & - & - \\
\emph{Chandra (ACIS-S)} & 7081 &  2007-02-20 & 3 & 112.1 & 110.7 & 7591 & -0.66 $^+_-$ 0.01 & - \\
\emph{Chandra (ACIS-S)} & 7080 &  2007-04-20 & 3 & 56.5 & 55.8 & 3379 & -0.69 $^+_-$ 0.01 & - \\
\emph{Chandra (ACIS-S)}* & 11269 & 2010-03-15 &  3 &  83.0 & 81.9 & 2091 & -0.82 $^+_-$ 0.01 & - \\
\emph{Chandra (ACIS-S)}* & 12124 & 2010-03-20 & 3 &  26.2 & 25.8 & 576 & -0.92 $^+_-$ 0.01 & - \\
\emph{XMM}-Newton*$^c$ & 0205010101 & 2004-05-23 & 25 & 35.9 & 20.8 & 34516 & -0.60 $^+_-$ 0.01 & Yes \\
 & & & & & \\
\multicolumn{7}{c}{NGC 4552}  \\ \hline
\emph{Chandra (ACIS-S)}$^c$ & 2072 & 2001-04-22 & 3 & 55.1 & 54.4 & 2288 & -0.75 $^+_-$ 0.01 & - \\
\emph{XMM}-Newton$^c$ & 0141570101 & 2003-07-10 & 25 & 44.8 & 17.3 & 12180 & -0.81 $^+_-$ 0.01 & Yes \\
 & & & & & \\
\multicolumn{7}{c}{NGC 5846}  \\ \hline
\emph{XMM}-Newton$^c$ & 0021540101 & 2001-01-25 & 25 & 30.0 & 25.6 & 25905 & -0.94 $^+_-$ 0.01 & Yes \\
\emph{XMM}-Newton$^d$ & 0021540501 & 2001-08-26 & 25 & 19.7 & 10.0 & 10027 & -0.95 $^+_-$ 0.01 & Yes \\
\emph{Chandra (ACIS-S)}$^c$ &  788 & 2000-05-24 & 7 & 30.3 & 29.9 & 1839 & -0.94 $^+_-$ 0.01 & - \\
\emph{Chandra (ACIS-I)} & 7923 & 2007-06-12 & 7 & 91.2 & 90.0 & 2307 & -0.88 $^+_-$ 0.01 & - \\
\hline
\end{tabular} 
\end{center}
\noindent
\begin{scriptsize}
* Observations were not used for the simultaneous fittings.

$^c$ Observations used to compare \emph{XMM}-Newton and \emph{Chandra} data.

$^d$ Observation with only optical OM data (UV not available).
\end{scriptsize}
\end{table*}

\begin{table*}
\begin{center}
\caption{\label{ftestcol1} F-test and $\chi^2/d.o.f$ applied to the SMF0. When no variation in one parameter is needed, we mark it as ``-". Name (Col. 1), instrument (Col. 2), best fit model (Col. 3), statistical test (Col. 4), parameter varying in respect to SMF0 (Col. 5, 6, 7, 8, 9 and 10) and the parameter that varies in SMF1 (Col. 11). } 
\begin{tabular}{lcccccccccr} \hline
\hline
Name & Instrument & Best fit & Test & vs($N_{H1}$) & vs($N_{H2}$) & vs(kT) & vs($\Gamma$)  & vs($Norm_1$) & vs($Norm_2$) & Var.  \\
(1) & (2) & (3) & (4) & (5) & (6) & (7) & (8) & (9) & (10) & (11)     \\  \hline
NGC~1052 & \emph{XMM} & ME2PL & F-test & 1    & 2.29e-113 & 0.07 & 8.5071e-114 & - & 5.27e-128 & $Norm_2$ \\
                  & & & $\chi^2/d.o.f$ & 1.31 & 1.10      & 1.26 & 1.10      & -   & 1.07 & \\
  
NGC~3226 & \emph{XMM} & 2PL & F-test & 3.83e-158 & 1.3462e-194 & - & 0.09 & 2.58e-139 & 2.58e-139 & $N_{H2}$ \\
                & & & $\chi^2/d.o.f$ & 1.18      & 0.98        & - & 2.57 & 1.30      & 1.30 & \\
 
NGC~3627 & \emph{XMM}/\emph{Ch} & MEPL & F-test & 3.25e-20 & 8.24e-26 & 8.90e-10 & 0.05 & 2.17e-17 & 1.68e-25 & $N_{H2}$ \\
           & & & $\chi^2/d.o.f$ & 2.27     & 1.63     & 4.21     & 6.54 & 2.76     & 1.71 & \\

NGC~4261 & \emph{Chandra} & ME2PL & F-test & 0.90 & 0.12 & 0.40 & 0.94 & 0.30 & 0.96 & None \\
                      & & & $\chi^2/d.o.f$ & 1.27 & 1.26 & 1.26 & 1.27 & 1.26 & 1.27 & \\

\hspace*{0.5cm} & \emph{XMM} & ME2PL & F-test & 0.09 & 0.39 & 0.34 & 0.92 & - & 0.75 & None \\
                  & & & $\chi^2/d.o.f$ & 1.18 & 1.18 & 1.18 & 1.18 & - & 1.18 & \\

NGC~4278 & \emph{Chandra} & MEPL & F-test & 1.75e-07 & 1.03e-08 & 1.01e-07 & 4.94e-06 & 5.27e-18 & 6.73e-46 & $Norm_2$ \\
                     & & & $\chi^2/d.o.f$ & 1.53     & 1.51     & 1.53     & 1.56     & 1.36     & 1.00 & \\

NGC~4552 &  \emph{XMM}/\emph{Ch} & MEPL & F-test &  3.22e-128 & 4.13e-90 & 1.06e-53 & 1.45e-03 & 5.70e-133 & 6.76e-94 & $Norm_1$  \\
            & & & $\chi^2/d.o.f$ & 4.01       & 6.82     & 11.35    & 22.7     & 3.75      & 6.47 & \\

NGC~5846 & \emph{XMM} & MEPL & F-test & 1    & 1    & 4.02e-04 & 0.38     & 0.05 & 0.66 & None \\
                 & & & $\chi^2/d.o.f$ & 1.29 & 1.29 & 1.26     & 1.29     & 1.28 & 1.29 & \\

\hspace*{0.5cm} & \emph{Chandra} & MEPL & F-test & 1    & 0.12 & 0.46 & 0.63 & 0.56 & 0.31 & None \\
                     & & & $\chi^2/d.o.f$ & 0.97 & 0.95 & 0.97 & 0.97 & 0.97 & 0.96 & \\
 
\hline
\end{tabular}
\end{center}
\end{table*}

\begin{table*}
\begin{center}
\caption{\label{ftestcol2} F-test and $\chi^2/d.o.f$ applied to the SMF1. When no variation in one parameter is needed, we mark it as ``-". Name (Col. 1), instrument (Col. 2), the parameter varying in SMF1 (Col. 3), statistical test (Col. 4), parameter varying in respect to SMF1 (Col. 5, 6, 7, 8, 9 and 10) and the parameter that varies in SMF2 (Col. 11). }
\begin{tabular}{lcccccccccr} \hline
\hline
Name & Instrument & Var. & Test & vs($N_{H1}$) & vs($N_{H2}$) & vs(kT) & vs($\Gamma$)  & vs($Norm_1$) & vs($Norm_2$) & Var.  \\
(1) & (2) & (3) & (4) & (5) & (6) & (7) & (8) & (9) & (10)  & (11)   \\  \hline
NGC~1052 & \emph{XMM} & $Norm_2$ & F-test & 1    & 2.94e-10 & 0.03 & 4.12e-04 & - & & $N_{H2}$ \\
                     & & & $\chi^2/d.o.f$ & 1.07 & 1.05     & 1.07 & 1.06     & - & & \\
 
NGC~3226 & \emph{XMM} & $N_{H2}$ & F-test & 1.59e-03 & & - & 3.84e-07 & 1.75e-06 & 1.33e-06 & None \\
                     & & & $\chi^2/d.o.f$ & 0.97     & & - & 0.96     & 0.96     & 0.96 & \\

NGC~3627 & Both & $N_{H2}$ & F-test & 2.46e-03 &     & 0.35     & 0.01     & 1    & 2.92e-03  & None \\
               & & & $\chi^2/d.o.f$ & 1.46     &     & 1.63     & 1.52     & 2.44 & 1.47  & \\

NGC~4278 & \emph{Chandra} & $Norm_2$ & F-test & 1     & 1    & 0.13     & 0.62 & 0.96 & & None \\
                         & & & $\chi^2/d.o.f$ & 1.00  & 1.00 & 0.99     & 1.00 & 1.00 & & \\

NGC~4552 & Both &  $Norm_1$ & F-test & 0.93 & 0.94 & 0.02 & 4.17e-10 & & 4.15e-24 & $Norm_2$ \\
                & & & $\chi^2/d.o.f$ & 1.65 & 1.65 & 1.62 & 1.50     & & 1.28 & \\

\hline
\end{tabular}
\end{center}
\end{table*}

\begin{table*}
\begin{center}
\caption{\label{bestfit} Final compilation of the best fit models for the sample, including the individual best fit model for each observation, and the simultaneous best fit model with the varying parameters. Satellite (Col. 1), obsID (Col. 2), best fit model (Col. 3), parameters in the model (Col. 4, 5, 6, 7, 8 and 9) and $\chi^2/d.o.f$ (Col. 10).} 
\begin{tabular}{lccccccccr} \hline
\hline
Instrument & ObsID & Best fit & $N_{H1}$ & $N_{H2}$ & kT & $\Gamma$ & $Norm_1$ & $Norm_2$ & $\chi^2/d.o.f$  \\
  & & & ($10^{22} cm^{-2}$) & ($10^{22} cm^{-2}$) & keV & & ($10^{-4}$) & ($10^{-4}$) &  \\
(1) & (2) & (3) & (4) & (5) & (6) & (7) & (8) & (9) & (10)  \\
\hline
NGC~1052 & & & & & & & & & \\
\hline
 \emph{XMM}-Newton & 093630101 & ME2PL & - & 13.77$^{15.78}_{11.95}$ & 0.62$^{0.69}_{0.51}$ & 1.24$^{1.35}_{1.11}$ & 1.05$^{1.13}_{0.97}$ & 6.09$^{4.55}_{4.79}$ & 269.16/234  \\
 \emph{XMM}-Newton & 306230101 & ME2PL & - & 9.30$ _{ 8.79 }^{ 9.82 }$ & 0.50 $_{ 0.45 }^{ 0.55 }$ & 1.30$ _{ 1.24 }^{ 1.36 }$ & 1.05$ _{ 1.01 }^{ 1.09 }$ & 7.20$ _{ 6.34 }^{ 8.14 }$ & 799.19/822  \\
 \emph{XMM}-Newton & 553300301 & ME2PL & - & 8.96$ _{ 8.54 }^{ 9.39}$ & 0.61$ _{ 0.57 }^{ 0.64 }$ & 1.38$ _{ 1.32 }^{ 1.43}$ & 1.09$ _{ 1.06 }^{ 1.14 }$ & 10.16$ _{ 9.10 }^{ 11.31 }$ & 923.38/869  \\
 \emph{XMM}-Newton & 553300401 & ME2PL & - & 9.47$ _{ 9.09 }^{ 9.86 }$ & 0.53$ _{ 0.48 }^{ 0.61}$ & 1.43$ _{ 1.38 }^{ 1.48 }$ & 1.10$ _{ 1.06 }^{ 1.14 }$ & 12.02$ _{ 10.82 }^{ 13.32 }$ & 1006.47/937  \\
\emph{Chandra} (3$\arcsec$)* & 5910 &  ME2PL & - & 5.33$ _{ 3.84 }^{ 6.86 }$ & 0.64$ _{ 0.58 }^{ 0.69 }$ & 1.21$ _{ 1.05 }^{ 1.46 }$ & 0.37$ _{ 0.37 }^{ 0.47 }$ & 0.57$ _{ 0.35 }^{ 1.06 }$ & 261.04/226 \\
\emph{Chandra} (25$\arcsec$)* & 5910 &  ME2PL & - & 8.13$ _{ 5.36 }^{ 11.49 }$ & 0.61$ _{ 0.57 }^{ 0.65 }$ & 1.25$ _{ 1.00 }^{ 1.50 }$ & 0.49$ _{ 0.49 }^{ 0.66 }$ & 0.61$ _{ 0.19 }^{ 1.21 }$ & 319.40/269 \\
\hline
 \emph{XMM}-Newton & 093630101 & ME2PL & - & 14.14$ _{ 12.66 }^{ 15.77 }$ & 0.59$^{0.61}_{0.57}$ & 1.36$_{1.33}^{1.39}$ & 1.09$^{1.11}_{1.06}$ & 8.06$^{8.89}_{7.31}$ &  3043.69/2886 \\
Simultaneous   & 306230101 &  &            & 9.80$ _{ 9.36 }^{ 10.26 }$ &  &  & & 8.38$^{8.97}_{7.82}$ &    \\
    & 553300301 &  &                       & 8.75$ _{ 8.40 }^{ 9.11 }$ &  &  & & 9.74$^{10.40}_{9.11}$ &    \\
   & 553300401 &  &                        & 9.21$ _{ 8.88 }^{ 9.55 }$ &  &  & & 10.41$^{11.11}_{9.75}$ &    \\
\hline

NGC~3226 & & & & & & & & & \\
\hline
 \emph{XMM}-Newton & 0101040301 & 2PL & 0.11$ _{ 0.05}^{ 0.15 }$ & 1.03$ _{ 0.76}^{ 1.35 }$ & - & 1.69$ _{ 1.58}^{ 1.81 }$ & 0.71$ _{ 0.47 }^{ 0.92 }$ & 1.85$ _{ 1.50 }^{ 2.28 }$ & 245.87/252  \\
 \emph{XMM}-Newton & 0400270101 & PL & 0.17$ _{ 0.16 }^{ 0.18}$ & - & - & 1.74$ _{ 1.72 }^{ 1.77 }$ & - & - & 635.19/663 \\

\emph{Chandra} (3$\arcsec$)* & 860 & PL & 0.18$ _{ 0.00 }^{ 0.41}$ & - & - & 1.73$ _{ 1.47 }^{ 2.10 }$ & - & - & 9.72/17 \\
\emph{Chandra} (25$\arcsec$)* & 860 & 2PL & 0.29$^{0.70}_{0.10}$ & 15.80$^{114.58}_{0.00}$ & - & 1.71$^{3.00}_{1.34}$ & 2.03$^{4.48}_{0.00}$ & 1.43$^{16.49}_{0.00}$ & 26.30/28 \\

\hline
 \emph{XMM}-Newton & 0101040301 & 2PL & $0.09^{0.12}_{0.05}$ & 0.86$^{0.99}_{0.76}$ & - & 1.73$^{1.77}_{1.71}$ & 0.86$^{0.98}_{0.76}$ & 1.89$^{2.09}_{1.68}$ & 905.42/920  \\
 Simultaneous   & 0400270101 &  &                            & 0.22$^{0.24}_{0.19}$ &  &  & &  &    \\ 
\hline

NGC~3627 & & & & & & & & & \\
\hline 
\emph{XMM}-Newton & 0093641101 & MEPL & - & - & 0.33$^{0.52}_{0.29}$ & 1.78$^{2.09}_{1.57}$ & 0.00$_{0.00}^{0.02}$ & 0.68$_{0.60 }^{0.83 }$ & 54.91/35 \\
 \emph{Chandra} (8$\arcsec$) &  9548 & MEPL & - & 0.05$^{0.15}_{0.00}$ & 0.65$^{0.80}_{0.46}$ & 1.45$^{1.70}_{1.25}$ & 0.06$^{0.21}_{0.05}$ & 0.24$^{0.31}_{0.19}$ & 33.63/34 \\
 \emph{Chandra} (25$\arcsec$)* &  9548 & MEPL & - & - & 0.40$ _{ 0.35 }^{ 0.51 }$ & 1.49$ _{ 1.38 }^{ 1.63 }$ & 0.31$ _{ 0.24 }^{ 0.37 }$ & 0.55$ _{ 0.50 }^{ 0.61 }$ & 149.42/96 \\
\hline
Both & 0093641101 & MEPL & - & - & 0.63$^{0.68}_{0.58}$ & 2.33$^{2.42}_{2.25}$ & 0.14$^{0.15}_{0.12}$ & 0.90$^{0.96}_{0.85}$ & 125.5/77 \\
Simultaneous & 9548 & & &   1.28 $_{ 1.11 }^{ 1.48 }$    &  &  & &  & \\
\hline

NGC~4261 & & & & & & & & & \\
\hline
 \emph{Chandra} & 834 & ME2PL & 0.18$ _{ 0.06 }^{ 0.47}$ & 11.89$ _{ 9.31}^{ 16.01 }$ & 0.58$ _{ 0.56 }^{ 0.60}$ & 2.11$ _{ 1.52 }^{ 3.07} $ & 0.29$ _{ 0.19 }^{ 0.55 }$ & 4.72$ _{ 1.60 }^{ 25.76 }$ & 110.82/80  \\
 \emph{Chandra} (3$\arcsec$) & 9569 & ME2PL & 0.18$ _{ 0.08}^{ 0.47}$ & 8.40$ _{ 6.90 }^{ 10.20}$ & 0.57$ _{ 0.56 }^{ 0.58 }$ & 1.24$ _{ 1.07 }^{ 1.51 }$ & 0.26$ _{ 0.20 }^{ 0.37 }$ & 2.13$ _{ 1.18 }^{ 4.57 }$ & 198.74/157  \\
\emph{Chandra} (25$\arcsec$)* & 9569 & ME2PL & - & 9.03$ _{ 6.55 }^{ 12.90 }$ & 0.61$ _{ 0.60 }^{ 0.62 }$ & 1.16$ _{ 0.89 }^{ 1.39 }$ & 0.35$ _{ 0.35 }^{ 0.47 }$ & 0.64$ _{ 0.29 }^{ 1.11 }$ & 382.42/218 \\

\hline
 \emph{Chandra} & 834/9569 & ME2PL & $0.18^{0.31}_{0.09}$ & 9.38$^{10.84}_{7.96}$ & 0.57$^{0.58}_{0.56}$ & 1.87$^{2.19}_{1.55}$ & 0.27$^{0.34}_{0.21}$ & 2.70$^{4.86}_{1.49}$ & 312.57/247  \\
 Simultaneous & & & & & & & & & \\
\hline
 \emph{XMM}-Newton & 0056340101 & ME2PL & 0.08$ _{ 0.01}^{ 0.15 }$ & 11.97$ _{ 9.31 }^{ 14.73 }$ & 0.63$ _{ 0.61}^{ 0.64 }$ & 1.84$ _{ 1.45 }^{ 2.22 }$ & 0.72$ _{ 0.54 }^{ 0.94 }$ & 2.82$ _{ 1.11 }^{ 6.30 }$ & 278.12/255  \\
 \emph{XMM}-Newton & 0502120101 & ME2PL & 0.04$ _{ 0.00 }^{ 0.07 }$ & 9.71$ _{ 8.30}^{ 11.14 }$ & 0.64$ _{ 0.64}^{ 0.65 }$ & 1.75$ _{ 1.54 }^{ 1.97 }$ & 0.64$ _{ 0.55 }^{ 0.74 }$ & 2.30$ _{ 1.40 }^{ 3.65 }$ & 563.80/461  \\
\hline
 \emph{XMM}-Newton & 0056340101 & ME2PL & $0.04^{0.08}_{0.01}$ & 10.28$^{11.56}_{9.02}$ & 0.64$^{0.65}_{0.63}$ & 1.76$^{1.95}_{1.57}$ & 0.67$^{0.76}_{0.58}$ & 2.36$^{3.54}_{1.53}$ &  855.28/726 \\
 Simultaneous  & 0502120101 &  &  &  &  &  &  &  &   \\
\hline

NGC~4278 & & & & & & & & & \\
\hline
 \emph{Chandra}(3$\arcsec$) & 7077 & MEPL & 0.31$ _{ 0.05 }^{ 0.58}$ & 0.01$ _{ 0.00 }^{ 0.03}$ & 0.27$ _{ 0.19 }^{ 0.38 }$ &  2.06$ _{ 1.99 }^{ 2.15 }$ & 0.80$ _{ 0.18 }^{ 6.21 }$ & 0.98$ _{ 0.90 }^{ 1.06 }$ & 144.23/158  \\
\emph{Chandra}(25$\arcsec$)* & 7077 & MEPL &  0.16$ _{ 0.01}^{ 0.45 }$ & - & 0.30$ _{ 0.21 }^{ 0.37}$ & 1.88$ _{ 1.83 }^{ 1.95 }$ & 0.81$ _{ 0.28 }^{ 3.96 }$ & 1.32$ _{ 1.27 }^{ 1.39 }$ & 250.15/210 \\
 \emph{Chandra} & 7081 & MEPL & - & - & 0.63$_{ 0.57 }^{ 0.68 }$ & 2.03$ _{ 1.99 }^{ 2.10}$ & 0.12$ _{ 0.09 }^{ 0.14 }$ & 0.77$ _{ 0.75 }^{ 0.82 }$ & 155.54/145 \\
 \emph{Chandra} & 7080 & MEPL & 0.22$ _{ 0.00 }^{ 0.71 }$ & - & 0.47$ _{ 0.18 }^{ 0.61 }$ &  2.10$ _{ 2.03 }^{ 2.25 }$ & 0.24$ _{ 0.07 }^{ 13.76 }$ & 0.69$ _{ 0.65 }^{ 0.79 }$ & 102.53/96  \\
\emph{XMM}-Newton* &  0205010101 & PL & 0.02$ _{ 0.02 }^{ 0.03 }$ & - & - & 2.04$ _{ 2.01 }^{ 2.08 }$ & - & - & 562.25/529 \\
\hline
 \emph{Chandra} & 7077 & MEPL & - & - & 0.58$^{0.62}_{0.48}$ & 2.05$^{2.11}_{2.03}$ & 0.11$^{0.12}_{0.10}$ & 0.97$^{1.02}_{0.82}$ & 414.71/415  \\
 Simultaneous   & 7081 &  &  &  &  &  &                                                                    & 0.78$^{0.82}_{0.77}$ &   \\
    & 7080 &  &  &  &  &  &                                                                                & 0.68$^{0.72}_{0.65}$ &   \\
\hline

NGC~4552 & & & & & & & & & \\
\hline    
 \emph{Chandra} (3$\arcsec$) & 2072 & MEPL & 0.23$ _{ 0.00 }^{ 0.33 }$ & - & 0.65 $_{ 0.59 }^{ 0.75 }$ &  1.85$ _{ 1.74 }^{ 2.13 }$ & 0.36$ _{ 0.16 }^{ 0.56 }$ & 0.29$ _{ 0.26 }^{ 0.41 }$ & 72.49/67 \\  
\emph{Chandra} (25$\arcsec$)* & 2072 & MEPL & 0.03$ _{ 0.00 }^{ 0.05 }$ & 0.00 $_{ 0.00 }^{ 0.02 }$ & 0.60$ _{ 0.58 }^{ 0.62 }$ & 1.89$ _{ 1.82 }^{ 1.96 }$ & 1.64$ _{ 1.48 }^{ 1.82 }$ & 0.91$ _{ 0.85 }^{ 0.97 }$ & 161.64/133 \\
 \emph{XMM}-Newton & 0141570101 & MEPL & 0.01$ _{ 0.00 }^{ 0.05 }$ & - & 0.59 $_{ 0.58 }^{ 0.61 }$ & 1.92$ _{ 1.85 }^{ 1.98}$ & 2.12$ _{ 1.98 }^{ 2.37 }$ & 1.46$ _{ 1.38 }^{ 1.55 }$ & 282.04/254 \\ 
\hline
Both & 2072 & MEPL & - & - & 0.60$^{0.62}_{0.59}$ & 1.86$^{1.91}_{1.81}$ & 0.15$^{0.17}_{0.13}$ & 0.31$^{0.33}_{0.29}$ & 397.01/328 \\
Simultaneous & 0141570101  & & & & &                                     & 2.11$^{2.35}_{1.96}$ & 1.40$^{1.48}_{1.32}$ &  \\   
\hline

NGC~5846 & & & & & & & & & \\
\hline
 \emph{XMM}-Newton & 0021540101 & MEPL & - & - & 0.61$^{0.61}_{0.60}$ & 2.05$^{2.17}_{1.92}$ & 5.09$^{5.17}_{5.00}$ & 0.61$^{0.68}_{0.53}$ & 356.05/266  \\
 \emph{XMM}-Newton & 0021540501 & MEPL & 0.04$ _{ 0.01}^{ 0.07 }$ & - & 0.62$ _{ 0.60 }^{ 0.63 }$ & 2.20$ _{ 1.99 }^{ 2.40 }$ & 5.55$ _{ 5.10 }^{ 6.07 }$ & 0.69$ _{ 0.56 }^{ 0.81 }$ & 227.57/194  \\
\hline
 \emph{XMM}-Newton & 0021540101 & MEPL & - & - & 0.62$^{0.62}_{0.61}$ & 2.05$^{2.14}_{1.94}$ & 5.06$^{5.13}_{4.98}$ & 0.61$^{0.67}_{0.54}$ &  606.06/469 \\
 Simultaneous   & 0021540501 &  &  &  &  &  &  &  &   \\
  \hline
 \emph{Chandra}(7$\arcsec$) & 788 & MEPL & - & 0.21$ _{ 0.00}^{ 0.43 }$ & 0.61$ _{ 0.57 }^{ 0.64 }$ & 3.06$ _{ 2.24}^{ 0 }$ & 0.52$ _{ 0.48 }^{ 0.79 }$ & 0.40$ _{ 0.18 }^{ 0.76 }$ & 41.72/50  \\
 \emph{Chandra}(25$\arcsec$)* & 788 & MEPL & - & - & 0.62$ _{ 0.60 }^{ 0.63 }$ & 1.53$ _{ 1.27 }^{ 1.99 }$ & 2.83$ _{ 2.69 }^{ 2.97 }$ & 0.67$ _{ 0.53 }^{ 0.81 }$ & 188.51/137 \\
 \emph{Chandra} & 7923 & MEPL & $0.11^{0.18}_{0.00}$ & 0.08$^{0.33}_{0.00}$ & 0.60$^{0.64}_{0.57}$ & 2.73$^{3.20}_{2.27}$ & 0.77$^{0.98}_{0.54}$ & 0.21$^{0.38}_{0.12}$ & 63.24/59  \\
\hline
 \emph{Chandra} & 788/7923 & MEPL & - & 0.21$^{0.38}_{0.14}$ & 0.62$^{0.64}_{0.60}$ & 3.11$^{0}_{2.40}$ & 0.52$^{0.56}_{0.48}$ & 0.38$^{0.68}_{0.23}$ & 105.82/110  \\
 Simultaneous  & & & & & & & & & \\
\hline
\end{tabular}
\end{center}
\begin{scriptsize}
* Observations not used in the simultaneous fit.
\end{scriptsize}
\end{table*}

\begin{table*}
\begin{center}
\caption{\label{lumincorr} Soft (0.5-2 keV) and hard (2-10 keV) intrinsic luminosities for individual (Col. 3 and 5) and simultaneous (Col. 4 and 6) fitting.}
\begin{tabular}{crcccc} \hline \hline
& & & & & \\
\multicolumn{6}{c}{NGC 1052} \\ \hline
Satellite & ObsID              & log(L(0.5-2 keV)) & log(L(0.5-2 keV)) & log(L(2-10 keV)) & log(L(2-10 keV)) \\ 
                               & & Individual        & Simultaneous      &  Individual      & Simultaneous  \\ 
                               (1) & (2) & (3) & (4) & (5) & (6) \\ \hline
\emph{XMM}-Newton &  093630101 & 40.955    [40.941  -  40.969] & 40.992    [40.987 -   40.996] & 41.460    [41.447 -  41.473] & 41.473    [41.458  -  41.488]  \\
\emph{XMM}-Newton &  306230101 & 40.960    [40.953  -  40.967] & 41.006    [41.002  -  41.011] & 41.483    [41.477 -  41.489] & 41.488    [41.481  -  41.494] \\
\emph{XMM}-Newton &  553300301 & 41.104    [41.096  -  41.112] & 41.063    [41.058   - 41.068] & 41.555    [41.549  - 41.561] & 41.545    [41.539  -  41.551] \\
\emph{XMM}-Newton & 553300401  & 41.117    [41.110   - 41.125] & 41.089    [41.084  -  41.093] & 41.551    [41.545  - 41.556] & 41.571    [41.565  -  41.577] \\
\emph{Chandra} (3$\arcsec$)    &  5910      & 40.186    [40.169  -  40.202] &     -    &   41.290  [41.266   -   41.292]   & - \\    
\emph{Chandra} (25$\arcsec$)    &  5910      & 40.290    [40.274  -  40.306] & -  &  41.336    [41.299  -  41.370] & - \\
 & & & & & \\
\multicolumn{6}{c}{NGC 3226}  \\ \hline
\emph{XMM}-Newton & 0101040301 & 40.762    [40.750  -  40.774]  & 40.775    [40.763  -  40.786] & 41.057    [41.041  -  41.072] & 41.013    [41.004 -   41.021] \\
\emph{XMM}-Newton & 0400270101 & 40.782    [40.777  -  40.787]  & 40.775    [40.769  -  40.780] & 40.994    [40.986   - 41.003] & 41.013    [41.005  -  41.021] \\

\emph{Chandra} (3$\arcsec$) & 860 & 40.473    [40.378  -  40.552] & - &   40.685       [ -   41.311] & -  \\
\emph{Chandra} (25$\arcsec$) & 860 & 40.552[40.495 - 40.602] & - & 40.909[ - 41.313] & - \\

 & & & & & \\
\multicolumn{6}{c}{NGC 3627}  \\ \hline
\emph{XMM}-Newton & 0093641101 & 39.440    [39.417  -  39.462]  &  39.452    [39.429  -  39.474]  & 39.420    [39.362  -  39.472]  & 39.245    [39.197  -  39.288]  \\
\emph{Chandra } (8$\arcsec$) & 9548         & 38.900 [38.854 -   38.942]     &  39.452    [39.421  -  39.481]  & 39.169 [ - 41.072]             & 39.224    [39.179   - 39.264] \\
\emph{Chandra } (25$\arcsec$) & 9548         &   39.367    [39.348  -  39.386] & - &    39.536    [39.268  -  39.701] & - \\
 & & & & & \\
\multicolumn{6}{c}{NGC 4261}  \\ \hline
\emph{Chandra } & 834          & 41.180    [41.161  -  41.197] & 40.976    [40.967  -  40.984] & 41.107    [40.857  -  41.265] & 41.018    [40.971  -  41.060] \\
\emph{Chandra} (3$\arcsec$) & 9569          & 40.924    [40.913  -  40.935] & 40.976    [40.967  -  40.984] & 40.996  [41.084    -  40.887] & 41.018    [40.971  -  41.060] \\
\emph{Chandra} (25$\arcsec$) & 9569          & 40.794    [40.779  -  40.807]   & - & 40.967    [40.940  -  40.993] & - \\
\emph{XMM}-Newton & 0056340101 & 41.301    [41.293  -  41.309] & 41.142    [41.138  -  41.146] & 41.188    [41.166  -  41.209] & 41.132    [41.113  -  41.151] \\
\emph{XMM}-Newton & 0502120101 & 41.223    [41.219  -  41.227] & 41.142    [41.138  -  41.146]  & 41.148    [41.136  -  41.160] & 41.132    [41.113  -  41.151] \\
 & & & & & \\
\multicolumn{6}{c}{NGC 4278}  \\ \hline
\emph{Chandra } (3$\arcsec$) & 7077  & 39.873    [39.865  -  39.881]  & 39.863    [39.856  -  39.871] & 39.762    [39.319  -  39.977] & 39.822    [39.806  -  39.837] \\
\emph{Chandra } (25$\arcsec$) & 7077  & 40.137    [40.102  -  40.169]  & - &  40.061    [40.035  -  40.084] & - \\
\emph{Chandra } & 7081  & 39.314 [39.303   -   39.324]   & 39.783    [39.774  -  39.792] & 39.269 [39.244    -  39.293]  & 39.732    [39.716   - 39.748] \\
\emph{Chandra } & 7080  &  39.778    [39.741  -  39.813] & 39.731    [39.717  -  39.743] & 39.615       [-   40.728] & 39.672    [39.653  -  39.690] \\
\emph{XMM}-Newton & 0205010101 & 40.733    [40.729  -  40.736] &  - &  40.768    [40.759  -  40.777]  & - \\
 & & & & & \\
\multicolumn{6}{c}{NGC 4552}  \\ \hline
\emph{Chandra } (3$\arcsec$) & 2072         & 39.687 [39.625   -   39.742] & 39.490    [39.471 -   39.509]  & 39.463 [39.045    -  40.147] &  39.449    [39.411  -  39.483]  \\
\emph{Chandra } (25$\arcsec$) & 2072         & 40.252    [40.239  -  40.266]  & - &  39.904    [39.819  -  39.974] & - \\
\emph{XMM}-Newton & 0141570101 & 40.396    [40.389  -  40.403] &  40.388    [40.370  -  40.406] & 40.128    [40.103  -  40.151] &  40.128    [40.103  -  40.151] \\
 & & & & & \\
\multicolumn{6}{c}{NGC 5846}  \\ \hline
\emph{XMM}-Newton & 0021540101 & 41.083    [41.079  -  41.088] & 41.080    [41.076 -   41.084] & 40.196    [40.163   - 40.227] & 40.189    [40.160  -  40.216] \\
\emph{XMM}-Newton & 0021540501 & 41.122    [41.114 -   41.130] & 41.080    [41.076 -   41.084] & 40.165    [40.115  -  40.210] & 40.189    [40.160  -  40.216] \\
\emph{Chandra }(7$\arcsec$) &  788         & 40.159    [40.118  -  40.195] & 40.291    [40.244  -  40.334] & 38.586    [38.528  -  38.637] & 39.326    [39.067  -  39.486] \\
\emph{Chandra }(25$\arcsec$) &  788        & 40.867    [40.853 -   40.880]  & - &  40.552    [40.407  -  40.661] & - \\
\emph{Chandra} & 7923          & 40.490    [40.471   - 40.507] & 40.291    [40.244  -  40.334] & 38.809    [38.760 -   38.853] & 39.326    [39.067  -  39.486] \\
\hline
\end{tabular} 
\end{center}
\end{table*}

\begin{table*}
\begin{center}
\caption{\label{annulus} Results for the best fit of the annular region (ring) in \emph{Chandra} data, and the best fit obtained for the nucleus of \emph{XMM}-Newton data when the contribution from the annular region was removed. Name and obsID in parenthesis (Col. 1), extracted region (Col. 2), best fit model (Col. 3), parameters of the best fit model (Col. 4, 5, 6, 7 and 8), soft and hard intrinsic luminosities (Col. 9 and 10), and the percentage of the contribution from the ring to the 25$\arcsec$ aperture \emph{Chandra} data in the 0.5-10.0 keV band (Col. 11).}
\begin{tabular}{lcccccccccc} \hline \hline
Name (obsID) & Region & Model & $N_{H1}$ & $N_{H2}$ & kT & $\Gamma$ & $\chi^2_r$ & $logL_{soft}$ & $logL_{hard}$ & Cont. \\
 & & & ($10^{22} cm^{-2}$) & ($10^{22} cm^{-2}$) & (keV) & & & (0.5-2 keV) & (2-10 keV) & \%  \\
(1) & (2) & (3)          & (4) & (5) & (6) & (7) & (8) & (9) & (10) & (11) \\ \hline  
NGC\,1052 (5910) & Ring & ME2PL & - & 26.10$ _{ 16.93 }^{ 38.14 }$ & 0.31$ _{ 0.28 }^{ 0.39 }$ & 1.98$ _{ 1.71 }^{ 2.23 }$ & 1.47 & 40.185 & 40.563 & 10  \\
NGC\,1052 (093630101) & Nucleus & ME2PL & - & 7.54$ _{ 5.20 }^{ 9.98 }$ & 0.78$ _{ 0.68 }^{ 0.89 }$ & 1.69$ _{ 1.50 }^{ 1.86 }$ & 1.13 & 39.875 & 41.235 \\ \hline
NGC\,3226 (860) & Ring & PL & - & - & - & 1.42$^{1.07}_{1.85}$ & 1.89 & 40.482 & 40.557 & 20 \\
NGC\,3226 (0101040301) & Nucleus & 2PL & 0.35$^{1.88}_{0.00}$ & 1.02$^{1.88}_{0.00}$ & - & 1.72$^{1.58}_{1.88}$ & 1.00 & 40.623 & 40.912 \\ \hline
NGC\,3627 (9548) & Ring & MEPL & - & - &  0.41$_{ 0.35 }^{ 0.49 }$ & 1.71 $_{ 1.58 }^{ 1.85 }$ & 1.35 & 39.302  & 39.384 & 92  \\
NGC\,3627 (093641101) & Nucleus & MEPL & - & - & 0.71$_{ 0.40 }^{ 1.02 }$ & 3.37$_{ 2.97 }^{ 3.76 }$ & 1.16 & 38.990 & 38.125 \\ \hline
NGC\,4261 (9569) & Ring & MEPL & 0.06$ _{ 0.03}^{ 0.11 }$ & 0.00$ _{ 0.00 }^{ 0.04 }$ & 0.61$ _{ 0.59 }^{ 0.62 }$ & 1.87$ _{ 1.70 }^{ 2.07 }$ & 2.07 & 40.663  & 40.252 & 37  \\
NGC\,4261 (0502120101) & Nucleus & ME2PL & - & 8.29$ _{ 7.28}^{ 9.89 }$ & 0.67$ _{ 0.65 }^{ 0.68 }$ & 1.56$ _{ 1.41 }^{ 1.72 }$ & 1.21 & 40.857 & 41.051 \\ \hline
NGC\,4278 (7077) & Ring & MEPL & - & - & 0.29$_{ 0.19 }^{ 0.36 }$ & 1.61$_{ 1.51 }^{ 1.71 }$ & 1.12 & 39.551  & 39.707 & 38  \\
NGC\,4278 (0205010101) & Nucleus & PL & 0.02$ _{ 0.02}^{ 0.03 }$ & & & 2.05$ _{ 2.03}^{ 2.10 }$ & 1.05 & 40.681 & 40.736 \\ \hline
NGC\,4552 (2072) & Ring & MEPL & - & - &  0.60$ _{ 0.57 }^{ 0.62 }$ & 1.92$ _{ 1.81 }^{ 2.06 }$ & 1.47 & 39.648  & 39.304 & 23  \\
NGC\,4552 (0141570101) & Nucleus & MEPL & - & - & 0.58$ _{ 0.52 }^{ 0.62 }$ & 1.90$ _{ 1.80 }^{ 2.06 }$ & 1.10 & 40.053 & 39.878 \\ \hline
NGC\,5846 (788) & Ring & MEPL & - & - & 0.61$ _{ 0.59 }^{ 0.62 }$ & 1.87$ _{ 1.61 }^{ 2.46 }$ & 1.63 & 40.815  & 40.217 & 73 \\
NGC\,5846 (0021540101) & Nucleus & MEPL & - & - & 0.63$ _{ 0.61 }^{ 0.64 }$ & 4.00$ _{ 3.56 }$ & 1.36 & 40.753 & 39.171 \\
\hline
\end{tabular} 
\end{center}
\end{table*}

\begin{table*}
\begin{center}
\caption{\label{simultanillo} Simultaneous fittings taking into account the contribution from the annular region given in Table \ref{annulus}. Name and obsID in parenthesis (Col. 1), parameters of the best fit model (Col. 2, 3, 4, 5, 6 and 7), $\chi^2/d.o.f$ (Col. 8) and soft and hard intrinsic luminosities (Col. 9 and 10). } 
\begin{tabular}{lccccccccr} \hline
\hline
 ObsID & $N_{H1}$ & $N_{H2}$ & kT & $\Gamma$ & $Norm_1$ & $Norm_2$ & $\chi^2/d.o.f$ & $logL_{soft}$ & $logL_{hard}$ \\
  & ($10^{22} cm^{-2}$) & ($10^{22} cm^{-2}$) & keV & & ($10^{-4}$) & ($10^{-4}$) & & (0.5-2 keV) & (2-10 keV)  \\
(1) & (2) & (3) & (4) & (5) & (6) & (7) & (8) & (9) & (10)  \\
\hline
NGC~1052 & & & & & & & & & \\
\hline
0093630101 & - & 6.11$_{5.07}^{7.18}$ & 0.68$_{0.63}^{0.73}$ & 1.58$_{1.47}^{1.70}$ & 0.48$_{0.44}^{0.53}$ & 1.28$_{0.95}^{1.69}$ & 577.98/477 & 40.35$_{40.34}^{40.36}$ & 41.33$_{41.32}^{41.35}$ \\
5910 & & & & & & & & 40.35$_{40.34}^{40.36}$ & 41.27$_{41.26}^{41.28}$ \\
\hline
NGC~3226 & & & & & & & & & \\
\hline
0101040301 & 0.34$^{0.01}_{0.54}$ & 1.03$^{1.42}_{0.68}$ & - & 1.72$^{1.86}_{1.60}$ & 0.55$^{0.93}_{0.03}$ & 1.59$^{1.23}_{2.11}$ & 268.434/279 & 40.710$^{40.732}_{40.688}$ & 41.681$^{41.725}_{41.633}$ \\
860 & & 0.07$^{0.00}_{0.32}$ & & & & 0.68$^{0.92}_{0.34}$ & & 40.510$^{40.581}_{40.425}$ & 40.958$^{40.978}_{40.938}$ \\

\hline
NGC~4261 & & & & & & & & & \\
\hline
0502120101 & 0.03$_{0.00}^{0.08}$ & 7.64$_{6.58}^{8.74}$ & 0.60$_{0.59}^{0.61}$ & 1.55$_{1.36}^{1.75}$ & 0.22$_{0.19}^{0.26}$ & 1.53$_{1.04}^{2.25}$ & 972.83/627 & 40.84$_{40.84}^{40.84}$ & 41.03$_{41.01}^{41.05}$ \\
9569 & & & & & & & & 40.84$_{40.83}^{40.85}$ & 40.99$_{40.97}^{41.01}$ \\
\hline
NGC~4278 & & & & & & & & & \\
\hline
0205010101 & 0.03$_{0.02}^{0.03}$ & - & - & 2.11$_{2.08}^{2.14}$ & 7.74$_{7.54}^{7.95}$ & & 844.93/690 & 40.71$_{40.71}^{40.72}$ & 40.71$_{40.70}^{40.72}$  \\
7077 & & & & & 1.15$_{1.11}^{1.18}$ & & & 39.89$_{39.88}^{39.89}$ & 39.86$_{39.84}^{39.87}$ \\
\hline
NGC~4552 & & & & & & & & & \\
\hline
0141570101 & 0.02$_{0.00}^{0.04}$ & 0.01$_{0.00}^{0.05}$ & 0.64$_{0.61}^{0.66}$ & 1.85$_{1.78}^{2.00}$ & 0.77$_{0.70}^{0.90}$ & 0.84$_{0.76}^{1.01}$ & 380.50/327 & 40.05$_{40.04}^{40.05}$ & 39.90$_{39.88}^{39.93}$ \\
2072 & & & & & 0.16$_{0.14}^{0.18}$ & 0.31$_{0.28}^{0.37}$ & & 39.51$_{39.49}^{39.53}$ & 39.47$_{39.42}^{39.51}$ \\
\hline
\end{tabular}
\end{center}
\end{table*}

\begin{table*}
\begin{center}
\caption{\label{estcurvas} Statistics for the light curves. Name (Col. 1), obsID (Col. 2), $\chi^2/d.o.f$ and the probability of being variable (Col. 3 and 4) and normalized excess variance with errors (Col. 5 and 6).} 
\begin{tabular}{crrccc} \hline \hline
Name & ObsID & $\chi^2/d.o.f$ & Prob. & $\sigma^2$ & $err(\sigma^2)$ \\
 & & & (\%) & ($10^{-2}$) & ($10^{-2}$) \\
(1) & (2) & (3)        & (4) & (5) & (6)  \\ \hline  
NGC~1052 & 0093630101 & 18.6/11 & 93 &   0.189 & 0.107 \\
         & 0306230101 & 57.1/49 & 80 & 0.043 & 0.046 \\
         & 0553300301 & 49.5/46 & 66 & 0.022 & 0.043 \\
         & 0553300401 & 39.3/49 & 16 & -0.042 & 0.042 \\
         &  5910      & 48.6/59 & 17 & -0.157 & 0.173 \\
NGC~3226 & 860        & 51.2/45 & 76 & -0.047 & 0.028 \\
         & 0101040301 & 35.0/32 & 67 & 0.074 & 0.195 \\
         & 0400270101 & 474.1/100 & 100 & 1.781 & 0.079 \\
NGC~3627 & 0093641101 & 2.6/5   & 24 & -0.312 & 0.605 \\
         &    9548    & 63.8/49 & 92 & 0.360 & 0.351 \\
NGC~4261 & 834        & 35.8/34 & 62 & 0.077 & 0.267 \\
         & 9569       & 85.7/100 & 15 & -0.243 & 0.195 \\
         & 0056340101 & 25.4/19 & 85 & 0.109 & 0.099 \\
         & 0502120101 & 79.6/64 & 91 & 0.075 & 0.057 \\
NGC~4278 &   7077     & 124.3/111 & 82 & 0.149 & 0.184 \\
         &   7081     & 91.4/107 & 14 & -0.299 & 0.227 \\
         &  7080      & 41.0/55   & 8 & -0.445 & 0.354 \\
         & 0205010101 & 2.2/11   & 0 & -0.074 & 0.041 \\
NGC~4552 &     2072   & 57.4/45 & 90 & 0.089 & 0.570 \\
         & 0141570101 & 12.8/14 & 46 & 0.000 & 0.081 \\
NGC~5846 & 0021540101 & 20.1/28 & 14 & -0.034 & 0.037 \\
         & 0021540501 & 7.9/8   & 56 & 0.014 & 0.070 \\
         &    788     & 20.4/16 & 80 & 0.513 & 0.640 \\
         & 7923       & 85.2/86 & 50 & -0.136 & 0.560 \\
\hline
\end{tabular} 
\end{center}
\end{table*}

\begin{table*}
\begin{center}
\caption{\label{luminUV} UV luminosities derived from the OM observations. Name (Col. 1), obsID (Col. 2), filter (Col. 3) and luminosity (Col. 4).} 
\begin{tabular}{ccccc} \hline \hline
Name & ObsID & Filter & logL & $\alpha_{ox}$  \\
 & & & (erg/s) & \\
(1) & (2) & (3)        & (4) & (5)   \\ \hline  
NGC\,1052 & 0306230101 & UVM2 &  41.038 $^+_-$ 0.013 & -0.94 $^+_-$ 0.03 \\
 & 0306230101 & UVW2 & 40.938 $^+_-$ 0.028 & \\
 & 0553300301 & UVM2 & 41.139 $^+_-$ 0.012 & -0.95 $^+_-$ 0.03 \\
 & 0553300301 & UVW2 & 40.947 $^+_-$ 0.026 & \\
 & 0553300401 & UVM2 & 41.041 $^+_-$ 0.014 & -0.91 $^+_-$ 0.03 \\
 & 0553300401 & UVW2 & 40.885 $^+_-$ 0.025 & \\
NGC\,3226 & 0101040301 & UVW1 & 41.375 $^+_-$ 0.006 & -1.15 $^+_-$ 0.70 \\
 & 0400270101 & UVW1 & 41.327 $^+_-$ 0.001 & * \\
NGC\,3627 & 0093641101 & UVW1 & 41.367 $^+_-$ 0.007 & -1.66 $^+_-$ 0.05 \\
NGC\,4261 & 0056340101 & UVM2 & 41.431 $^+_-$ 0.024 & -1.04 $^+_-$ 0.05 \\
 & 0056340101 & UVW1 & 42.169 $^+_-$ 0.005 & \\
 & 0502120201 & UVM2 & 41.609 $^+_-$ 0.014 & -1.11 $^+_-$ 0.03 \\
 & 0502120201 & UVW1 & 42.214 $^+_-$ 0.001 & \\
NGC\,4278 & 0205010101 & UVW1 & 40.903 $^+_-$ 0.020 & -1.11 $^+_-$ 0.02 \\
NGC\,4552 & 0141570101 & UVW2 & 41.273 $^+_-$ 0.033 & -1.28 $^+_-$ 0.04 \\
NGC\,5846 & 0021540101 & UVW2 & 40.561 $^+_-$ 0.148 & -0.81 $^+_-$ 0.14 \\ 
\hline
\begin{scriptsize}
* The observation was not used (see text).
\end{scriptsize}
\end{tabular} 
\end{center}
\end{table*}

\begin{table*}
\begin{center}
\caption{\label{variability} Summary. Name, and the instrument in parenthesis (Col. 1), type (Col. 2), logarithm of the soft (0.5-2 keV) and hard (2-10 keV) X-ray luminosities, where the mean was calculated or \emph{Chandra} luminosity was given when both instruments were used, and percentages in flux variations (Col. 3 and 4), black hole mass in logarithmical scale, from \cite{omaira2009b} (Col. 5), Eddington ratio, calculated from \cite{eracleous2010b} using $L_{bol}=33L_{2-10 keV}$ (Col. 6), best-fit for the SMF0 (Col. 7), parameter varying in SMF1, with the percentage of variation (Col. 8), parameter varying in SMF2, with the percentage of variation (Col. 9), the sampling time-scale for variations (Col. 10), and variations in the hardness ratios (Col. 11).} 
\begin{tabular}{lccccccccccc} \hline \hline
Name & Type & log $L_{soft}$ & log $L_{hard}$ & log $M_{BH}$ & $L_{bol}/L_{Edd}$ & & Variability & & T & HR  \\ 
 & & (0.5-2 keV) & (2-10 keV) & & & SMF0 &  SMF1 & SMF2 & (Years) & \\
(1) & (2) & (3)          & (4) & (5) & (6) & (7) & (8) & (9) & (10) & (11) \\ \hline
NGC~1052 (X) & AGN       & 41.04  & 41.52  & 8.07 &  7.2 $\times 10^{-4}$   &  ME2PL & $Norm_2$  & $N_{H2}$  &   8 & &   \\
 & 1.9 & 20\% & 20\% & & & & 49\% & 31\% & & 33\%  \\
NGC~3226 (X,C) & AGN       & 40.78  & 41.01  & 8.22 &  1.6 $\times 10^{-4}$     &  2PL   & $N_{H2}$  & $Norm_2$ & 1 & &   \\
 & 1.9 & 37\% & 81\% & & & & 93\% & 57\% & &   \\
NGC~4261 (X) & AGN       & 40.98  & 41.02  & 8.96 &  2.9 $\times 10^{-5}$   &  ME2PL &  - &  - & 8 & &  \\
 & 2 & 0\% & 0\% & & & & & & &  0\%  \\
\hspace*{1.4cm}  (C) & &  & &  &     &  ME2PL &  - &  - & 6 &  \\
 &  & 0\% & 0\% & & & & & & &  19\%  \\
NGC~4278 (C) & AGN       & 39.80  & 39.75  & 8.46 &  5.0 $\times 10^{-6}$  &  MEPL    &  $Norm_2$ & - & 1  & & \\
 & 1.9 & 26\% & 29\% & & & & 30\% & & & 40\%*  \\
NGC~4552 (X,C) & AGN     & 39.49  & 39.45  & 8.84 &  1.0 $\times 10^{-6}$   &  MEPL   &  $Norm_1$  &  $Norm_2$  & 2  & &   \\
 & 2 & 73\% & 63\% & & & & 93\% & 78\% & &   \\
NGC~5846 (X) & Non-AGN   & 40.29  & 39.33  & 8.49 &  1.8  $\times 10^{-6}$  &  MEPL & - & - & 7 &     \\
 & 2 & 0\% & 0\% & & & & & & & 1\%  \\
\hspace*{1.4cm} (C) & &  & &  &  &  MEPL & - & - & 0.6    \\
 &  & 0\% & 0\% & & & & & & & 6\%  \\
\hline
\end{tabular} 
\end{center}
\begin{scriptsize}
* For all \emph{Chandra} data. With useful spectroscopic data variations are 4\% .
\end{scriptsize}
\end{table*}

\clearpage
\newpage

\appendix

\section{ \label{previous} Notes and comparisons with previous results for individual objects}

\subsection{NGC\,1052}

The brightest elliptical galaxy in the Cetus~I group, NGC\,1052, already classed as a LINER in the pioneering work by \cite{heckman1980}, was classified as LINER type 1.9 by \cite{ho1997}.

NGC\,1052 was observed twice with \emph{Chandra} and five times with the \emph{XMM}-Newton satellite, what makes it a good candidate for studying variability. The general characteristics of the two \emph{Chandra} observations ObsID 385 (taken in 2000) and ObsID 5910 (2005) were reported by \cite{omaira2009a} and \cite{boroson2011}, respectively, showing quite different spectral behaviour with a flatter spectral index, lower $N_{H}$ and lower luminosity in 2005. We analyzed the observation from 2005 and found a different spectral fit from \cite{boroson2011}, since they used a PL model to fit the spectrum. However, similar luminosities were found between \cite{omaira2009a}, \cite{boroson2011} and this work (with log(L(2-10 keV)=41.4$^{+0.1}_{-0.8}$, 41.4$^+_-$0.01 and 41.3$^{+0.00}_{-0.02}$, respectively. 

Only one \emph{XMM}-Newton observation (ObsID 306230101 taken in 2006) was previously analyzed by \cite{omaira2009a} and \cite{brightman2011}, showing quite similar results, with \cite{brightman2011} in spite of a more absorbed spectrum in the latter. Our results are in good agreement with those provided by \cite{omaira2009a}. The most recent observation at X-rays reported so far is a 100 ks observation taken with \emph{SUZAKU} in 2007. The derived spectral characteristics reported by \cite{brenneman2009} appear to be similar to those from \emph{XMM}-Newton which is compatible with the values in \cite{omaira2009a}, \cite{brightman2011} and this paper (intrinsic luminosity of log(L(2-10 keV)) $\sim$ 41.5).

In the UV range, \cite{maoz2005} studied this galaxy with \emph{HST} ACS and found a decrease in the flux of the source of a factor of 2 between the 1997 data reported by \cite{pogge2000} and their 2002 dataset. We found UV flux variations of a factor of 1.3 from \emph{XMM}-OM data in seven months.

\subsection{NGC\,3226}

NGC\,3226 is a dwarf elliptical galaxy, which is strongly interacting with the Seyfert 1.5 galaxy NGC\,3227, located at 2$\arcmin$ in proyected distance \citep[see Figure C.19 in][]{omaira2009a}.  NGC\,3226 was classified by \cite{ho1997} as a type 1.9 LINER.

This galaxy was observed two times with \emph{Chandra}-ACIS in 1999 and 2001 and four times with \emph{XMM}-Newton from 2000 to 2006.

\emph{Chandra} data taken in 1999 were analyzed by \cite{george2001} and those taken in 2001 by \cite{terashimawilson2003}. In both cases the X-ray spectra were fitted to a single power law, but with differing column densities, the data in 2001 being a factor 2 more absorbed; this leads to a difference in the X-ray luminosity in the hard (2-10 keV) band of 70\%. The reanalysis of these data performed by \cite{younes2011} shows spectral parameters consistent with the previous studies although their estimated difference in luminosity is lower (40\%). We did not use these data because obsID 1616 does not match our criteria of the minimum number of counts.

The analysis by \cite{gondoin2004} of snapshot \emph{XMM}-Newton observations taken in 2000 also showed an X-ray spectrum consistent with a power law with log(L(2-10 keV)) = 40.38$^+_-$0.01, close to the value obtained by \cite{george2001}, log(L(2-10 keV)) = 40.26$^+_-$0.01, for the same epoch. \cite{binder2009} studied the \emph{XMM}-Newton observation from 2006 and found significant short time scale flux variation, with a fractional variability amplitude of 11.7. They compare their measurements, log (L(0.4-2 keV)) = 40.47$^{+0.01}_{-0.00}$ and log (L(2-10 keV)) = 40.55$^{+0.02}_{-0.08}$, with those from \cite{gondoin2004} (see above), finding variability on time-scales of years. \cite{younes2011} analyzed the longest exposure data from 2000 and 2006 and found a difference in the hard luminosity of 40\%. Considering both, \emph{Chandra} and \emph{XMM}-Newton data (obsID. 860 and 1616, 0101040301 and 0400270101, respectively), they concluded that the source variability is due to modifications in the $\rm{N_H}$. We find spectral parameters in agreement to what \cite{younes2011} found for the \emph{XMM}-Newton and \emph{Chandra} data, and we obtained a $\rm{\chi^2_r = 4.7}$ for obsID 0400270101, but this observation seemed to be affected by the presence of the rapid variable Seyfert 1.5 NGC\,3227. \cite{binder2009} and \cite{younes2011} found similar results for this galaxy, interpretation of such variability was related to outflows or feedback processes.

\subsection{NGC\,3627}

Together with NGC\,3628 and NGC\,3623, these three galaxies form the Leo Triplet (see Figure C.27 in \cite{omaira2009a}). \cite{cappellari1999} classified it as type 2.0 LINER.

This galaxy was observed twice with \emph{Chandra}, in 1999 and 2008, and once with \emph{XMM}-Newton, in 2001.

\cite{ho2001} and \cite{panessa2006} studied \emph{Chandra} ObsID 394 from 1999, using a PL model with $\Gamma=1.8$, deriving similar luminosities, log (L(2-10 keV)) = 37.6 and 37.9, respectively. \emph{Chandra} ObsID 9548 was analyzed by \cite{grier2011}, who used a PL with $\Gamma=2$ to get log (L(0.3-8 keV)) = 38.51$\pm$0.03. We used a MEPL model to fit this spectrum and obtained greater luminosity log (L(2-10) keV)) = 39.2$^{+1.9}$, the difference most probably due to the different models used for the analysis. The \emph{XMM}-Newton observation was analyzed by \cite{omaira2009a} and \cite{brightman2011} who used PL and MEPL models to calculate luminosities, and derived log (L(2-10 keV)) = 39.2$\pm$0.1 and 39.5, respectively, with which our results agree (39.4$\pm$0.1).

No information on UV was found in the literature.

\subsection{NGC\,4261}

NGC\,4261 contains a pair of symmetric kpc-scale jets \citep{birkinshawdavies1985} and a nuclear disc of dust roughly perpendicular to the radio jet \citep{ferrarese1996}. \cite{ho1997} classified it as type 2.0 LINER.

It has been observed twice with \emph{Chandra}, in 2000 and 2008, and with \emph{XMM}-Newton in other three epochs from 2001 to 2007. However, only the analyses on \emph{Chandra} ObsID 834 and \emph{XMM}-Newton ObsID 56340101 are published.

\cite{satyapal2005}, \cite{rinn2005} and \cite{donato2004} reported quite consistent spectral parameters by fitting the \emph{Chandra} spectra with a thermal and a power law component but with a range of variation in the reported luminosities log(L(2-10 keV)) between 40.5 and 41.0. On the other side \cite{omaira2009a} and \cite{zezas2005} fitted the spectra with a ME2PL, using the same value for the spectral index for the two PL in the case of \cite{omaira2009a} and varying the spectral index in the case of \cite{zezas2005}. The comparison of the spectral parameters is quite consistent with the largest difference being the $\rm{N_H}$, $16.45\times 10^{22} cm^{-2}$ for \cite{omaira2009a} and $8.4\times 10^{22} cm^{-2}$ for \cite{zezas2005}; this leads to a higher luminosity in \cite{zezas2005}, log(L(2-10keV)) = 42.0, against log(L(2-10keV)) = 41.1$^{+0.1}_{-0.7}$ in \cite{omaira2009a}. We used the ME2PL model to fit this spectrum and found spectral parameters and luminosities in agreement with \cite{omaira2009a}.

The \emph{XMM}-Newton observation ObsID 056340101 taken in 2001 was analyzed by \cite{omaira2009a}, \cite{sambruna2003} and \cite{gliozzi2003}. Different models were used for the three works: an absorbed PL \citep{gliozzi2003}, an absorbed MEPL \citep{sambruna2003} and a ME2PL \citep{omaira2009a}. This could explain, in principle, the different luminosities reported. Our results (log L (2-10 keV) = 41.13$^+_-$0.02) are in good agreement with the spectral parameters and luminosities (41.2$^{+0.0}_{-0.7}$) reported by \cite{omaira2009a}, but not with those obtained by \cite{sambruna2003} and \cite{gliozzi2003} (41.9).

Information from the UV was not found in the literature.

\subsection{NGC\,4278}

NGC\,4278 is an elliptical galaxy classified as type 1.9 LINER by \cite{ho1997}, who found a relatively weak, broad $H_{\alpha}$ line. The north-northwest side of the galaxy is heavily obscured by large-scale dust-lanes, whose distribution shows several dense knots interconnected by filaments \citep{carollo1997}.

This galaxy has been observed in nine occasions with \emph{ Chandra} from 2000 to 2010 and once with \emph{XMM}-Newton in 2004. \cite{brassington2009} used six \emph{Chandra} observations and found 97 variable sources within NGC\,4278, in an elliptical area of 4$\arcmin$ centered in the nucleus. None of them are within the aperture of 3$\arcsec$ we used for the nuclear extraction. \emph{Chandra} observations were taken by G. Fabbiano to study the plethora of sources detected in this galaxy ($\sim$ 250) (see \cite{brassington2009} and \cite{boroson2011}). The nuclear source was studied by \cite{omaira2009a} using \emph{Chandra} obsID 7077 taken in 2006 and \emph{XMM}-Newton data. They found a hard X-ray luminosity difference of a factor of 10 between the two observations, that were attributed to the contamination of the numerous sources around the nucleus.

The nuclear variability of this source has been already studied by \cite{younes2010}. They conclude that long time-scale (months) variability is detected with a flux increase of a factor of $\sim$ 3 on a time-scale of a few months and a factor of 5 between the faintest and the brightest observation separated by $\sim$ 3 years. We used three of these observations, our spectral fittings being in good agreement with theirs, although we found smaller variations in luminosities. Whereas the different \emph{Chandra} observations did not show short time-scale (minutes to hours) variability, during the \emph{XMM}-Newton observation, where the highest flux level was detected, they found a 10\% flux increase on a short time-scale of a few hours. With the same dataset we got a 3\% variation in the same time range, the difference being most probably due to the different apertures used for the analysis.

In the UV, \cite{cardullo2008} found that the luminosity increased a factor of 1.6 in around six months using data from \emph{HST} WFPC2/F218W.

\subsection{NGC\,4552}

This Virgo elliptical galaxy has been classified as LINER 2.0 \citep{cappellari1999}. A radio jet was detected with \emph{VLBI} observations \citep{nagar2005}.

It has been observed four times with \emph{Chandra} from 2001 to 2012 and with \emph{XMM}-Newton in a single epoch in 2003. However, three of the \emph{Chandra} observations are not public yet, so the only reported results came from \emph{Chandra} ObsID 2072 \citep{filho2004, xu2005, omaira2009a, grier2011, boroson2011}. \cite{filho2004} and \cite{omaira2009a} fitted the spectra to a MEPL obtaining compatible results with log (L(2-10keV)) = 39.4 and 39.2$^{+0.2}_{-0.4}$, respectively. \cite{xu2005} and \cite{grier2011} by fitting a single PL obtained log (L(0.3-10 keV))= 39.6 and 40.0$^{+0.01}_{-0.01}$, respectively. Our results (log L(2-10 keV) = 39.5$^{+0.7}_{-0.4}$) are in good agreement with all of them. 

In the UV, \cite{cappellari1999} studied this LINER with both \emph{HST} imaging (FOC) and spectroscopy (FOS), with images taken in 1991, 1993 and 1996 showing long term variability. This is in agreement with \cite{maoz2005}, who used \emph{HST-ACS} observations with its HRC mode and found a 20\% variation of the nuclear flux in both F250W and F330W bands.

\subsection{NGC\,5846}

NGC\,5846 is a giant elliptical galaxy at the center of a small compact group of galaxies. The inner region of the galaxy contains dust and a radio core \citep{moellenhoff1992}. \cite{ho1997} classified this galaxy as an ambiguous case like transient 2.0 object but the revision made by  \cite{omaira2009a} located this object into the LINERs 2.0 cathegory.

Three \emph{Chandra} observations are archived for this galaxy between 2000 and 2007 and two observations with \emph{XMM}-Newton in January and August 2001. \emph{Chandra} ObsID 788 was analyzed by \cite{filho2004}, \cite{satyapal2005} and \cite{trinchieri2002}. Three different models were fitted (PL, APEC and MEPL), and an order of magnitude difference in luminosity was found between \cite{filho2004} (log L(2-10 keV) = 38.4) and the other two works (39.4 and 39.6$^+_-$0.4, respectively), maybe entirely due to the different models used. Our luminosity (log (L(2-10 keV))=39.3$^{+0.2}_{-0.3}$) is in good agreement with those from \cite{satyapal2005} and \cite{trinchieri2002}. \cite{omaira2009a} reported \emph{Chandra} ObsID 4009 from 2003, but we have noticed that obsID 4009 corresponds to the galaxy NGC\,5845, not to NGC\,5846. The \emph{XMM}-Newton observations reported by \cite{omaira2009a} based in data taken on 2001 were fitted with a ME2PL model resulting in a log L(2-10 keV) = 40.8$^{+0.0}_{-2.4}$, in agreement with our results (40.2$^+_-$0.0).

\section{X-ray Images}

In this appendix we present the images from \emph{Chandra} (left) and \emph{XMM}-Newton (right) that we used to compare data from these instruments in the 0.5-10 keV band. Big circles represent 25$\arcsec$ apertures. Small circles in the left figures represent the nuclear extraction aperture used with \emph{Chandra} observations (see Table \ref{obs}). In all cases, the grey levels go from two times the value of the background dispersion, to the maximum value at the center of each galaxy.

\begin{figure*}[H]
\centering
\includegraphics[width=\textwidth]{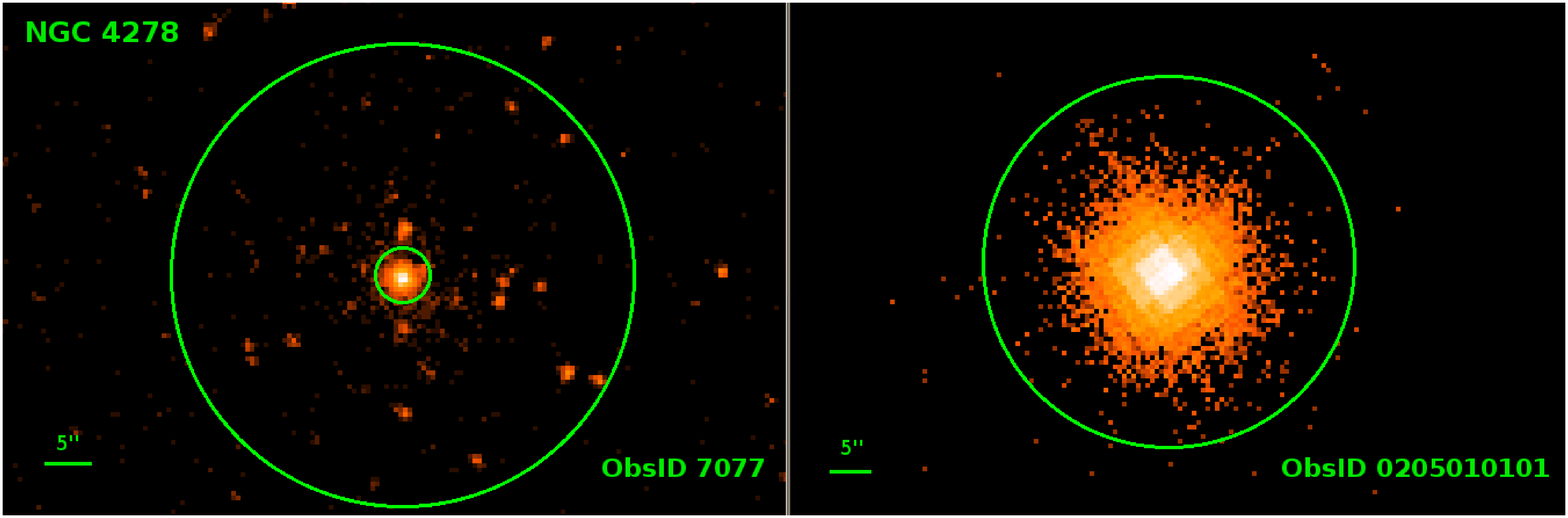}

\vspace*{1cm}
\includegraphics[width=\textwidth]{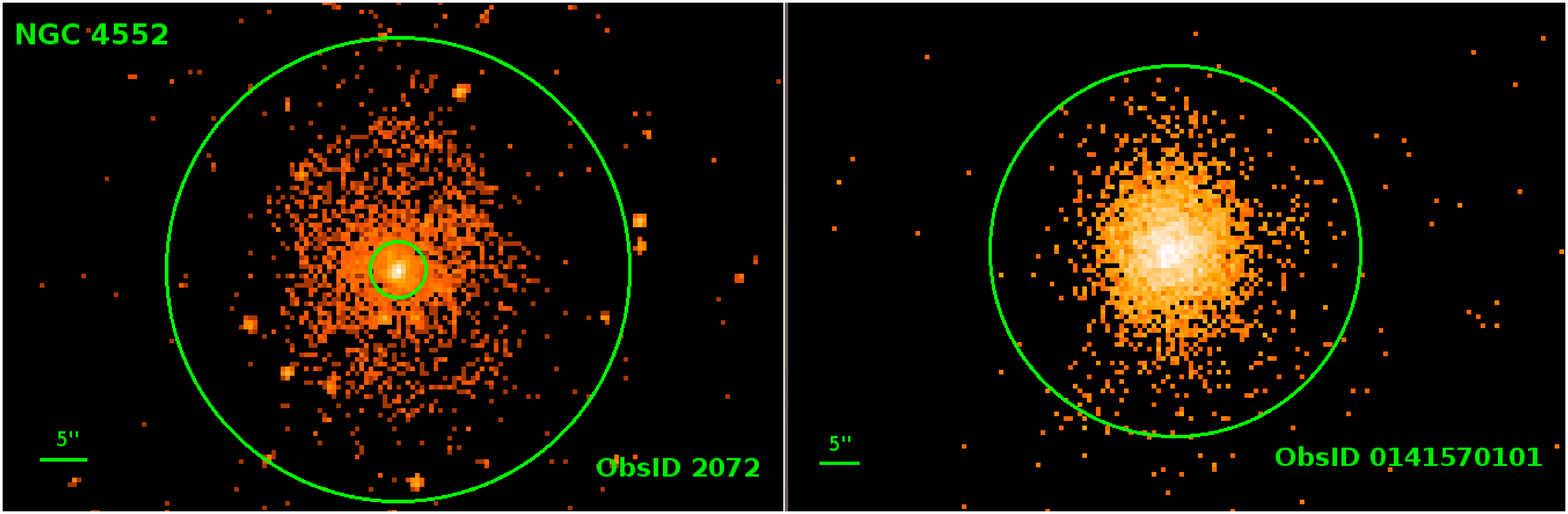}

\vspace*{1cm}
\includegraphics[width=\textwidth]{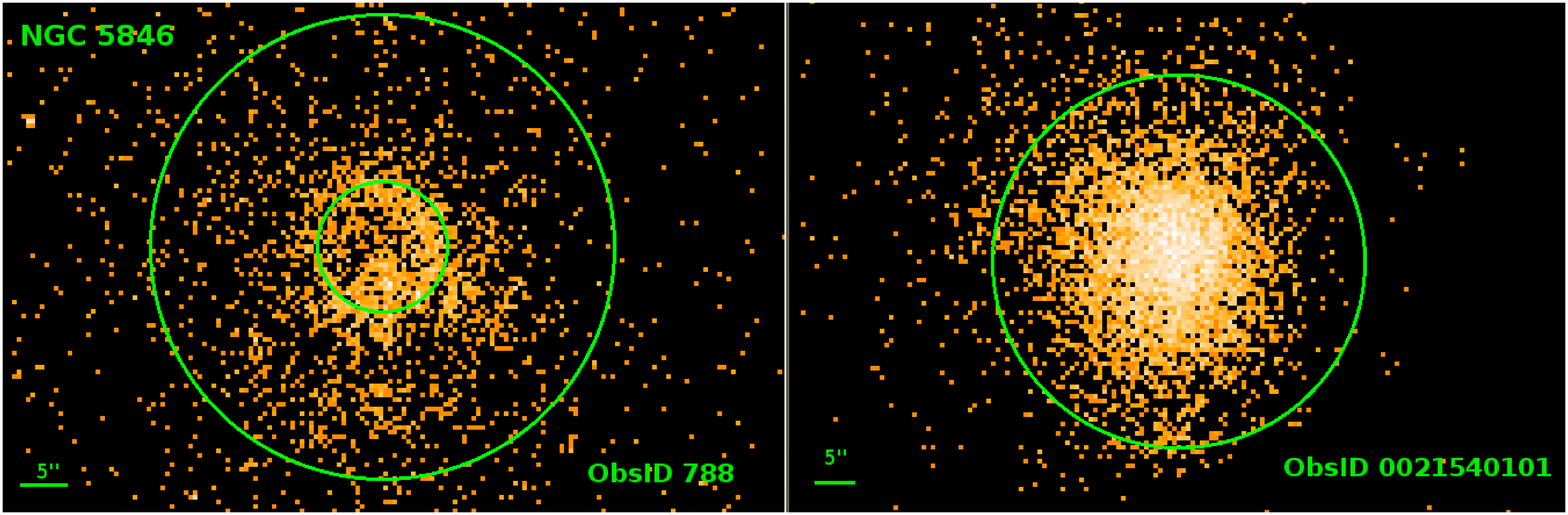}
\end{figure*}

\section{Light Curves}

The plots corresponding to the light curves are provided.

\begin{figure*}[H]
\caption{ \label{lightcurves3226} Lightcurves for NGC\,3226 (up). Light curve from NGC\,3227 (middle-right), where the red solid line represents a linear regresion, and the light curve from the background (middle-left). \emph{XMM}-Newton obsID 0400270101 image (down). The background follows the same behaviour that NGC\,3227.}
\centering
\includegraphics[width=0.4\textwidth]{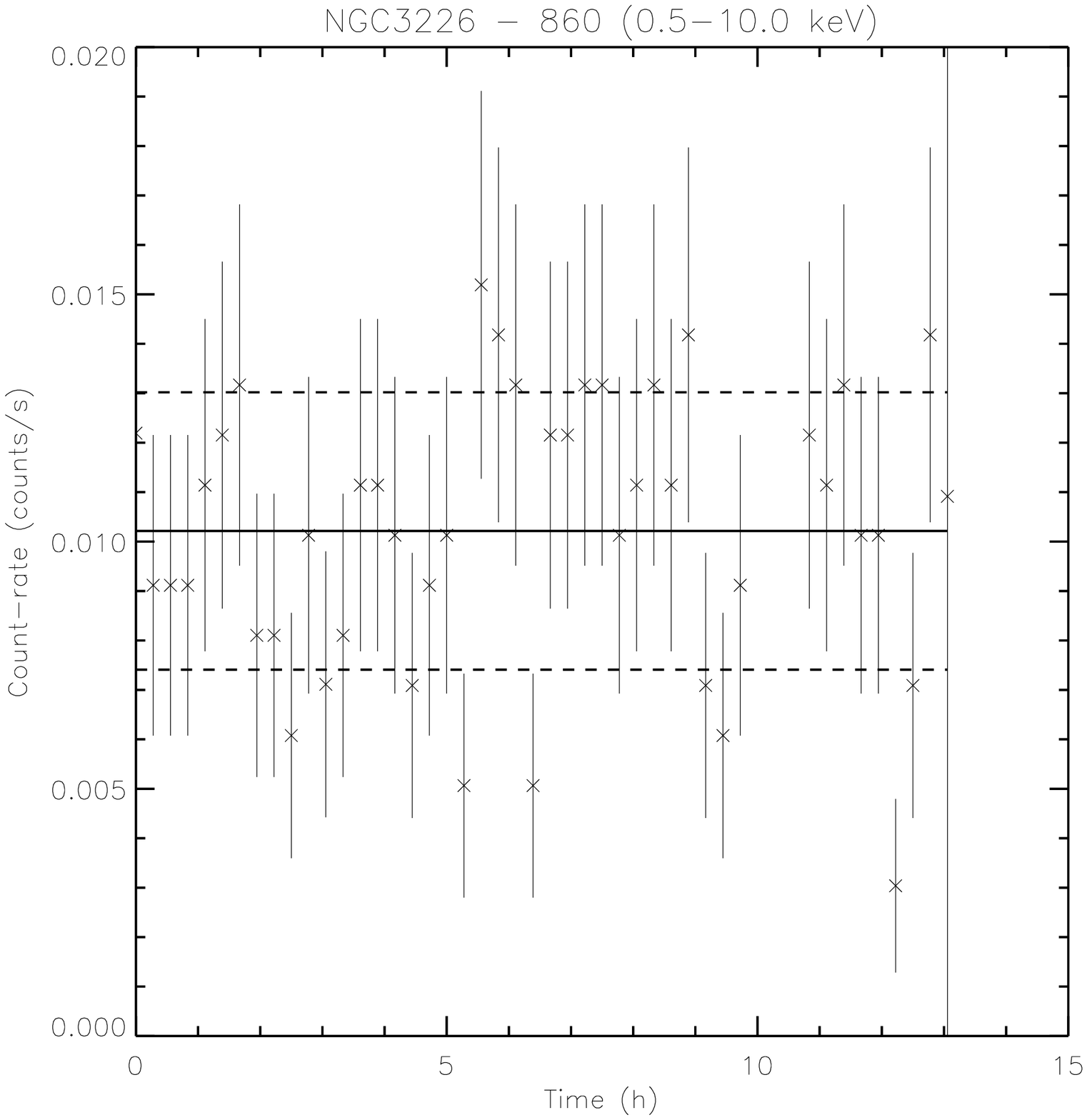}

\includegraphics[width=0.4\textwidth]{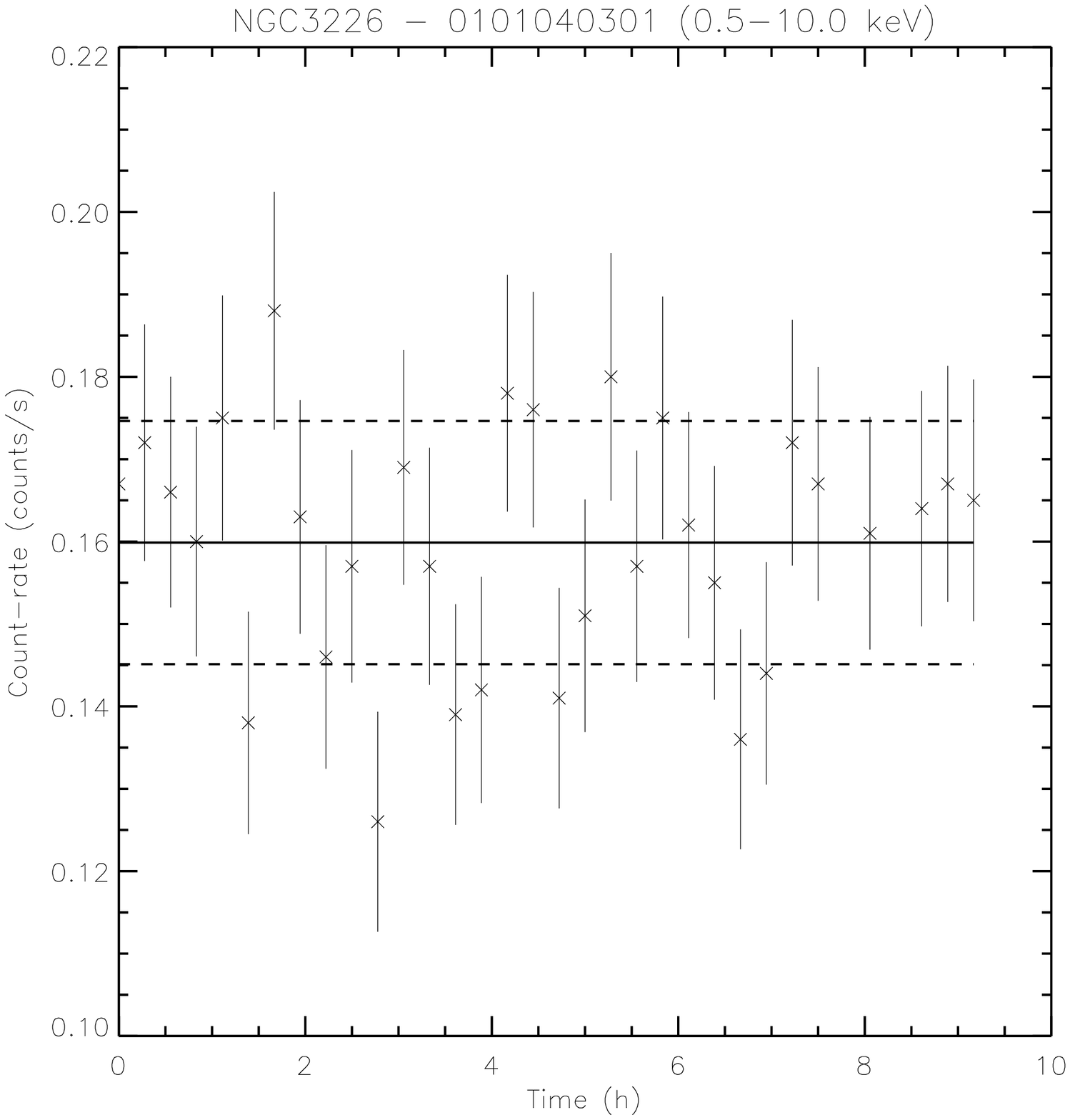} \includegraphics[width=0.4\textwidth]{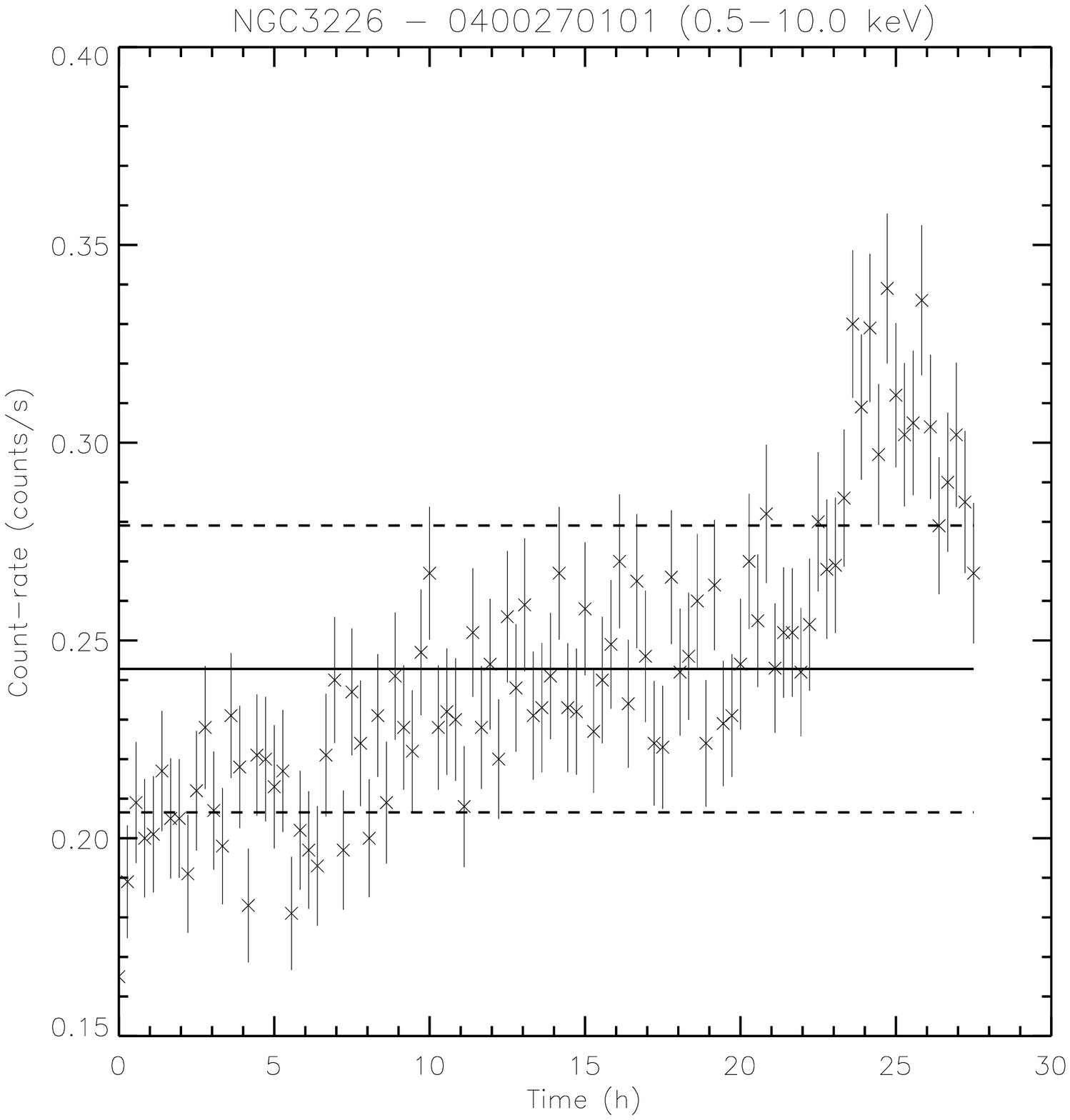}

\includegraphics[width=0.4\textwidth]{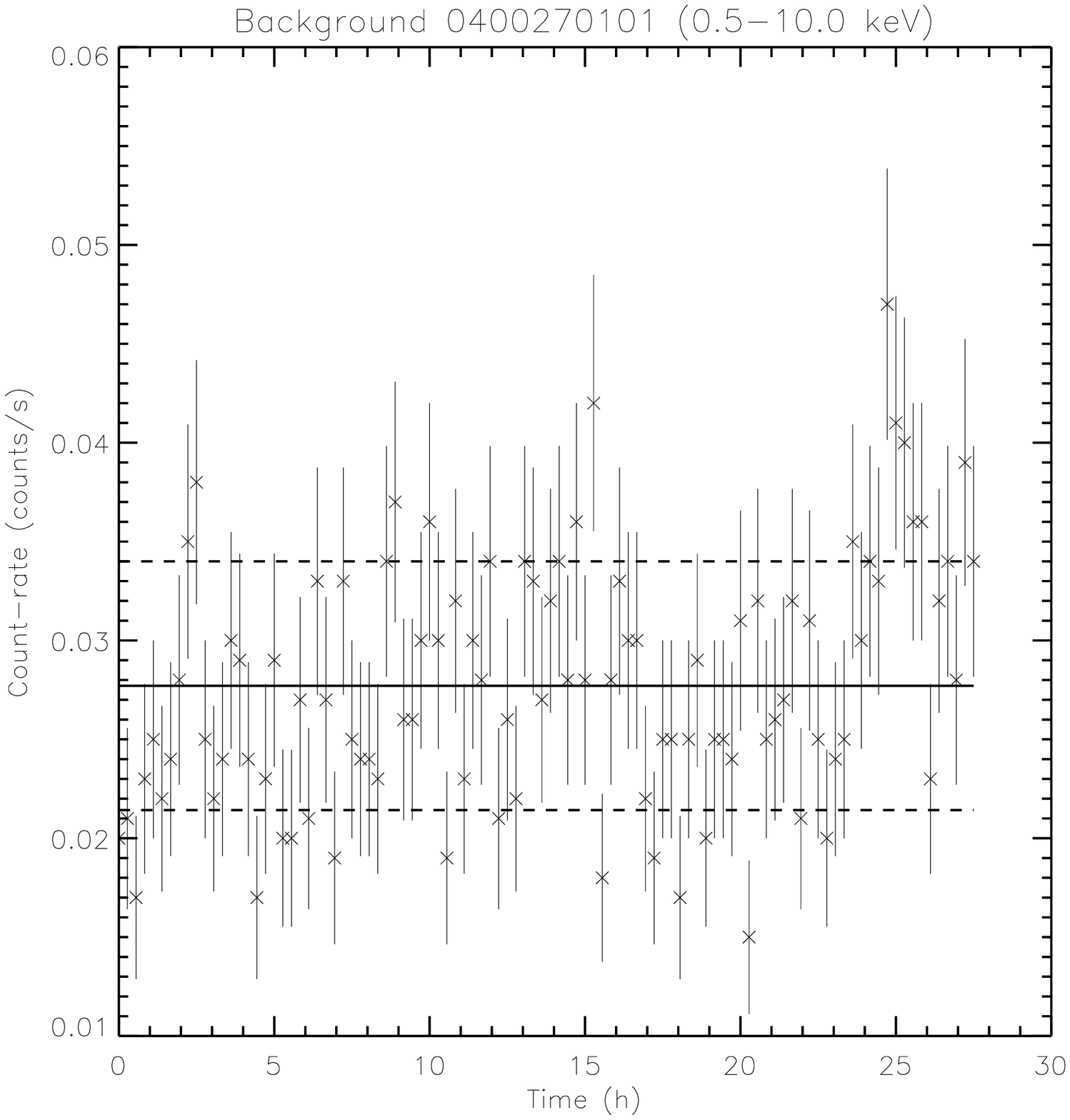} \includegraphics[width=0.4\textwidth]{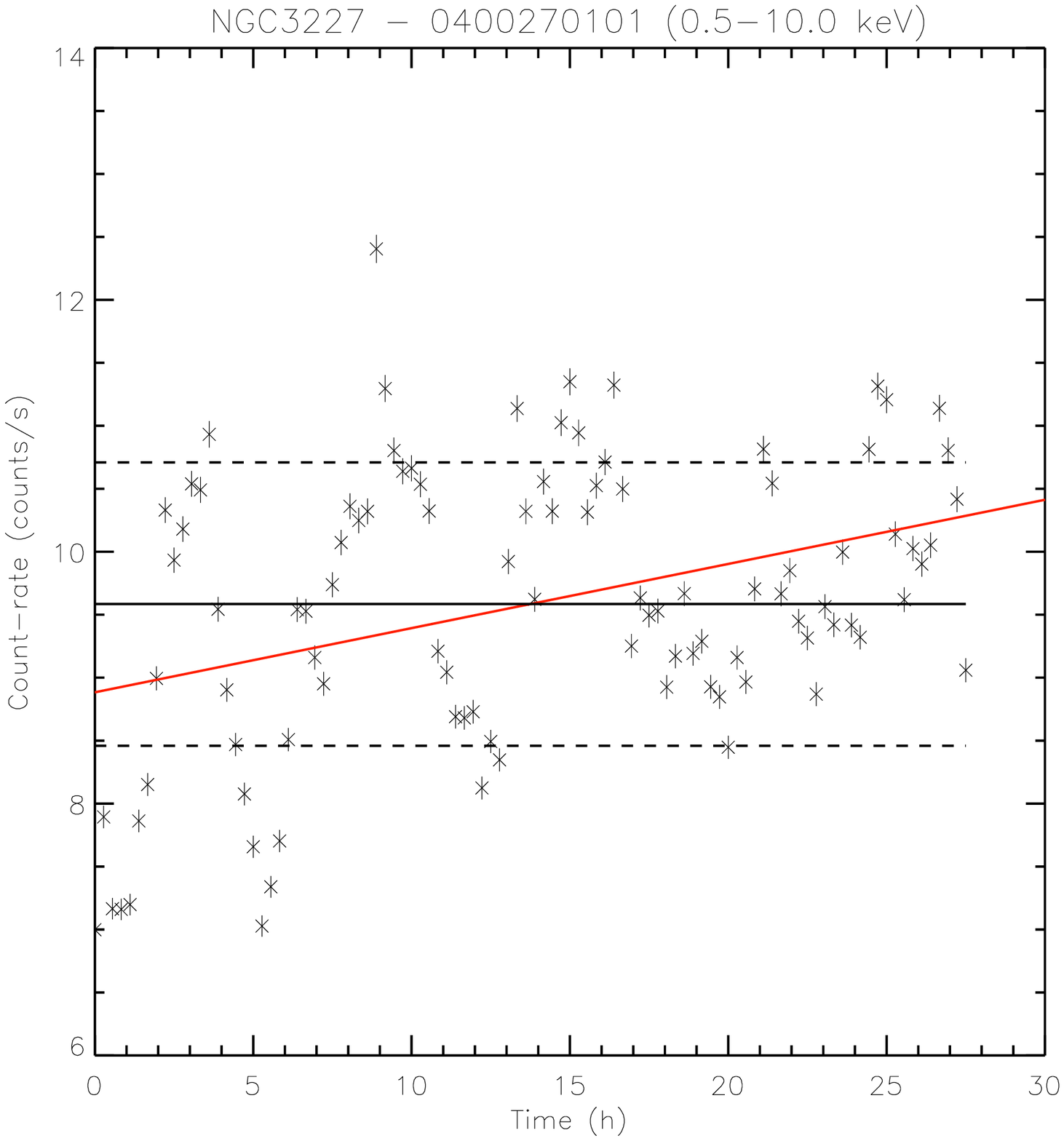}
\end{figure*}

\begin{figure*}[H]
\centering
\setcounter{figure}{1}
\caption{Cont.}
\includegraphics[width=0.57\textwidth]{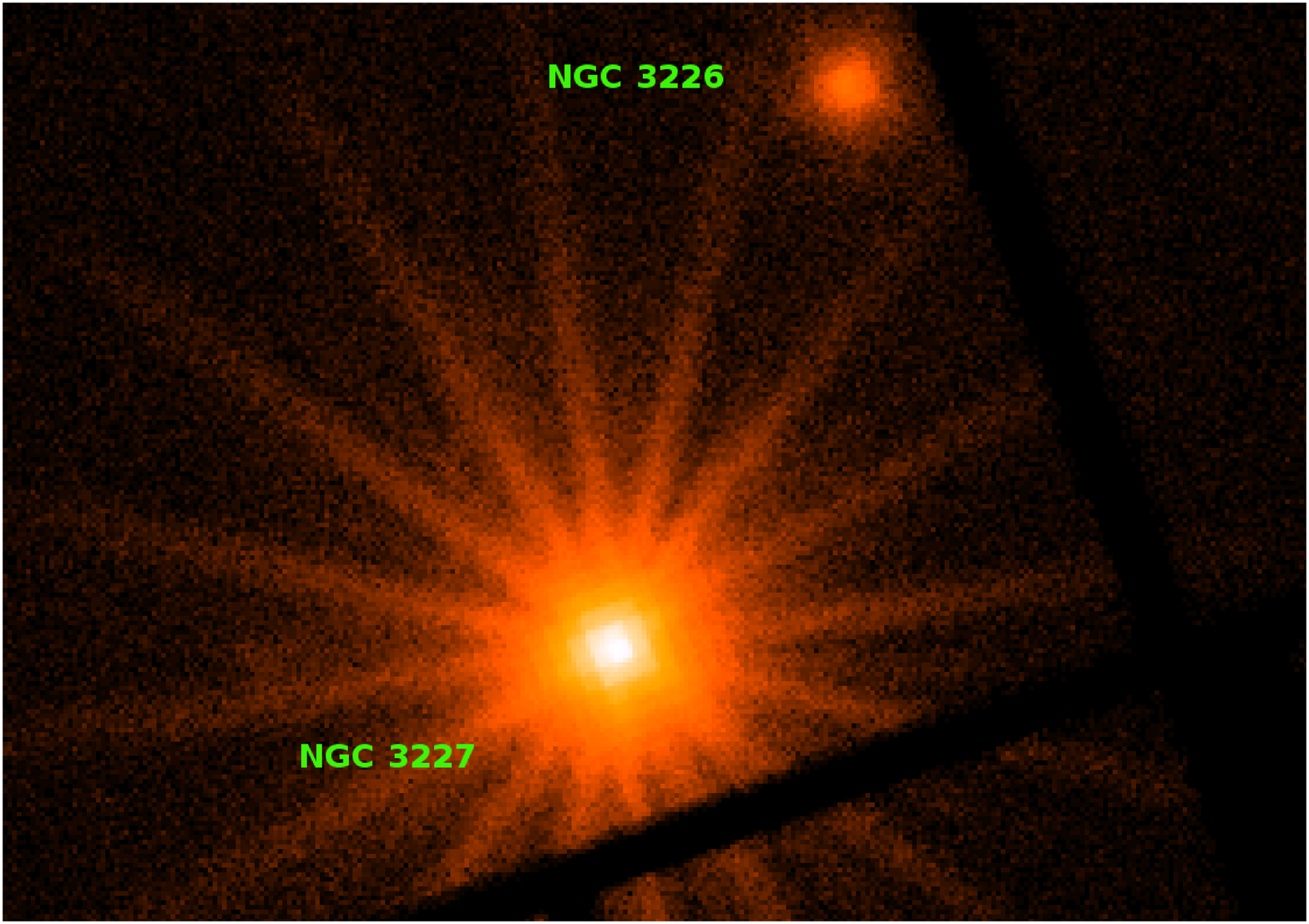}
\end{figure*}

\end{document}